\documentclass[longauth]{aa_mod}

\usepackage{mathptmx}
\usepackage{subfig}  
\usepackage{url}
\usepackage[breaklinks=true]{hyperref}
\hypersetup{citecolor=green, linkcolor=red}
\usepackage{amsmath}
\usepackage[normalem]{ulem}
\usepackage{natbib}
\bibpunct{(}{)}{;}{a}{}{,} 

\usepackage{fixltx2e}

\def\gtorder{\mathrel{\raise.3ex\hbox{$>$}\mkern-14mu
             \lower0.6ex\hbox{$\sim$}}}
\def\ltorder{\mathrel{\raise.3ex\hbox{$<$}\mkern-14mu
             \lower0.6ex\hbox{$\sim$}}}

\newcommand{\swtool}[1]{\textit{#1}}
\newcommand{\swpkg}[1]{\textsc{#1}}
\newcommand{\batblip}{BAT Blip}
\newcommand{\xray}{\mbox{X-ray}}

\newcommand{\Fermi}{\emph{Fermi}\xspace}

\usepackage{txfonts}

\begin{document}

\title{Multiwavelength follow-up of a rare IceCube neutrino multiplet}

\author{
{\tiny
{\bf IceCube:}
M.~G.~Aartsen\inst{\ref{Adelaide}}
\and M.~Ackermann\inst{\ref{Zeuthen}}
\and J.~Adams\inst{\ref{Christchurch}}
\and J.~A.~Aguilar\inst{\ref{BrusselsLibre}}
\and M.~Ahlers\inst{\ref{MadisonPAC}}
\and M.~Ahrens\inst{\ref{StockholmOKC}}
\and I.~Al~Samarai\inst{\ref{Geneva}}
\and D.~Altmann\inst{\ref{Erlangen}}
\and K.~Andeen\inst{\ref{Marquette}}
\and T.~Anderson\inst{\ref{PennPhys}}
\and I.~Ansseau\inst{\ref{BrusselsLibre}}
\and G.~Anton\inst{\ref{Erlangen}}
\and M.~Archinger\inst{\ref{Mainz}}
\and C.~Arg\"uelles\inst{\ref{MIT}}
\and J.~Auffenberg\inst{\ref{Aachen}}
\and S.~Axani\inst{\ref{MIT}}
\and X.~Bai\inst{\ref{SouthDakota}}
\and S.~W.~Barwick\inst{\ref{Irvine}}
\and V.~Baum\inst{\ref{Mainz}}
\and R.~Bay\inst{\ref{Berkeley}}
\and J.~J.~Beatty\inst{\ref{Ohio},\ref{OhioAstro}}
\and J.~Becker~Tjus\inst{\ref{Bochum}}
\and K.-H.~Becker\inst{\ref{Wuppertal}}
\and S.~BenZvi\inst{\ref{Rochester}}
\and D.~Berley\inst{\ref{Maryland}}
\and E.~Bernardini\inst{\ref{Zeuthen}}
\and A.~Bernhard\inst{\ref{Munich}}
\and D.~Z.~Besson\inst{\ref{Kansas}}
\and G.~Binder\inst{\ref{LBNL},\ref{Berkeley}}
\and D.~Bindig\inst{\ref{Wuppertal}}
\and E.~Blaufuss\inst{\ref{Maryland}}
\and S.~Blot\inst{\ref{Zeuthen}}
\and C.~Bohm\inst{\ref{StockholmOKC}}
\and M.~B\"orner\inst{\ref{Dortmund}}
\and F.~Bos\inst{\ref{Bochum}}
\and D.~Bose\inst{\ref{SKKU}}
\and S.~B\"oser\inst{\ref{Mainz}}
\and O.~Botner\inst{\ref{Uppsala}}
\and J.~Braun\inst{\ref{MadisonPAC}}
\and L.~Brayeur\inst{\ref{BrusselsVrije}}
\and H.-P.~Bretz\inst{\ref{Zeuthen}}
\and S.~Bron\inst{\ref{Geneva}}
\and A.~Burgman\inst{\ref{Uppsala}}
\and T.~Carver\inst{\ref{Geneva}}
\and M.~Casier\inst{\ref{BrusselsVrije}}
\and E.~Cheung\inst{\ref{Maryland}}
\and D.~Chirkin\inst{\ref{MadisonPAC}}
\and A.~Christov\inst{\ref{Geneva}}
\and K.~Clark\inst{\ref{Toronto}}
\and L.~Classen\inst{\ref{Munster}}
\and S.~Coenders\inst{\ref{Munich}}
\and G.~H.~Collin\inst{\ref{MIT}}
\and J.~M.~Conrad\inst{\ref{MIT}}
\and D.~F.~Cowen\inst{\ref{PennPhys},\ref{PennAstro}}
\and R.~Cross\inst{\ref{Rochester}}
\and M.~Day\inst{\ref{MadisonPAC}}
\and J.~P.~A.~M.~de~Andr\'e\inst{\ref{Michigan}}
\and C.~De~Clercq\inst{\ref{BrusselsVrije}}
\and E.~del~Pino~Rosendo\inst{\ref{Mainz}}
\and H.~Dembinski\inst{\ref{Bartol}}
\and S.~De~Ridder\inst{\ref{Gent}}
\and P.~Desiati\inst{\ref{MadisonPAC}}
\and K.~D.~de~Vries\inst{\ref{BrusselsVrije}}
\and G.~de~Wasseige\inst{\ref{BrusselsVrije}}
\and M.~de~With\inst{\ref{Berlin}}
\and T.~DeYoung\inst{\ref{Michigan}}
\and V.~di~Lorenzo\inst{\ref{Mainz}}
\and H.~Dujmovic\inst{\ref{SKKU}}
\and J.~P.~Dumm\inst{\ref{StockholmOKC}}
\and M.~Dunkman\inst{\ref{PennPhys}}
\and B.~Eberhardt\inst{\ref{Mainz}}
\and T.~Ehrhardt\inst{\ref{Mainz}}
\and B.~Eichmann\inst{\ref{Bochum}}
\and P.~Eller\inst{\ref{PennPhys}}
\and S.~Euler\inst{\ref{Uppsala}}
\and P.~A.~Evenson\inst{\ref{Bartol}}
\and S.~Fahey\inst{\ref{MadisonPAC}}
\and A.~R.~Fazely\inst{\ref{Southern}}
\and J.~Feintzeig\inst{\ref{MadisonPAC}}
\and J.~Felde\inst{\ref{Maryland}}
\and K.~Filimonov\inst{\ref{Berkeley}}
\and C.~Finley\inst{\ref{StockholmOKC}}
\and S.~Flis\inst{\ref{StockholmOKC}}
\and C.-C.~F\"osig\inst{\ref{Mainz}}
\and A.~Franckowiak\inst{\ref{Zeuthen}}
\and E.~Friedman\inst{\ref{Maryland}}
\and T.~Fuchs\inst{\ref{Dortmund}}
\and T.~K.~Gaisser\inst{\ref{Bartol}}
\and J.~Gallagher\inst{\ref{MadisonAstro}}
\and L.~Gerhardt\inst{\ref{LBNL},\ref{Berkeley}}
\and K.~Ghorbani\inst{\ref{MadisonPAC}}
\and W.~Giang\inst{\ref{Edmonton}}
\and L.~Gladstone\inst{\ref{MadisonPAC}}
\and T.~Glauch\inst{\ref{Aachen}}
\and T.~Gl\"usenkamp\inst{\ref{Erlangen}}
\and A.~Goldschmidt\inst{\ref{LBNL}}
\and J.~G.~Gonzalez\inst{\ref{Bartol}}
\and D.~Grant\inst{\ref{Edmonton}}
\and Z.~Griffith\inst{\ref{MadisonPAC}}
\and C.~Haack\inst{\ref{Aachen}}
\and A.~Hallgren\inst{\ref{Uppsala}}
\and F.~Halzen\inst{\ref{MadisonPAC}}
\and E.~Hansen\inst{\ref{Copenhagen}}
\and T.~Hansmann\inst{\ref{Aachen}}
\and K.~Hanson\inst{\ref{MadisonPAC}}
\and D.~Hebecker\inst{\ref{Berlin}}
\and D.~Heereman\inst{\ref{BrusselsLibre}}
\and K.~Helbing\inst{\ref{Wuppertal}}
\and R.~Hellauer\inst{\ref{Maryland}}
\and S.~Hickford\inst{\ref{Wuppertal}}
\and J.~Hignight\inst{\ref{Michigan}}
\and G.~C.~Hill\inst{\ref{Adelaide}}
\and K.~D.~Hoffman\inst{\ref{Maryland}}
\and R.~Hoffmann\inst{\ref{Wuppertal}}
\and K.~Hoshina\inst{\ref{MadisonPAC},\ref{a}}
\and F.~Huang\inst{\ref{PennPhys}}
\and M.~Huber\inst{\ref{Munich}}
\and K.~Hultqvist\inst{\ref{StockholmOKC}}
\and S.~In\inst{\ref{SKKU}}
\and A.~Ishihara\inst{\ref{Chiba}}
\and E.~Jacobi\inst{\ref{Zeuthen}}
\and G.~S.~Japaridze\inst{\ref{Atlanta}}
\and M.~Jeong\inst{\ref{SKKU}}
\and K.~Jero\inst{\ref{MadisonPAC}}
\and B.~J.~P.~Jones\inst{\ref{MIT}}
\and W.~Kang\inst{\ref{SKKU}}
\and A.~Kappes\inst{\ref{Munster}}
\and T.~Karg\inst{\ref{Zeuthen}}
\and A.~Karle\inst{\ref{MadisonPAC}}
\and U.~Katz\inst{\ref{Erlangen}}
\and M.~Kauer\inst{\ref{MadisonPAC}}
\and A.~Keivani\inst{\ref{PennPhys}}
\and J.~L.~Kelley\inst{\ref{MadisonPAC}}
\and A.~Kheirandish\inst{\ref{MadisonPAC}}
\and J.~Kim\inst{\ref{SKKU}}
\and M.~Kim\inst{\ref{SKKU}}
\and T.~Kintscher\inst{\ref{Zeuthen}}
\and J.~Kiryluk\inst{\ref{StonyBrook}}
\and T.~Kittler\inst{\ref{Erlangen}}
\and S.~R.~Klein\inst{\ref{LBNL},\ref{Berkeley}}
\and G.~Kohnen\inst{\ref{Mons}}
\and R.~Koirala\inst{\ref{Bartol}}
\and H.~Kolanoski\inst{\ref{Berlin}}
\and R.~Konietz\inst{\ref{Aachen}}
\and L.~K\"opke\inst{\ref{Mainz}}
\and C.~Kopper\inst{\ref{Edmonton}}
\and S.~Kopper\inst{\ref{Wuppertal}}
\and D.~J.~Koskinen\inst{\ref{Copenhagen}}
\and M.~Kowalski\inst{\ref{Berlin},\ref{Zeuthen}}
\and K.~Krings\inst{\ref{Munich}}
\and M.~Kroll\inst{\ref{Bochum}}
\and G.~Kr\"uckl\inst{\ref{Mainz}}
\and C.~Kr\"uger\inst{\ref{MadisonPAC}}
\and J.~Kunnen\inst{\ref{BrusselsVrije}}
\and S.~Kunwar\inst{\ref{Zeuthen}}
\and N.~Kurahashi\inst{\ref{Drexel}}
\and T.~Kuwabara\inst{\ref{Chiba}}
\and A.~Kyriacou\inst{\ref{Adelaide}}
\and M.~Labare\inst{\ref{Gent}}
\and J.~L.~Lanfranchi\inst{\ref{PennPhys}}
\and M.~J.~Larson\inst{\ref{Copenhagen}}
\and F.~Lauber\inst{\ref{Wuppertal}}
\and M.~Lesiak-Bzdak\inst{\ref{StonyBrook}}
\and M.~Leuermann\inst{\ref{Aachen}}
\and L.~Lu\inst{\ref{Chiba}}
\and J.~L\"unemann\inst{\ref{BrusselsVrije}}
\and J.~Madsen\inst{\ref{RiverFalls}}
\and G.~Maggi\inst{\ref{BrusselsVrije}}
\and K.~B.~M.~Mahn\inst{\ref{Michigan}}
\and S.~Mancina\inst{\ref{MadisonPAC}}
\and M.~Mandelartz\inst{\ref{Bochum}}
\and R.~Maruyama\inst{\ref{Yale}}
\and K.~Mase\inst{\ref{Chiba}}
\and R.~Maunu\inst{\ref{Maryland}}
\and F.~McNally\inst{\ref{MadisonPAC}}
\and K.~Meagher\inst{\ref{BrusselsLibre}}
\and M.~Medici\inst{\ref{Copenhagen}}
\and M.~Meier\inst{\ref{Dortmund}}
\and T.~Menne\inst{\ref{Dortmund}}
\and G.~Merino\inst{\ref{MadisonPAC}}
\and T.~Meures\inst{\ref{BrusselsLibre}}
\and S.~Miarecki\inst{\ref{LBNL},\ref{Berkeley}}
\and J.~Micallef\inst{\ref{Michigan}}
\and G.~Moment\'e\inst{\ref{Mainz}}
\and T.~Montaruli\inst{\ref{Geneva}}
\and M.~Moulai\inst{\ref{MIT}}
\and R.~Nahnhauer\inst{\ref{Zeuthen}}
\and U.~Naumann\inst{\ref{Wuppertal}}
\and G.~Neer\inst{\ref{Michigan}}
\and H.~Niederhausen\inst{\ref{StonyBrook}}
\and S.~C.~Nowicki\inst{\ref{Edmonton}}
\and D.~R.~Nygren\inst{\ref{LBNL}}
\and A.~Obertacke~Pollmann\inst{\ref{Wuppertal}}
\and A.~Olivas\inst{\ref{Maryland}}
\and A.~O'Murchadha\inst{\ref{BrusselsLibre}}
\and T.~Palczewski\inst{\ref{LBNL},\ref{Berkeley}}
\and H.~Pandya\inst{\ref{Bartol}}
\and D.~V.~Pankova\inst{\ref{PennPhys}}
\and P.~Peiffer\inst{\ref{Mainz}}
\and \"O.~Penek\inst{\ref{Aachen}}
\and J.~A.~Pepper\inst{\ref{Alabama}}
\and C.~P\'erez~de~los~Heros\inst{\ref{Uppsala}}
\and D.~Pieloth\inst{\ref{Dortmund}}
\and E.~Pinat\inst{\ref{BrusselsLibre}}
\and P.~B.~Price\inst{\ref{Berkeley}}
\and G.~T.~Przybylski\inst{\ref{LBNL}}
\and M.~Quinnan\inst{\ref{PennPhys}}
\and C.~Raab\inst{\ref{BrusselsLibre}}
\and L.~R\"adel\inst{\ref{Aachen}}
\and M.~Rameez\inst{\ref{Copenhagen}}
\and K.~Rawlins\inst{\ref{Anchorage}}
\and R.~Reimann\inst{\ref{Aachen}}
\and B.~Relethford\inst{\ref{Drexel}}
\and M.~Relich\inst{\ref{Chiba}}
\and E.~Resconi\inst{\ref{Munich}}
\and W.~Rhode\inst{\ref{Dortmund}}
\and M.~Richman\inst{\ref{Drexel}}
\and B.~Riedel\inst{\ref{Edmonton}}
\and S.~Robertson\inst{\ref{Adelaide}}
\and M.~Rongen\inst{\ref{Aachen}}
\and C.~Rott\inst{\ref{SKKU}}
\and T.~Ruhe\inst{\ref{Dortmund}}
\and D.~Ryckbosch\inst{\ref{Gent}}
\and D.~Rysewyk\inst{\ref{Michigan}}
\and L.~Sabbatini\inst{\ref{MadisonPAC}}
\and S.~E.~Sanchez~Herrera\inst{\ref{Edmonton}}
\and A.~Sandrock\inst{\ref{Dortmund}}
\and J.~Sandroos\inst{\ref{Mainz}}
\and S.~Sarkar\inst{\ref{Copenhagen},\ref{Oxford}}
\and K.~Satalecka\inst{\ref{Zeuthen}}
\and P.~Schlunder\inst{\ref{Dortmund}}
\and T.~Schmidt\inst{\ref{Maryland}}
\and S.~Schoenen\inst{\ref{Aachen}}
\and S.~Sch\"oneberg\inst{\ref{Bochum}}
\and L.~Schumacher\inst{\ref{Aachen}}
\and D.~Seckel\inst{\ref{Bartol}}
\and S.~Seunarine\inst{\ref{RiverFalls}}
\and D.~Soldin\inst{\ref{Wuppertal}}
\and M.~Song\inst{\ref{Maryland}}
\and G.~M.~Spiczak\inst{\ref{RiverFalls}}
\and C.~Spiering\inst{\ref{Zeuthen}}
\and J.~Stachurska\inst{\ref{Zeuthen}}
\and T.~Stanev\inst{\ref{Bartol}}
\and A.~Stasik\inst{\ref{Zeuthen}}
\and J.~Stettner\inst{\ref{Aachen}}
\and A.~Steuer\inst{\ref{Mainz}}
\and T.~Stezelberger\inst{\ref{LBNL}}
\and R.~G.~Stokstad\inst{\ref{LBNL}}
\and A.~St\"o{\ss}l\inst{\ref{Chiba}}
\and R.~Str\"om\inst{\ref{Uppsala}}
\and N.~L.~Strotjohann\inst{\ref{Zeuthen}}
\and G.~W.~Sullivan\inst{\ref{Maryland}}
\and M.~Sutherland\inst{\ref{Ohio}}
\and H.~Taavola\inst{\ref{Uppsala}}
\and I.~Taboada\inst{\ref{Georgia}}
\and J.~Tatar\inst{\ref{LBNL},\ref{Berkeley}}
\and F.~Tenholt\inst{\ref{Bochum}}
\and S.~Ter-Antonyan\inst{\ref{Southern}}
\and A.~Terliuk\inst{\ref{Zeuthen}}
\and G.~Te{\v{s}}i\'c\inst{\ref{PennPhys}}
\and S.~Tilav\inst{\ref{Bartol}}
\and P.~A.~Toale\inst{\ref{Alabama}}
\and M.~N.~Tobin\inst{\ref{MadisonPAC}}
\and S.~Toscano\inst{\ref{BrusselsVrije}}
\and D.~Tosi\inst{\ref{MadisonPAC}}
\and M.~Tselengidou\inst{\ref{Erlangen}}
\and C.~F.~Tung\inst{\ref{Georgia}}
\and A.~Turcati\inst{\ref{Munich}}
\and E.~Unger\inst{\ref{Uppsala}}
\and M.~Usner\inst{\ref{Zeuthen}}
\and J.~Vandenbroucke\inst{\ref{MadisonPAC}}
\and N.~van~Eijndhoven\inst{\ref{BrusselsVrije}}
\and S.~Vanheule\inst{\ref{Gent}}
\and M.~van~Rossem\inst{\ref{MadisonPAC}}
\and J.~van~Santen\inst{\ref{Zeuthen}}
\and M.~Vehring\inst{\ref{Aachen}}
\and M.~Voge\inst{\ref{Bonn}}
\and E.~Vogel\inst{\ref{Aachen}}
\and M.~Vraeghe\inst{\ref{Gent}}
\and C.~Walck\inst{\ref{StockholmOKC}}
\and A.~Wallace\inst{\ref{Adelaide}}
\and M.~Wallraff\inst{\ref{Aachen}}
\and N.~Wandkowsky\inst{\ref{MadisonPAC}}
\and A.~Waza\inst{\ref{Aachen}}
\and Ch.~Weaver\inst{\ref{Edmonton}}
\and M.~J.~Weiss\inst{\ref{PennPhys}}
\and C.~Wendt\inst{\ref{MadisonPAC}}
\and S.~Westerhoff\inst{\ref{MadisonPAC}}
\and B.~J.~Whelan\inst{\ref{Adelaide}}
\and S.~Wickmann\inst{\ref{Aachen}}
\and K.~Wiebe\inst{\ref{Mainz}}
\and C.~H.~Wiebusch\inst{\ref{Aachen}}
\and L.~Wille\inst{\ref{MadisonPAC}}
\and D.~R.~Williams\inst{\ref{Alabama}}
\and L.~Wills\inst{\ref{Drexel}}
\and M.~Wolf\inst{\ref{StockholmOKC}}
\and T.~R.~Wood\inst{\ref{Edmonton}}
\and E.~Woolsey\inst{\ref{Edmonton}}
\and K.~Woschnagg\inst{\ref{Berkeley}}
\and D.~L.~Xu\inst{\ref{MadisonPAC}}
\and X.~W.~Xu\inst{\ref{Southern}}
\and Y.~Xu\inst{\ref{StonyBrook}}
\and J.~P.~Yanez\inst{\ref{Edmonton}}
\and G.~Yodh\inst{\ref{Irvine}}
\and S.~Yoshida\inst{\ref{Chiba}}
\and M.~Zoll\inst{\ref{StockholmOKC}}\\ \smallskip
{\bf ASAS-SN:}
K.~Z.~Stanek\inst{\ref{OhioAstro}, \ref{Ohio}}
\and B.~J.~Shappee\inst{\ref{pasadena}, \ref{hubble}}
\and C.~S.~Kochanek\inst{\ref{OhioAstro}, \ref{Ohio}}
\and T.~W.-S.~Holoien\inst{\ref{OhioAstro}, \ref{Ohio}}
\and J.~L.~Prieto\inst{\ref{santiago1}, \ref{santiago2}}\\ \smallskip
{\bf The Astrophysical Multimessenger Observatory Network:}
D.~B.~Fox\inst{\ref{PennAstro},\ref{PennParticle},\ref{PennCosmology}}
\and J.~J.~DeLaunay\inst{\ref{PennPhys},\ref{PennParticle}}
\and C.~F.~Turley\inst{\ref{PennPhys},\ref{PennParticle}}
\and S.~D.~Barthelmy\inst{\ref{goddard}}
\and A.~Y.~Lien\inst{\ref{goddard}}
\and P.~M\'{e}sz\'{a}ros\inst{\ref{PennPhys},\ref{PennAstro},\ref{PennParticle},\ref{PennCosmology}}
\and K.~Murase\inst{\ref{PennPhys},\ref{PennAstro},\ref{PennParticle},\ref{PennCosmology}}\\ \smallskip
{\bf Fermi:}
D.~Kocevski\inst{\ref{goddard}}
\and R.~Buehler\inst{\ref{Zeuthen}} 
\and M.~Giomi\inst{\ref{Zeuthen}}
\and J.~L.~Racusin\inst{\ref{goddard}} \\ \smallskip
{\bf HAWC:}
{A.~Albert}\inst{\ref{LANL}}
\and {R.~Alfaro}\inst{\ref{IF-UNAM}}
\and {C.~Alvarez}\inst{\ref{UNACH}}
\and {J.~D.~\'Alvarez}\inst{\ref{UMSNH}}
\and {R.~Arceo}\inst{\ref{UNACH}}
\and {J.~C.~Arteaga-Vel\'azquez}\inst{\ref{UMSNH}}
\and {H.~A.~Ayala~Solares}\inst{\ref{MTU}}
\and {A.~S.~Barber}\inst{\ref{Utah}}			
\and {N.~Baustista-Elivar}\inst{\ref{UPP}}		
\and {A.~Becerril}\inst{\ref{IF-UNAM}}			
\and {E.~Belmont-Moreno}\inst{\ref{IF-UNAM}}
\and {A.~Bernal}\inst{\ref{IA-UNAM}}
\and {C.~Brisbois}\inst{\ref{MTU}}
\and {K.~S.~Caballero-Mora}\inst{\ref{UNACH}}
\and {T.~Capistr\'an}\inst{\ref{INAOE}}
\and {A.~Carrami\~nana}\inst{\ref{INAOE}}
\and {S.~Casanova}\inst{\ref{IFJ-PAN}}
\and {M.~Castillo}\inst{\ref{UMSNH}}
\and {U.~Cotti}\inst{\ref{UMSNH}}
\and {S.~Couti\~no~de~Le\'on}\inst{\ref{INAOE}}
\and {E.~de~la~Fuente}\inst{\ref{UdG}}
\and {C.~De~Le\'on}\inst{\ref{FCFM-BUAP}}
\and {R.~Diaz~Hernandez}\inst{\ref{INAOE}}
\and {J.~C.~D\'iaz-V\'elez}\inst{\ref{UdG},\ref{MadisonPAC}}
\and {B.~L.~Dingus}\inst{\ref{LANL}}
\and {M.~A.~DuVernois}\inst{\ref{MadisonPAC}}
\and {R.~W.~Ellsworth}\inst{\ref{GMU}}			
\and {K.~Engel}\inst{\ref{Maryland}}			
\and {D.~W.~Fiorino}\inst{\ref{Maryland}}
\and {N.~Fraija}\inst{\ref{IA-UNAM}}
\and {J.~A.~Garc\'ia-Gonz\'alez}\inst{\ref{IF-UNAM}}
\and {M.~Gerhardt}\inst{\ref{MTU}}
\and {A.~Gonz\'alez~Mu\~noz}\inst{\ref{IF-UNAM}}
\and {M.~M.~Gonz\'alez}\inst{\ref{IA-UNAM}}
\and {J.~A.~Goodman}\inst{\ref{Maryland}}
\and {Z.~Hampel-Arias}\inst{\ref{MadisonPAC}}
\and {J.~P.~Harding}\inst{\ref{LANL}}
\and {S.~Hernandez}\inst{\ref{IF-UNAM}}
\and {C.~M.~Hui}\inst{\ref{MSFC}}
\and {P.~H\"untemeyer}\inst{\ref{MTU}}
\and {A.~Iriarte}\inst{\ref{IA-UNAM}}
\and {A.~Jardin-Blicq}\inst{\ref{MPIK}}
\and {V.~Joshi}\inst{\ref{MPIK}}
\and {S.~Kaufmann}\inst{\ref{UNACH}}
\and {A.~Lara}\inst{\ref{IGeof-UNAM}}
\and {R.~J.~Lauer}\inst{\ref{UNM}}
\and {W.~H.~Lee}\inst{\ref{IA-UNAM}}
\and {D.~Lennarz}\inst{\ref{Atlanta}}
\and {H.~Le\'on~Vargas}\inst{\ref{IF-UNAM}}
\and {J.~T.~Linnemann}\inst{\ref{Michigan}}
\and {G.~Luis~Raya}\inst{\ref{UPP}}
\and {R.~Luna-Garc\'ia}\inst{\ref{CIC-IPN}}
\and {R.~L\'opez-Coto}\inst{\ref{MPIK}}
\and {K.~Malone}\inst{\ref{PennPhys}}
\and {S.~S.~Marinelli}\inst{\ref{Michigan}}
\and {O.~Martinez}\inst{\ref{FCFM-BUAP}}
\and {I.~Martinez-Castellanos}\inst{\ref{Maryland}}
\and {J.~Mart\'inez-Castro}\inst{\ref{CIC-IPN}}
\and {H.~Mart\'inez-Huerta}\inst{\ref{CINVESTAV}}
\and {J.~A.~Matthews}\inst{\ref{UNM}}
\and {P.~Miranda-Romagnoli}\inst{\ref{UAEH}}		
\and {E.~Moreno}\inst{\ref{FCFM-BUAP}}
\and {M.~Mostaf\'a}\inst{\ref{PennPhys}}
\and {L.~Nellen}\inst{\ref{ICN-UNAM}}
\and {M.~Newbold}\inst{\ref{Utah}}
\and {M.~U.~Nisa}\inst{\ref{Rochester}}
\and {R.~Noriega-Papaqui}\inst{\ref{UAEH}}		
\and {R.~Pelayo}\inst{\ref{CIC-IPN}}
\and {J.~Pretz}\inst{\ref{PennParticle}}
\and {E.~G.~P\'erez-P\'erez}\inst{\ref{UPP}}
\and {Z.~Ren}\inst{\ref{UNM}}
\and {C.~D.~Rho}\inst{\ref{Rochester}}
\and {C.~Rivi\`ere}\inst{\ref{Maryland}}
\and {D.~Rosa-Gonz\'alez}\inst{\ref{INAOE}}
\and {M.~Rosenberg}\inst{\ref{PennParticle}}
\and {F.~Salesa~Greus}\inst{\ref{IFJ-PAN}}
\and {A.~Sandoval}\inst{\ref{IF-UNAM}}
\and {M.~Schneider}\inst{\ref{SantaCruz}}
\and {H.~Schoorlemmer}\inst{\ref{MPIK}}
\and {G.~Sinnis}\inst{\ref{LANL}}
\and {A.~J.~Smith}\inst{\ref{Maryland}}
\and {R.~W.~Springer}\inst{\ref{Utah}}
\and {P.~Surajbali}\inst{\ref{MPIK}}
\and {O.~Tibolla}\inst{\ref{UNACH}}
\and {K.~Tollefson}\inst{\ref{Michigan}}
\and {I.~Torres}\inst{\ref{INAOE}}
\and {T.~N.~Ukwatta}\inst{\ref{LANL}}			
\and {L.~Villase\~nor}\inst{\ref{UMSNH}}
\and {T.~Weisgarber}\inst{\ref{MadisonPAC}}
\and {I.~G.~Wisher}\inst{\ref{MadisonPAC}}
\and {J.~Wood}\inst{\ref{MadisonPAC}}
\and {T.~Yapici}\inst{\ref{Michigan}}
\and {A.~Zepeda}\inst{\ref{CINVESTAV}}
\and {H.~Zhou}\inst{\ref{LANL}}\\ \smallskip
{\bf LCO:}
I.~Arcavi\inst{\ref{lcogt},\ref{kavli},\ref{santabarbara},\ref{einstein}}
\and G.~Hosseinzadeh\inst{\ref{lcogt},\ref{santabarbara}}
\and D.~A.~Howell\inst{\ref{lcogt},\ref{santabarbara}}
\and S.~Valenti\inst{\ref{davis}}
\and C.~McCully\inst{\ref{lcogt},\ref{santabarbara}}\\ \smallskip
{\bf MASTER:}
V.~M.~Lipunov\inst{\ref{lomonosov1},\ref{lomonosov2}}
\and E.~S.~Gorbovskoy\inst{\ref{lomonosov2}}
\and N.~V.~Tiurina\inst{\ref{lomonosov2}}
\and P.~V.~Balanutsa\inst{\ref{lomonosov2}}
\and A.~S.~Kuznetsov\inst{\ref{lomonosov2}}
\and V.~G.~Kornilov\inst{\ref{lomonosov1},\ref{lomonosov2}}
\and V.~Chazov\inst{\ref{lomonosov2}}
\and N.~M.~Budnev\inst{\ref{irkutsk}}
\and O.~A.~Gress\inst{\ref{irkutsk}}
\and K.~I.~Ivanov\inst{\ref{irkutsk}}
\and A.~G.~Tlatov\inst{\ref{kislovodsk}}
\and R.~Rebolo~Lopez\inst{\ref{tenerife}}
\and M.~Serra-Ricart\inst{\ref{tenerife}}\\ \smallskip
{\bf Swift:} 
P.~A.~Evans\inst{\ref{leicester}}
\and J.~A.~Kennea\inst{\ref{PennAstro}}
\and N.~Gehrels\inst{\ref{goddard}}\thanks{Deceased: 6 Feb 2017}
\and J.~P.~Osborne\inst{\ref{leicester}}
\and K.~L.~Page\inst{\ref{leicester}}\\ \smallskip
{\bf VERITAS:}
A.~U.~Abeysekara\inst{\ref{Utah}} \and
A.~Archer\inst{\ref{WashU}} \and
W.~Benbow\inst{\ref{FLWO}} \and
R.~Bird\inst{\ref{UCLA}} \and
T.~Brantseg\inst{\ref{IowaState}} \and
V.~Bugaev\inst{\ref{WashU}} \and
J.~V~Cardenzana\inst{\ref{IowaState}} \and
M.~P.~Connolly\inst{\ref{Galway}} \and
W.~Cui\inst{\ref{Purdue},\ref{Tsinghua}} \and
A.~Falcone\inst{\ref{PennAstro}} \and
Q.~Feng\inst{\ref{McGill}} \and
J.~P.~Finley\inst{\ref{Purdue}} \and
H.~Fleischhack\inst{\ref{Zeuthen}} \and
L.~Fortson\inst{\ref{Minnesota}} \and
A.~Furniss\inst{\ref{CSUEastBay}} \and
S.~Griffin\inst{\ref{McGill},\ref{WashU}} \and
J.~Grube\inst{\ref{Stevens}} \and
M.~H\"utten\inst{\ref{Zeuthen}} \and
O.~Hervet\inst{\ref{SantaCruz}} \and
J.~Holder\inst{\ref{Bartol}} \and
G.~Hughes\inst{\ref{FLWO}} \and
T.~B.~Humensky\inst{\ref{Columbia}} \and
C.~A.~Johnson\inst{\ref{SantaCruz}} \and
P.~Kaaret\inst{\ref{UIowa}} \and
P.~Kar\inst{\ref{Utah}} \and
N.~Kelley-Hoskins\inst{\ref{Zeuthen}} \and
M.~Kertzman\inst{\ref{DePauw}} \and
M.~Krause\inst{\ref{Zeuthen}} \and
S.~Kumar\inst{\ref{Bartol}} \and
M.~J.~Lang\inst{\ref{Galway}} \and
T.~T.Y.~Lin\inst{\ref{McGill}} \and
S.~McArthur\inst{\ref{Purdue}} \and
P.~Moriarty\inst{\ref{Galway}} \and
R.~Mukherjee\inst{\ref{Barnard}} \and
D.~Nieto\inst{\ref{Columbia}} \and
R.~A.~Ong\inst{\ref{UCLA}} \and
A.~N.~Otte\inst{\ref{Georgia}} \and
M.~Pohl\inst{\ref{Potsdam},\ref{Zeuthen}} \and
A.~Popkow\inst{\ref{UCLA}} \and
E.~Pueschel\inst{\ref{UCD}} \and
J.~Quinn\inst{\ref{UCD}} \and
K.~Ragan\inst{\ref{McGill}} \and
P.~T.~Reynolds\inst{\ref{Cork}} \and
G.~T.~Richards\inst{\ref{Georgia}} \and
E.~Roache\inst{\ref{FLWO}} \and
C.~Rulten\inst{\ref{Minnesota}} \and
I.~Sadeh\inst{\ref{Zeuthen}} \and
M.~Santander\inst{\ref{Barnard}} \and
G.~H.~Sembroski\inst{\ref{Purdue}} \and
D.~Staszak\inst{\ref{UChicago}} \and
S.~Tr\'{e}panier\inst{\ref{McGill}} \and
J.~Tyler\inst{\ref{McGill}} \and
S.~P.~Wakely\inst{\ref{UChicago}} \and
A.~Weinstein\inst{\ref{IowaState}} \and
P.~Wilcox\inst{\ref{UIowa}} \and
A.~Wilhelm\inst{\ref{Potsdam},\ref{Zeuthen}} \and
D.~A.~Williams\inst{\ref{SantaCruz}} \and
B.~Zitzer\inst{\ref{McGill}}
\\ \smallskip
E.~Bellm\inst{\ref{caltech}}
\and Z.~Cano\inst{\ref{granada}}
\and A.~Gal-Yam\inst{\ref{weizmann}}
\and D.~A.~Kann\inst{\ref{tautenburg}}
\and E.~O.~Ofek\inst{\ref{weizmann}}
\and M.~Rigault\inst{\ref{Berlin}}
\and M.~Soumagnac\inst{\ref{weizmann}}
}}
\offprints{nora.linn.strotjohann@desy.de}
\date{Received 14 February 2017 /
Accepted 30 July 2017}
\institute{
{\tiny
~$^1$ III. Physikalisches Institut, RWTH Aachen University, D-52056 Aachen, Germany\label{Aachen}
\and Department of Physics, University of Adelaide, Adelaide, 5005, Australia\label{Adelaide}
\and {Dept of Physics and Astronomy, University of New Mexico, Albuquerque, NM, USA }\label{UNM}
\and Fred Lawrence Whipple Observatory, Harvard-Smithsonian Center for Astrophysics, Amado, AZ 85645, USA \label{FLWO}
\and Department of Physics and Astronomy, Iowa State University, Ames, IA 50011, USA \label{IowaState}
\and Dept.~of Physics and Astronomy, University of Alaska Anchorage, 3211 Providence Dr., Anchorage, AK 99508, USA\label{Anchorage}
\and CTSPS, Clark-Atlanta University, Atlanta, GA 30314, USA\label{Atlanta}
\and School of Physics and Center for Relativistic Astrophysics, Georgia Institute of Technology, Atlanta, GA 30332, USA\label{Georgia}
\and Dept.~of Physics, Southern University, Baton Rouge, LA 70813, USA\label{Southern}
\and Department of Physics and Center for Astrophysics, Tsinghua University, Beijing 100084, China. \label{Tsinghua}
\and Dept.~of Physics, University of California, Berkeley, CA 94720, USA\label{Berkeley}
\and Lawrence Berkeley National Laboratory, Berkeley, CA 94720, USA\label{LBNL}
\and Institut f\"ur Physik, Humboldt-Universit\"at zu Berlin, D-12489 Berlin, Germany\label{Berlin}
\and Fakult\"at f\"ur Physik \& Astronomie, Ruhr-Universit\"at Bochum, D-44780 Bochum, Germany\label{Bochum}
\and Physikalisches Institut, Universit\"at Bonn, Nussallee 12, D-53115 Bonn, Germany\label{Bonn}
\and Universit\'e Libre de Bruxelles, Science Faculty CP230, B-1050 Brussels, Belgium\label{BrusselsLibre}
\and Vrije Universiteit Brussel, Dienst ELEM, B-1050 Brussels, Belgium\label{BrusselsVrije}
\and Dept.~of Physics, Massachusetts Institute of Technology, Cambridge, MA 02139, USA\label{MIT}
\and {Instituto de Astronom\'{i}a, Universidad Nacional Autónoma de México, Ciudad de Mexico, Mexico }\label{IA-UNAM}
\and {Instituto de F\'{i}sica, Universidad Nacional Autónoma de México, Ciudad de Mexico, Mexico }\label{IF-UNAM}
\and {Instituto de Geof\'{i}sica, Universidad Nacional Autónoma de México, Ciudad de Mexico, Mexico }\label{IGeof-UNAM}
\and {Centro de Investigaci\'on en Computaci\'on, Instituto Polit\'ecnico Nacional, M\'exico City, M\'exico.}\label{CIC-IPN}
\and {Physics Department, Centro de Investigacion y de Estudios Avanzados del IPN, Mexico City, DF, Mexico }\label{CINVESTAV}
\and {Instituto de Ciencias Nucleares, Universidad Nacional Autónoma de Mexico, Ciudad de Mexico, Mexico }\label{ICN-UNAM}
\and {Universidad Autónoma de Chiapas, Tuxtla Gutiérrez, Chiapas, México}\label{UNACH}
\and Dept. of Physics and Institute for Global Prominent Research, Chiba University, Chiba 263-8522, Japan\label{Chiba}
\and Enrico Fermi Institute, University of Chicago, Chicago, IL 60637, USA \label{UChicago}
\and Dept.~of Physics and Astronomy, University of Canterbury, Private Bag 4800, Christchurch, New Zealand\label{Christchurch}
\and Dept.~of Physics, University of Maryland, College Park, MD 20742, USA\label{Maryland}
\and Dept.~of Physics and Center for Cosmology and Astro-Particle Physics, Ohio State University, Columbus, OH 43210, USA\label{Ohio}
\and Dept.~of Astronomy, Ohio State University, Columbus, OH 43210, USA\label{OhioAstro}
\and Niels Bohr Institute, University of Copenhagen, DK-2100 Copenhagen, Denmark\label{Copenhagen}
\and Department of Physical Sciences, Cork Institute of Technology, Bishopstown, Cork, Ireland \label{Cork}
\and Department of Physics, University of California, 1 Shields Ave, Davis, CA 95616-5270, USA\label{davis}
\and Dept.~of Physics, TU Dortmund University, D-44221 Dortmund, Germany\label{Dortmund}
\and School of Physics, University College Dublin, Belfield, Dublin 4, Ireland \label{UCD}
\and Dept.~of Physics and Astronomy, Michigan State University, East Lansing, MI 48824, USA\label{Michigan}
\and Dept.~of Physics, University of Alberta, Edmonton, Alberta, Canada T6G 2E1\label{Edmonton}
\and Einstein Fellow\label{einstein}
\and Erlangen Centre for Astroparticle Physics, Friedrich-Alexander-Universit\"at Erlangen-N\"urnberg, D-91058 Erlangen, Germany\label{Erlangen}
\and School of Physics, Astronomy, and Computational Sciences, George Mason University, Fairfax, VA, USA \label{GMU}
\and School of Physics, National University of Ireland Galway, University Road, Galway, Ireland \label{Galway}
\and D\'epartement de physique nucl\'eaire et corpusculaire, Universit\'e de Gen\`eve, CH-1211 Gen\`eve, Switzerland\label{Geneva}
\and Dept.~of Physics and Astronomy, University of Gent, B-9000 Gent, Belgium\label{Gent}
\and Las Cumbres Observatory, 6740 Cortona Dr Ste 102, Goleta, CA 93117-5575, USA\label{lcogt}
\and Instituto de Astrof\'isica de Andaluc\'ia (IAA-CSIC), Glorieta de la Astronom\'ia s/n, E-18008, Granada, Spain\label{granada}
\and NASA Goddard Space Flight Center, 8800 Greenbelt Road, Greenbelt, MD 20771, USA\label{goddard}
\newpage
\and Department of Physics and Astronomy, DePauw University, Greencastle, IN 46135-0037, USA \label{DePauw}
\and {Departamento de F\'{i}sica, Centro Universitario de Ciencias Exactase Ingenierias, Universidad de Guadalajara, Guadalajara, Mexico }\label{UdG}
\and Department of Physics, California State University - East Bay, Hayward, CA 94542, USA \label{CSUEastBay}
\and {Max-Planck Institute for Nuclear Physics, 69117 Heidelberg, Germany}\label{MPIK}
\and {Universidad Politecnica de Pachuca, Pachuca, Hgo, Mexico }\label{UPP}
\and Department of Physics, Stevens Institute of Technology, Hoboken, NJ 07030, USA \label{Stevens}
\and {Department of Physics, Michigan Technological University, Houghton, MI, USA }\label{MTU}
\and Hubble and Carnegie-Princeton Fellow\label{hubble}
\and {NASA Marshall Space Flight Center, Astrophysics Office, Huntsville, AL 35812, USA}\label{MSFC}
\and Department of Physics and Astronomy, University of Iowa, Van Allen Hall, Iowa City, IA 52242, USA \label{UIowa}
\and Institute of Applied Physics, Irkutsk State University 20, Gagarin Blvd, Irkutsk, 664003, Russia\label{irkutsk}
\and Dept.~of Physics and Astronomy, University of California, Irvine, CA 92697, USA\label{Irvine}
\and Kislovodsk Solar Station of Main Pulkovo Observatory P.O.Box 45, ul. Gagarina 100, Kislovodsk 357700, Russia\label{kislovodsk}
\and {Instytut Fizyki Jadrowej im Henryka Niewodniczanskiego Polskiej Akademii Nauk, IFJ-PAN, Krakow, Poland }\label{IFJ-PAN}
\and Dept.~of Physics and Astronomy, University of Kansas, Lawrence, KS 66045, USA\label{Kansas}
\and University of Leicester, X-ray and Observational Astronomy Research Group, Leicester Institute for Space and Earth Observation, Dept of Physics \& Astronomy, University Road, Leicester, LE1 7RH UK\label{leicester}
\and {Physics Division, Los Alamos National Laboratory, Los Alamos, NM, USA }\label{LANL}
\and Department of Physics and Astronomy, University of California, Los Angeles, CA 90095, USA \label{UCLA}
\and Dept.~of Astronomy, University of Wisconsin, Madison, WI 53706, USA\label{MadisonAstro}
\and Dept.~of Physics and Wisconsin IceCube Particle Astrophysics Center, University of Wisconsin, Madison, WI 53706, USA\label{MadisonPAC}
\and Institute of Physics, University of Mainz, Staudinger Weg 7, D-55099 Mainz, Germany\label{Mainz}
\and Department of Physics, Marquette University, Milwaukee, WI, 53201, USA\label{Marquette}
\and School of Physics and Astronomy, University of Minnesota, Minneapolis, MN 55455, USA \label{Minnesota}
\and Universit\'e de Mons, 7000 Mons, Belgium\label{Mons}
\and Physics Department, McGill University, Montreal, QC H3A 2T8, Canada \label{McGill}
\and {Universidad Michoacana de San Nicolás de Hidalgo, Morelia, Mexico }\label{UMSNH}
\and Lomonosov Moscow State University, Physics Department, Leninskie gory, GSP-1, Moscow, 119991, Russia\label{lomonosov1}
\and Lomonosov Moscow State University, Sternberg Astronomical Institute, Universitetsky Prospekt 13, Moscow, 119192, Russia\label{lomonosov2}
\and Physik-department, Technische Universit\"at M\"unchen, D-85748 Garching, Germany\label{Munich}
\and Institut f\"ur Kernphysik, Westf\"alische Wilhelms-Universit\"at M\"unster, D-48149 M\"unster, Germany\label{Munster}
\and Bartol Research Institute and Dept.~of Physics and Astronomy, University of Delaware, Newark, DE 19716, USA\label{Bartol}
\and Dept.~of Physics, Yale University, New Haven, CT 06520, USA\label{Yale}
\and Physics Department, Columbia University, New York, NY 10027, USA \label{Columbia}
\and Department of Physics and Astronomy, Barnard College, Columbia University, NY 10027, USA \label{Barnard}
\and Dept.~of Physics, University of Oxford, 1 Keble Road, Oxford OX1 3NP, UK\label{Oxford}
\and Universidad Aut\'{o}noma del Estado de Hidalgo, Pachuca, Mexico \label{UAEH}
\and Carnegie Observatories, 813 Santa Barbara Street, Pasadena, CA 91101, USA\label{pasadena}
\and Cahill Center for Astronomy and Astrophysics, California Institute of Technology, Pasadena, CA 91125\label{caltech}
\and Dept.~of Physics, Drexel University, 3141 Chestnut Street, Philadelphia, PA 19104, USA\label{Drexel}
\and Institute of Physics and Astronomy, University of Potsdam, 14476 Potsdam-Golm, Germany \label{Potsdam}
\and {Instituto Nacional de Astrof\'{i}sica, Óptica y Electrónica, Puebla, Mexico }\label{INAOE}
\and {Facultad de Ciencias F\'{i}sico Matemáticas, Benemérita Universidad Autónoma de Puebla, Puebla, Mexico }\label{FCFM-BUAP}
\and Physics Department, South Dakota School of Mines and Technology, Rapid City, SD 57701, USA\label{SouthDakota}
\and Department of Particle Physics \& Astrophysics, Weizmann Institute of Science, Rehovot 7610001, Israel\label{weizmann}
\and Dept.~of Physics, University of Wisconsin, River Falls, WI 54022, USA\label{RiverFalls}
\and Dept.~of Physics and Astronomy, University of Rochester, Rochester, NY 14627, USA\label{Rochester}
\and Department of Physics and Astronomy, University of Utah, Salt Lake City, UT 84112, USA \label{Utah}
\newpage
\and Department of Physics, University of California, Santa Barbara, CA 93106-9530, USA\label{santabarbara}
\and Kavli Institute for Theoretical Physics, University of California, Santa Barbara, CA 93106-4030, USA\label{kavli}
\and Santa Cruz Institute for Particle Physics and Department of Physics, University of California, Santa Cruz, CA 95064, USA \label{SantaCruz}
\and Nucleo de Astronomia de la Facultad de Ingenieria, Universidad Diego Portales, Av. Ejercito 441, Santiago, Chile\label{santiago1}
\and Millennium Institute of Astrophysics, Santiago, Chile\label{santiago2}
\and Department of Physics, Washington University, St. Louis, MO 63130, USA \label{WashU}
\and Oskar Klein Centre and Dept.~of Physics, Stockholm University, SE-10691 Stockholm, Sweden\label{StockholmOKC}
\and Dept.~of Physics and Astronomy, Stony Brook University, Stony Brook, NY 11794-3800, USA\label{StonyBrook}
\and Dept.~of Physics, Sungkyunkwan University, Suwon 440-746, Korea\label{SKKU}
\and Th\"uringer Landessternwarte Tautenburg, Sternwarte 5, 07778 Tautenburg, Germany\label{tautenburg}
\and Instituto de Astrof\'isica de Canarias, Observatorio del Teide C/Via Lactea, s/n. E38205, La Laguna, Tenerife, Spain\label{tenerife}
\and Earthquake Research Institute, University of Tokyo, Bunkyo, Tokyo 113-0032, Japan\label{a}
\and Dept.~of Physics, University of Toronto, Toronto, Ontario, Canada, M5S 1A7\label{Toronto} 
\and Dept.~of Physics and Astronomy, University of Alabama, Tuscaloosa, AL 35487, USA\label{Alabama} 
\and Dept.~of Astronomy and Astrophysics, Pennsylvania State University, University Park, PA 16802, USA\label{PennAstro}
\and Dept.~of Physics, Pennsylvania State University, University Park, PA 16802, USA\label{PennPhys}
\and Center for Particle \& Gravitational Astrophysics, Institute for Gravitation and the Cosmos, Pennsylvania State University, University Park, PA 16802, USA\label{PennParticle}
\and Center for Theoretical \& Observational Cosmology, Institute for Gravitation and the Cosmos, Pennsylvania State University, University Park, PA 16802, USA\label{PennCosmology}
\and Dept.~of Physics and Astronomy, Uppsala University, Box 516, S-75120 Uppsala, Sweden\label{Uppsala}
\and Department of Physics and Astronomy, Purdue University, West Lafayette, IN 47907, USA \label{Purdue}
\and Dept.~of Physics, University of Wuppertal, D-42119 Wuppertal, Germany\label{Wuppertal}
\and DESY, D-15735 Zeuthen, Germany\label{Zeuthen}
}}

\titlerunning{Follow-up of a neutrino multiplet}
\authorrunning{IceCube et al.}
\maketitle

\twocolumn[\begin{center}
{\small\sffamily\bfseries\MakeUppercase ABSTRACT}
\end{center}
{\small On February 17 2016, the IceCube real-time neutrino search identified, for the first time, three muon neutrino candidates arriving within 100\,s of one another, consistent with coming from the same point in the sky. Such a triplet is expected once every 13.7\,years as a random coincidence of background events. 
However, considering the lifetime of the follow-up program the probability of detecting at least one triplet from atmospheric background is 32\%. 
Follow-up observatories were notified in order to search for an electromagnetic counterpart. Observations were obtained by
\emph{Swift}'s X-ray telescope, by ASAS-SN, LCO and MASTER at optical wavelengths, and by VERITAS in the very-high-energy gamma-ray regime. Moreover, the \emph{Swift} BAT serendipitously observed the location 100\,s after the first neutrino was detected, and data from the \emph{Fermi} LAT and HAWC observatory were analyzed. We present details of the neutrino triplet and the follow-up observations. No likely electromagnetic counterpart was detected, and we discuss the implications of these constraints on candidate neutrino sources such as gamma-ray bursts, core-collapse supernovae and active galactic nucleus flares. This study illustrates the potential of and challenges for future follow-up campaigns.}\\
\vspace{5mm}\\
{\small\sffamily\bfseries\MakeUppercase Key words.}
{\small astroparticle physics --- neutrinos --- Gamma-ray burst: general --- supernovae: general --- Galaxies: active --- X-rays: bursts}\\
\vspace{15mm}
]

\section{Introduction}
\label{sec:introduction}

In 2013, the IceCube neutrino observatory presented the first evidence for a high-energy flux of cosmic neutrinos \citep{icecube2013,icecube2015}. While the evidence for their existence continues to mount, no explicit sources have been identified (see e.g., \citealt{icecube2014, icecube2016b}). The arrival directions of the events are distributed isotropically which likely implies that many events are of extragalactic origin.

High-energy neutrinos are produced when cosmic rays interact with ambient matter ($pp$ interactions) or photon fields ($p\gamma$ interactions). These interactions are expected to happen mainly within cosmic-ray sources where the target photon and/or matter densities are high. The detection of a neutrino source would imply that this source also accelerates cosmic rays.

Cosmic rays can be accelerated at collisionless shock fronts which are expected in a wide variety of astrophysical objects. Among those are gamma-ray bursts (GRBs; see e.g., \citealt{baerwald2015, bustamante2015, meszaros2015}), as well as the related class of low-luminosity GRBs (LLGRBs) or core-collapse supernovae (CCSNe) containing a choked jet \citep{murase2013, fraija2014, tamborra2016, senno2016}. CCSNe could in  addition produce cosmic rays when their ejecta interact with circumstellar medium emitted by the star prior to the explosion \citep{murase2011, murase2014, katz2011}. Other potential neutrino sources are active galactic nuclei (AGN; see \citealt{murase2015b}, for a review), tidal disruption events \citep{farrar2014, pfeffer2015, wang2016} and starburst galaxies \citep{tamborra2014, waxman2015}.

Thus far dedicated searches for correlations with specific source classes have not yielded a significant detection. At 90\% confidence level, GRBs can at most account for 1\% of the detected flux \citep{icecube2015c} and the contribution from blazars has been limited to at most 30\% \citep{icecube2016e}. The non-detection of any neutrino sources implies that the astrophysical flux must originate from a large number of relatively faint neutrino sources \citep{ahlers2014, kowalski2015, murase2016}.

Several coincidences of neutrino events with astrophysical sources have been reported in the literature. For example a supernova of Type IIn was detected in follow-up observations of a neutrino doublet \citep{icecube2015b}. It is however likely unrelated given the large implied neutrino luminosity. \citet{padovani2016} observe a correlation between extreme blazars and high-energy neutrino events and \citet{kadler2016} found a bright gamma-ray outburst of a blazar which was aligned with a multi PeV neutrino event. However, all of these associations have a chance-coincidence probability of a few percent and are hence not significant detections.

The most energetic neutrino candidate detected so far, with a deposited energy of 2.6\,PeV, was observed in June 2014 \citep{icecube2015d, icecube2016c}. The probability that this event was produced in the Earth's atmosphere is smaller than 1\% and the angular uncertainty is $0.27^\circ$ (at 50\% confidence) which makes it one of the best localized events observed with IceCube. However, no timely follow-up observations were triggered and a transient counterpart could have gone unnoticed. Since mid-2016, such events are identified, reconstructed, and published within minutes \citep{icecube2016d} to allow quick follow-up observations (see \citealt{blaufuss2016}, as an example for the first published event). 

In addition to the publicly announced high-energy neutrino alerts, IceCube has a real-time program that searches for multiple neutrinos from a similar direction \citep{icecube2012, icecube2016d}. When two or more muon neutrino candidates are detected within 100\,s of each other optical and X-ray observations can be triggered automatically \citep{evans2015, icecube2015b}. Real-time follow-up observations are also triggered by the ANTARES neutrino telescope, but have not lead to the discovery of an electromagnetic counterpart \citep{antares2012, antares2016}.

In February 2016, we found -- for the first time -- three events within this 100\,s time window. The detection of such a triplet from atmospheric background is not unlikely considering that the search has been running since December 2008 (compare Sect.~\ref{sec:SignficanceCalc}). However, since it is the most significant neutrino multiplet detected so far, multiwavelength follow-up observations were triggered to search for a potential electromagnetic counterpart.

In this paper we present details of the neutrino triplet and results of the follow-up observations. In Sect. \ref{sec:ofu} we introduce the follow-up program. The properties of the triplet are given in Sect. \ref{sec:alert}. The follow-up observations, covering optical wavelengths up to very-high-energy (VHE) gamma rays, are presented in Sect. \ref{sec:observations}. Finally, in Sect. \ref{sec:discussion} we draw conclusions from the various observations and discuss the sensitivity of our program to candidate neutrino source classes.

\section{The IceCube follow-up program}
\label{sec:ofu}
\subsection{The IceCube neutrino telescope}
IceCube is a cubic-kilometer-sized neutrino detector installed in the ice at the geographic South Pole between a depth of 1,450\,m and 2,450\,m \citep{icecube2016g}. An array of 5,160 digital optical modules (DOMs; \citealt{icecube2009, icecube2010b}), which are deployed in the ice, detects the Cherenkov radiation from secondary particles produced in neutrino interactions \citep{Achterberg2006}. Based on the pattern of the Cherenkov light, both the direction and energy of the neutrinos can be measured. The detector has been running in its full configuration since May 2011.

Neutrinos can interact and produce secondary particles through neutral current (NC) interactions or through charged current (CC) interactions. CC interactions induced by electron or tau neutrinos, as well as NC interactions induced by any neutrino flavor, produce localized, almost spherical light patterns inside the detector (see \citealt{icecube2013}, for examples), which makes directional reconstructions challenging. Muons produced in $\nu_\mu$ CC interactions, on the other hand, can travel up to several kilometers in the ice and emit Cherenkov light along their trajectories. These events are called tracks and their source directions can be reconstructed to better than one degree if their energy is $>1\text{\,TeV}$ \citep{icecube2016b}. Track events often extend beyond the detector volume which means that the detected energy is a lower limit on the neutrino energy. Due to their superior angular resolution, track events are preferred for neutrino astronomy and the real-time system only uses $\nu_\mu$ CC events.

\subsection{Real-time event selection}
\label{sec:event_selection}
IceCube has several real-time follow-up programs which select events and generate alerts in different ways \citep{icecube2016d}. The neutrino alert described in this paper was found by the optical follow-up program (see also \citealt{icecube2012, evans2015, icecube2015b}) which searches for short transient neutrino sources and triggers optical telescopes as well as the \emph{Swift} X-ray telescope.

Event selection starts from the online \textit{Muon Filter} selection that identifies high-quality muon tracks with a rate of about 40\,Hz. This rate is dominated by muons produced in cosmic-ray air showers. To increase the neutrino purity of the sample, more advanced and time-consuming reconstructions are required. Since computing power at the South Pole is limited, these reconstructions can only be applied to a subset of events. At the South Pole, the \textit{Online Level 2 Filter} uses the outcome of a maximum likelihood reconstruction to further reduce contamination from atmospheric muons. This reconstruction takes into account how photons propagate to the optical modules in the detector. Selection criteria are, for example, the quality of the likelihood fit and the total number of modules that detected a photon. After application of these criteria, the event rate is reduced to 5\,Hz, which is low enough to apply more sophisticated and time-consuming reconstruction algorithms (see \citealt{icecube2015b}, for a more detailed description). Based on the results of these reconstructions, the most signal-like events are selected using a multivariate classifier (see \citealt{icecube2016d}, for more details on the event selection and data transmission).

To avoid the background of atmospheric muons entering the detector from above, the follow-up program only uses events coming from below and is hence only sensitive to sources in the Northern sky. The final event rate is 3\,mHz and has a neutrino purity of $\sim\!80\%$. 
Most selected neutrino candidates are produced in atmospheric showers and out of $\sim\!10^5$ detected events per year only several hundreds are expected to be of cosmic origin (see Sect.~\ref{sec:triplet_source}). To overcome this background we restrict our search to short transient sources which are detected with several neutrinos.

\subsection{Alert generation}
\label{sec:multi_filter}

The IceCube optical follow-up program has been running since December 2008 ~\citep{icecube2012}. After selecting a stream dominated by upward-going neutrino events, it searches for coincident events. A multiplet alert is generated whenever two or more tracks arrive within 100\,s with an angular separation of less than $3.5^\circ$\footnote{While IceCube was running in the 40 and 59 string configuration the required angular separation was 4$^\circ$ (2008-12-16 to 2009-12-31).}. The length of the time window was chosen such that it covers the typical duration of a SN core-collapse and the lifetime of a jet in a GRB (compare \citealt{icecube2012}). To measure the significance of a neutrino doublet, a quality parameter is calculated using Eq.~1 in~\citet{icecube2015b}. Based on this parameter, we select the doublets that are the least likely to be chance coincidences of background events (i.e., the reconstructed directions of the two events are consistent within the errors, they are detected within a short time, and both events are well localized). Follow-up observations are triggered automatically for doublets above a fixed significance threshold. Multiplets consisting of more than two events are rare (compare Sect. \ref{sec:SignficanceCalc}) and no additional significance cut is applied.

We use simulated neutrino events following an $E^{-2.5}$ spectrum to quantify the efficiency of the multiplet selection process. If three neutrinos from a transient source pass the event selection within less than 100\,s, a triplet or two doublets with one common event are detected in 79\% of the cases. One doublet would be detected if one of the three events was separated by more than $3.5^\circ$ from the two other events, which happens with a probability of 18\%. There is a 3\% chance that the reconstructed directions of all three neutrinos would be separated by more than $3.5^\circ$ and no alert would be issued.

\section{The alert}
\label{sec:alert}
Two neutrino doublets, which have one event in common, were found on 2016-02-17 19:21:31.65 (detection time of the first neutrino event, referred to as T0 in the following; all dates are in UTC). All three events arrived within less than 100\,s. They were not automatically identified as a triplet because the second and third events were separated by $3.6^\circ$, while our cut is an at angular distance of $3.5^\circ$. However, for convenience we refer to the alert as a triplet in the following.

\begin{figure}[t]
\begin{center}
\includegraphics[width=88mm]{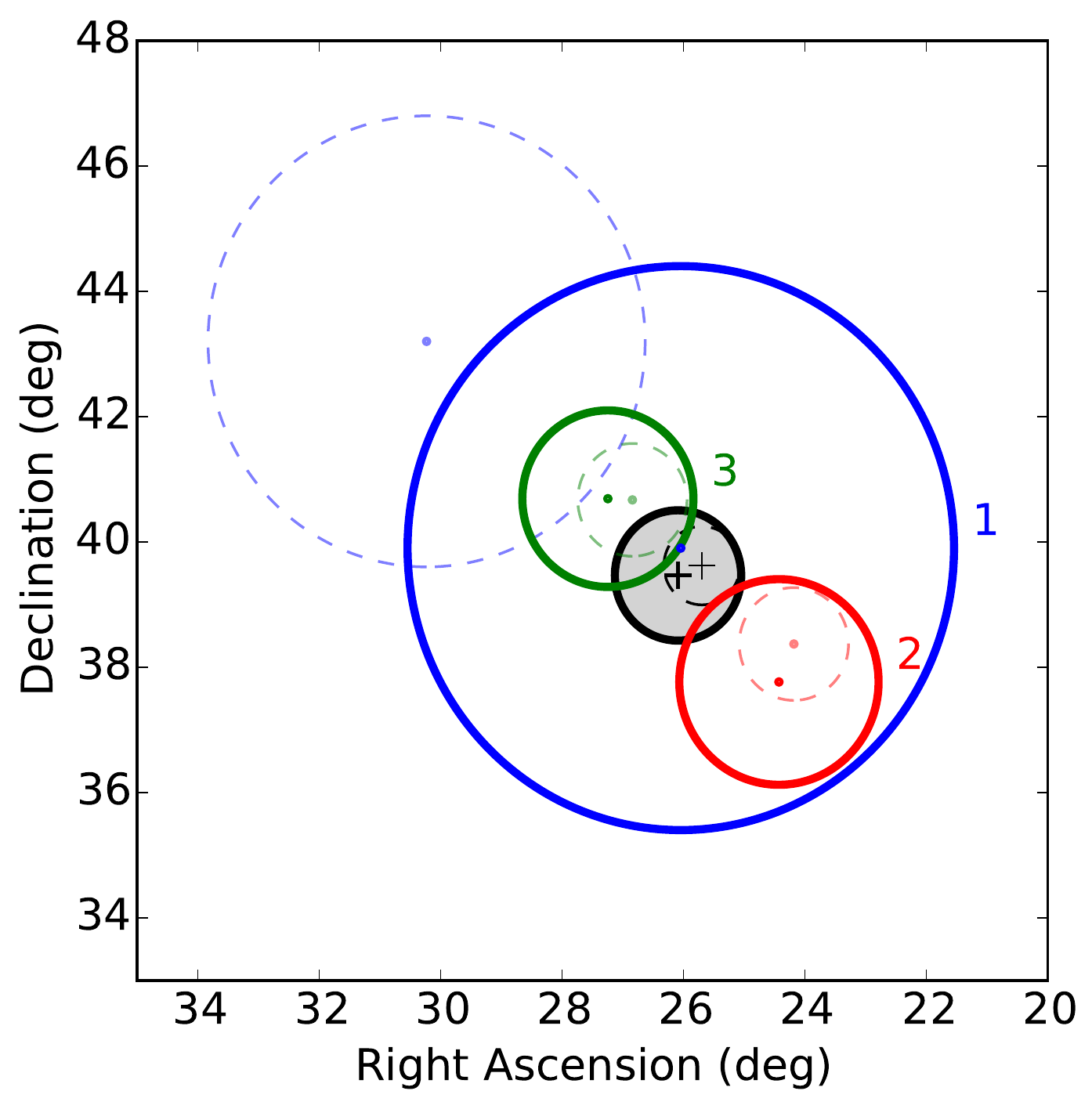}
\noindent
\caption{\small Location of the three neutrino candidates in the triplet with their 50\% error circles. The plus sign shows the combined direction and the shaded circle is the combined 50\% error circle. The solid circles show the results of the MPE reconstruction and the thin dashed circles correspond to the results of the Spline MPE reconstruction (compare Table~\ref{tab:events}). All further results are based on the MPE reconstruction which was the reconstruction used for the follow-up program until May 2016.}
\label{fig:AlertErrorCircle}
\end{center}
\end{figure}

Neither doublet passed the required significance cut for individual doublets to be automatically forwarded to the Palomar Transient Factory (PTF; \citealt{law2009, rau2009}) or to the \emph{Swift} satellite~\citep{swift2004}. More details on the individual events are given in Table~\ref{tab:events} and the projection of the events on the sky is shown in Fig. \ref{fig:AlertErrorCircle}.

\begin{table*}
{\small
\hfill{}
\caption{Details on IceCube events}
\label{tab:events}     
\begin{center}
\begin{tabular}{llllllll}        
\hline\hline       
ID & IceCube Event ID & Alert ID& Time & RA          & Dec       & Error          & Deposited energy\\
        &  &      & (s) & ($^\circ$)  & ($^\circ$)      & ($^\circ$) & (TeV)\\
\hline
1 & 62474825 & 7, 8 &  0    & 26.0 [30.2] & 39.9 [43.2] & 4.5 [3.6] & 0.26\\
2 & 62636100 & 7    & +55.4 & 24.4 [24.2] & 37.8 [38.4] & 1.6 [0.9] & 1.1\\
3 & 62729180 & 8    & +87.3 & 27.2 [26.8] & 40.7 [40.7] & 1.4 [0.9] & 0.52\\
\hline                                  
\end{tabular}
\end{center}
}
\tablefoot{The directions are the result of the reconstruction algorithm that was used in the follow-up program at the time of the alert (MPE fit),
while the values in brackets result from an alternative reconstruction algorithm with an improved ice model (Spline MPE fit).
The error on the direction is the radius of the 50\% error circle.
The last column shows an estimate of the energy deposited by the muons in the detector, which is a lower limit on the neutrino energy.
All times are relative to 2016-02-17 19:21:31.65 UTC.}
\hfill{}
\end{table*}

The combined average neutrino direction is RA~$=26.1^\circ$ and Dec~$=39.5^\circ$ J2000 with a 50\% error circle of $1.0^\circ$ and a 90\% error circle of 3.6$^\circ$. This direction corresponds to the weighted arithmetic mean position taking into account the angular uncertainties of the individual events, $\sigma_i$. The error on the combined direction is defined as $\sigma_w = (\sum_{i=1}^{N} \sigma_i^{-2})^{-1/2}$, where $N=3$ is the number of events. To estimate the 90\% error circle of the detected events we use simulated neutrino events which deposited a similar amount of energy in the detector. We determine by what factor the 50\% error circle has to be increased such that it contains the true neutrino direction for 90\% of the simulated events.

All quoted directions were obtained with the multi-photoelectron (MPE) fit (see \citealt{amanda2004}) which was used for the follow-up program at the time of the alert. An improved version of this algorithm, called Spline MPE, uses a more realistic model of light propagation in ice and on average reaches a more precise reconstruction of the direction \citep{icecube2014}. The Spline MPE reconstruction has been used for the follow-up program since May 2016. The Spline MPE fit yields shifted coordinates which are shown in brackets in Table~\ref{tab:events}. The reconstructed direction changes the most for the first event, which deposited light in a relatively small number of DOMs due to its low energy. Based on the Spline MPE fit, the average direction of all three events is RA~$=25.7^\circ$, Dec~$=39.6^\circ$ with error circles of $0.6^\circ$ (50\%) and $1.9^\circ$ (90\%).

Based on the Spline MPE reconstruction, events 1 and 2 (see Table~\ref{tab:events}) would no longer form a doublet, while events 2 and 3 would have formed a doublet. We expect the detection of 66 doublets per year due to background, and the $\sim\!5$ most significant doublets are followed up (see Sect.~\ref{sec:multi_filter}). The doublet consisting of events 2 and 3 does not pass the significance threshold (compare Sect. \ref{sec:multi_filter}). Hence, the alert would not have been considered interesting and no follow-up observations would have been triggered even if our program had been running with the Spline MPE reconstruction at the time of the alert.

We used simulated neutrino events following an $E^{-2.5}$ neutrino spectrum (compare Sect.~\ref{sec:triplet_source}) to calculate the probability that three events from a point source form a triplet based on the MPE reconstruction, which is not recovered when using the Spline MPE algorithm. The resulting probability is 8\%. For background triplets (i.e., events that are aligned by chance but do not stem from a point source) we evaluate scrambled data (compare Sect.~\ref{sec:SignficanceCalc}) and find that the probability is 36\%. The fact that the triplet is not re-detected when using the Spline MPE algorithm is therefore a slight indication that it might not be of astrophysical origin, but a coincidence of aligned background events.

To test more precisely whether the three events are consistent with a single point source origin we simulated events from a similar zenith range. The true direction of the events is shifted to the same position and we select events with comparable estimated angular errors. We then check how often they are reconstructed further from their true direction than the three detected events. We quantify this by defining a test statistic equivalent to the spatial term used in the standard point source analysis (Eq.~3 in Ref.~\citealt{icecube2016b}) and find that this happens in $\sim$75\% ($\sim$50\% using the SplineMPE results) of all cases. Therefore, the detected events are consistent with a point source origin when considering their errors and the detector properties for this zenith direction.

All following analyses are based on the MPE position and error estimate which are shown as solid lines in Fig.~\ref{fig:AlertErrorCircle}. Compared to the angular separations between the neutrino candidates the mean position only changes slightly and the 50\% error circle of the MPE reconstruction fully contains the 50\% error circle of the Spline MPE fit.

\subsection{Detector stability}

\begin{figure}[t]
\begin{center}
\includegraphics[width=88mm]{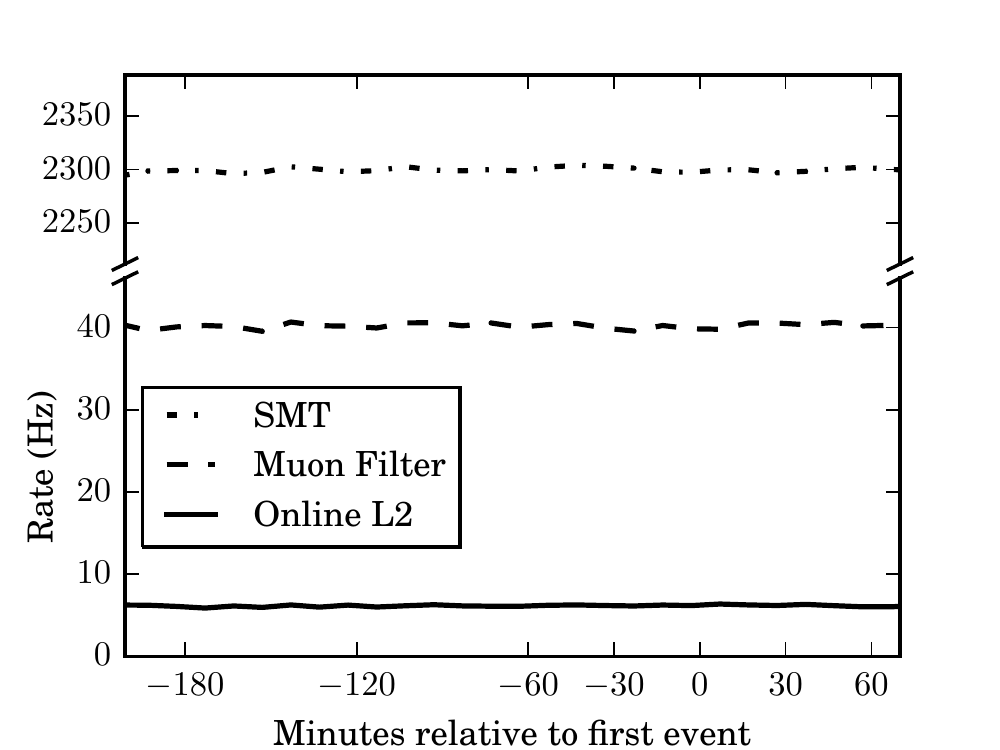}
\caption{\small Temporal behavior of different filter rates: The \textit{Simple Multiplicity Trigger}, \textit{Muon Filter}, and \textit{Online Level 2} rate. No significant deviation from normal detector behavior was observed around the time of the alert.}
\label{fig:rateStability}
\end{center}
\end{figure}

Before triggering follow-up observations we examined
the status of the detector carefully.
A set of selected trigger and filter rates
related to the analysis are monitored in real-time.
Figure~\ref{fig:rateStability} shows the rate of the
\textit{Simple Multiplicity Trigger}, the \textit{Muon Filter,}
and the \textit{Online Level 2 Filter} (see Sect. \ref{sec:event_selection}) near the time of the events.
A \textit{Simple Multiplicity} consists of eight
DOMs forming at least four pairs in close temporal and spatial coincidence which trigger
within $5\,\mu$s.

These quantities are sensitive to 
disturbances in the data-collection process \citep{icecube2016d}. These disturbances are
classified as either internal,
such as interrupted connections to a segment of the detector,
or external,
such as interference from other experiments at the South Pole.
Periods of
bad operating conditions can be flagged by monitoring the moving average of the rates and comparing it to 
expected statistical fluctuations.
This system has operated for several years and has reliably identified
occasional internal and external disturbances during that period.
No significant deviation from normal detector behavior was
observed for a time period
spanning several hours around the events in the triplet.

In addition we generated test alerts which consisted of two events within 100\,s that are separated by more than 3.5\,$^\circ$, but less than 7.5\,$^\circ$. The test alert rate did not show any anomalies around the time of the alert. We hence conclude that the detector was stable when the neutrino triplet was detected.

\subsection{Significance calculation}
\label{sec:SignficanceCalc}

To quantify the significance of the neutrino detection, we calculate how often triplets are expected from chance coincidences of background events. We use the data obtained during the previous IceCube season from 2014-05-06 to 2015-05-18 when the follow-up program was running in the same configuration. Considering only the time when the follow-up program was running stably, the uptime of this season was 359\,days, during which 100\,799 neutrino candidates passed the event selection of the follow-up program.

To estimate the multiplet false positive rate from atmospheric backgrounds, we randomly exchanged the detection times of all events during this data-taking season. The event directions in detector coordinates remained the same, but the equatorial coordinates were recalculated using the newly assigned detection time. 
This method preserves both the temporal variations in the data (e.g., seasonal variations; see \citealt{icecube2010}) and directional effects caused by the detector geometry. At the same time, any potential signal from a transient or steady source is smeared out.

To the generated background data, we applied our {\it a priori} cuts and searched for neutrino doublets (two events arriving within 100\,s and with an angular separation of at most $3.5^\circ$). We then counted how many doublets had at least one neutrino event in common and found that such overlapping doublets or triplets are expected 0.0732 $\pm$ 0.0009 times per full year of live time, hence one every 13.7\,yr assuming the configuration in which the program was running at the time of the alert\footnote{We emphasize that our definition of a triplet only requires that one of the three events forms a doublet with the two other ones. The two other events can therefore be separated by more than $3.5^\circ$ and do not have to arrive within 100\,s.}. The expected number of background alerts is calculated for every season since the start of the follow-up program in December 2008. Within this time both the event selection and alert generation of the follow-up program were improved yielding different sensitivities. Moreover, we consider the down time of the follow-up program. Adding up the different contributions since 2008, the total number of expected triplets from background was 0.38 at the arrival time of the first triplet. The probability to detect one or more triplets from background is hence 32\%. The detected neutrino triplet may therefore be caused by a chance alignment of background events.


\section{Follow-up observations}
\label{sec:observations}

The neutrino triplet was not automatically forwarded to any follow-up observatory because it did not pass the required criteria (all events within $3.5^\circ$) and neither of the individual doublets reached the required significance threshold for triggering follow-up observations. As calculated in Sect.~\ref{sec:SignficanceCalc} the detection of a triplet from background is expected once every 13.7\,yr, which makes it a rare alert and the most significant neutrino multiplet detected so far. Therefore, the IceCube Collaboration decided to notify the partners providing electromagnetic follow-up observations. Our follow-up partners were informed 22\,h after the detection of the triplet. In case of automatic forwarding, the median latency for triggering follow-up observatories is $\sim\!1$\,min.

The triplet direction was $\sim\!70^\circ$ from the Sun and difficult to observe from ground-based observatories since it was located close to the horizon during night time and a large air mass impaired the image quality.

Several source classes have been suggested as potential transient neutrino sources. We therefore obtained multiwavelength observations at different times after the neutrino detection. We specifically search for GRBs, CCSNe (which might contain choked jets) and AGN flares. In this section we present reports on the observations obtained with optical (Sect. \ref{sec:optical}), X-ray (Sect. \ref{sec:xrays}) and gamma-ray (Sect. \ref{sec:gamma}) telescopes. The results are summarized and evaluated in Sect. \ref{sec:discussion}.

\subsection{Optical observations}
\label{sec:optical}

Optical follow-up observations were obtained with ASAS-SN, MASTER, and LCO. No observations could be obtained with the PTF P48 telescope which was undergoing engineering work. In addition to these follow-up observations, we also analyze archival data obtained within a period of 30\,days before the neutrino triplet.

\subsubsection{All-Sky Automated Survey for SuperNovae} 
The All-Sky Automated Survey for SuperNovae (ASAS-SN or ``Assassin'';
\citealt{shappee2014}) monitors the whole sky
down to a limiting magnitude of $V \sim 17$ mag. The focus of the survey is to find nearby supernovae (SNe)
and other bright transient sources. Currently, ASAS-SN consists of
two fully robotic units with four 14 cm telescopes each on Mount Haleakala in Hawaii and Cerro Tololo
in Chile. These eight telescopes allow
ASAS-SN to survey 20\,000~deg$^2$ per night, covering the entire
visible sky every two days. The pipeline is fully automatic and
discoveries are announced within hours of the data being collected.
The data are photometrically calibrated  using the AAVSO Photometric All-Sky
Survey (APASS; \citealt{henden2015}).

The ASAS-SN ``Brutus'' station in Hawaii has regularly observed the field
containing the triplet position since 2013-10-27,
obtaining 408 ninety-second V-band images on 178 separate nights. Before the
neutrino trigger, this field was last observed two weeks earlier, on
2016-02-03, as the observability of this field was limited due to the Sun angle. In Table \ref{tab:optical_obs} we list the dates on which this field was
observed during the 30 days before the trigger, and also the
typical $5\,\sigma$ $V$ band detection limit reached, in the
$3\times90$ s dithered exposures. The resulting limits are shown in Sect. \ref{sec:sn}.

Following the neutrino trigger, we scheduled $20\times 90$ s
exposures of the field containing the trigger position, which were
taken between UTC 2016-02-19.229 and 2016-02-19.253, that is, 34\,h after the neutrino detection. The ASAS-SN field contains
about 90\% of the final 50\% error circle of $1^\circ$. Because of the bright Moon, the combined
depth of $V\lesssim 18.0$ is relatively shallow while the $5\,\sigma$ depth
of the individual 90\,s exposures is $V\lesssim 16.5$. No transient sources were detected.

\subsubsection{Las Cumbres Observatory}

The Las Cumbres Observatory (LCO\footnote{\url{http://lco.global}}; \citealt{brown2013}) consists of seven 0.4 m, nine 1 m and two 2 m robotic telescopes situated in six sites around the world (two additional 1 m telescopes will be deployed in the near future to a seventh site). The network specializes in time domain astronomy, and has the capability of performing immediate target-of-opportunity observations of almost any point in the sky within minutes.

The error circle of the neutrino triplet was tiled with nine pointings that were observed with the LCO 1 m telescope at the McDonald observatory in Texas. The observations cover the inner $\sim\!60$\% of the 50\% error circle of the final triplet location. Observations started 30\,h after the neutrino detection and various combinations of UBVgri filters were used on different nights (Table \ref{tab:optical_obs_lcogt} and Sect. \ref{sec:sn}). The limiting magnitudes were calculated following calibration to the APASS catalog (see Appendix B of \citealt{valenti2016} for more details). 
Due to the proximity of the field to the sun, additional epochs could not be obtained in the weeks following the alert to determine whether or not any transient sources were present in the images.

\subsubsection{Mobile Astronomical System of the Telescope-Robots}

\begin{table*}
{\small
\hfill{}
\caption{Observing conditions at the MASTER telescopes at the time 2016-02-18 17:15:58 UTC}
\label{tab:MASTERCond}     
\begin{center}
\begin{tabular}{llll}     
\hline\hline       
MASTER node       &
Object altitude          &
Sun altitude     &
Notes\\ 
       &
$(^\circ)$          &
$(^\circ)$     &
\\ 
\hline
MASTER-Amur         & 3.98  & $-47.01$    & too close to the horizon for good observations\\
MASTER-Tunka        & 13.45 & $-49.91$    & cloudy and too close to the horizon for good observations \\
MASTER-Ural         & 37.06 & $-33.25$    & bad weather\\
MASTER-Kislovodsk   & 43.50 & $-28.31$    & good conditions, observations began\\
MASTER-SAAO         & $-8.36$ & 0.93      & below the horizon at night time\\
MASTER-IAC          & 78.22 & 20.25     & snow storm \\
MASTER-OAFA         & $-1.1$  & 69.06     & below the horizon at night time\\
\hline                                  
\end{tabular}
\end{center}
}
\hfill{}
\end{table*}

The Mobile Astronomical System of the Telescope-Robots (MASTER; \citealt{lipunov2010,kornilov2012,gorbovskoy2013}) Global Robotic Net consists of seven observatories in both hemispheres (see Table \ref{tab:MASTERCond}). All MASTER observatories include identical twin 40 cm wide-field telescopes with two $4$ square degree FoV which monitor the sky down to 21st magnitude. In divergent mode, the twin telescopes can cover 8 square degrees per exposure and the telescope mounts allow rapid pointing to follow up short transient sources. Each MASTER node is equipped with BVRI Johnson/Bessel filters, two orthogonal polarization filters and two white filters (called unfiltered). To collect as many photons as possible, the MASTER telescopes are usually operated without a filter when searching for transients. In addition, each observatory hosts very-wide-field cameras which cover 400 square degrees and are sensitive to sources brighter than 15th magnitude.

An important component of MASTER is its in-house detection software which provides photometric and astrometric information about all optical sources in the image within 1-2 minutes of the frame readout. The processing time includes primary reduction (bias, dark, flat field), source extraction with help of the SExtractor algorithm\footnote{\url{http://www.astromatic.net/software/sextractor}} \citep{bertin1996}, the identification of cataloged objects and the selection of unknown objects. New sources detected in two images at the same position are classified as optical transients \citep{lipunov2016}. The unfiltered magnitudes are calibrated using stars from the USNO-B1 catalog where the catalog magnitudes are converted to unfiltered magnitudes via $0.2\times B + 0.8\times R$. For each image, a limiting magnitude is calculated.

The MASTER network received the neutrino triplet coordinates by email at 2016-02-18 17:15:58 UTC. The altitudes and visibility constraints of the position at the different observatories are listed in Table \ref{tab:MASTERCond} for the time when the neutrino detection was communicated. Observations started at the MASTER-Kislovodsk telescopes within less than one hour and the position was monitored by MASTER-Kislovodsk, MASTER-Tunka, and MASTER-IAC for the following month (compare Table~\ref{tab:optical_obs}).

The majority of the observations listed in Table \ref{tab:optical_obs} are centered on the triplet position and include the complete 50\% error circle of the final position. Moreover, except for small gaps, the complete 90\% error circle was covered both before and after the neutrino detection. No transients were found above the $5\,\sigma$ limiting magnitudes given in Table \ref{tab:optical_obs} and shown in Sect. \ref{sec:sn}.
The very- wide-field cameras did not detect any transient brighter than 15th magnitude within the 400 square degrees surrounding the triplet location.



\subsection{X-ray observations}
\label{sec:xrays}

We triggered the X-Ray Telescope (XRT) on board the \emph{Swift} satellite \citep{swift2004} to search for GRB afterglows, AGN flares, or other X-ray transients (see Sect. \ref{sec:xrt}). By chance, the \emph{Swift} Burst Alert Telescope (BAT; \citealt{batfunction2005}) observed the triplet position within a minute after the neutrino detection as described in Sect. \ref{sec:bat}.

\subsubsection{Swift Burst Alert Telescope}
\label{sec:bat}

\emph{Swift} BAT detects hard X-rays in the energy range from 15 to 150\,keV. The FoV covers about 10\% of the sky and the detector is illuminated through a partially coded aperture mask.

Just 100\,s after the first neutrino was detected, the \emph{Swift} satellite completed a preplanned slew to RA~$ = 23.38^\circ$, Dec~$=+41.12^\circ$ which placed the triplet position within the BAT FoV, at a partial coding fraction of 60\%. We retrieved the BAT data for this pointing from the \emph{Swift} Quick Look website (ObsID 00085146016). No rate- or image-triggered transients were detected above the significance threshold of $\mathcal{S}>6.5\,\sigma$ during the pointing, so only survey mode data are available. Survey data for the pointing consist of three exposures of 59\,s, 10\,s, and 15\,s, with intervening gaps for maintenance operations. The BAT analysis was conducted using the \swpkg{heasoft}\footnote{\swpkg{heasoft} website: \url{http://heasarc.nasa.gov/lheasoft/}} (v.\ 6.18) software tools and calibration, closely following the analyses from \citet{swiftresults2005, swiftagns2008, swiftsurvey2010} and \citet{swiftsurvey2013}.

We used the \swpkg{heasoft} tool \swtool{batcelldetect} on the summed exposure as well as on the first exposure over the full bandpass ($15-150 \,\text{keV}$), with a detection threshold of $\mathcal{S} = 3.5\,\sigma$ (the lowest allowed setting). The most significant detection within the triplet 90\% confidence region was in the first exposure at RA~$=28.6083^\circ$, Dec~$=37.34583^\circ$ (henceforth referred to as the \batblip) with single-trial significance $\mathcal{S}= 4.6\,\sigma$.

To estimate the significance of the \batblip\ given the search area, we find the number of similar or more significant fluctuations in a rectangular region of the BAT image plane centered around the position of \batblip\ in 2655 BAT pointings with similar exposure times. We find an average of $0.13$ such candidate sources per pointing. Since the triplet 90\%-confidence region corresponds to 41\% of the rectangular region, this yields a p-value of $p=9.9\%$ for the \batblip. A trial factor penalty of two was included since both the summed and the first exposure were analyzed. The \batblip\ is hence consistent with a random fluctuation of the background.

Flux upper limits were derived from the summed exposure noise map, including the \batblip, over the triplet 90\%-confidence region, and we find a $4\,\sigma$ upper limit to the fluence of $3.3 \times10^{-7} \,\text{erg}\,\text{cm}^{-2}$ for the energy range of 15--150\,keV. This corresponds to a limit of $3.9\times10^{-9}\,\text{erg}\,\text{cm}^{-2}\,\text{s}^{-1}$ on the average flux between $100\,\text{s}$ and $256\,\text{s}$ after the detection of the first neutrino. BAT count limits are converted to fluences using the PIMMS\footnote{available at \url{https://heasarc.gsfc.nasa.gov/docs/software/tools/pimms.html}} online tool, assuming a power law with a spectral index of $\Gamma = -2$. This spectral index corresponds to a typical GRB spectrum in this energy range. It is moreover very close to the mean AGN spectral index which was measured to be $-1.95$ by \citet{burlon2011}. In Sect.~\ref{sec:grb} we compare the limit to typical prompt fluxes of GRBs detected by the BAT.

\begin{table*}
{\small
\hfill{}
\caption{XRT sources \label{tab:sources}}
\begin{center}
\begin{tabular}{lllllll}     
\hline\hline     
Name & RA & Dec & Exposure time & Rate & Alt. name & Object type \\
     &      &      & (s)           & (counts/s) &     & \\
\hline
X1 & 25.4909 & $+$39.3921 & 308 & $0.097\pm 0.020$ & B2 0138+39B & Seyfert 1 Galaxy \\
X2 & 25.6546 & $+$40.3788 & 285 & $0.047\pm 0.015$ & HD 10438 & Star\\
X3 & 25.5324 & $+$39.4129 & 324 & $0.035\pm 0.012$ & V* OQ And & Variable Star \\
X4 & 26.7475 & $+$39.2575 & 284 & $0.024\pm 0.011$ & 1RXS J-14658.4+391526 & Star \\
X5 & 25.0723 & $+$39.5886 & 221 & $0.029\pm 0.014$ & HD 10169 & Star\\
X6 & 25.0107 & $+$39.6033 & 506 & $0.017\pm 0.007$ & -- & unknown \\
\hline                                  
\end{tabular}
\end{center}
}
\tablefoot{Coordinates are provided in J2000.}
\hfill{}
\end{table*}


\subsubsection{Swift X-Ray Telescope}
\label{sec:xrt}
\begin{figure}
\centering
\includegraphics[width=88mm]{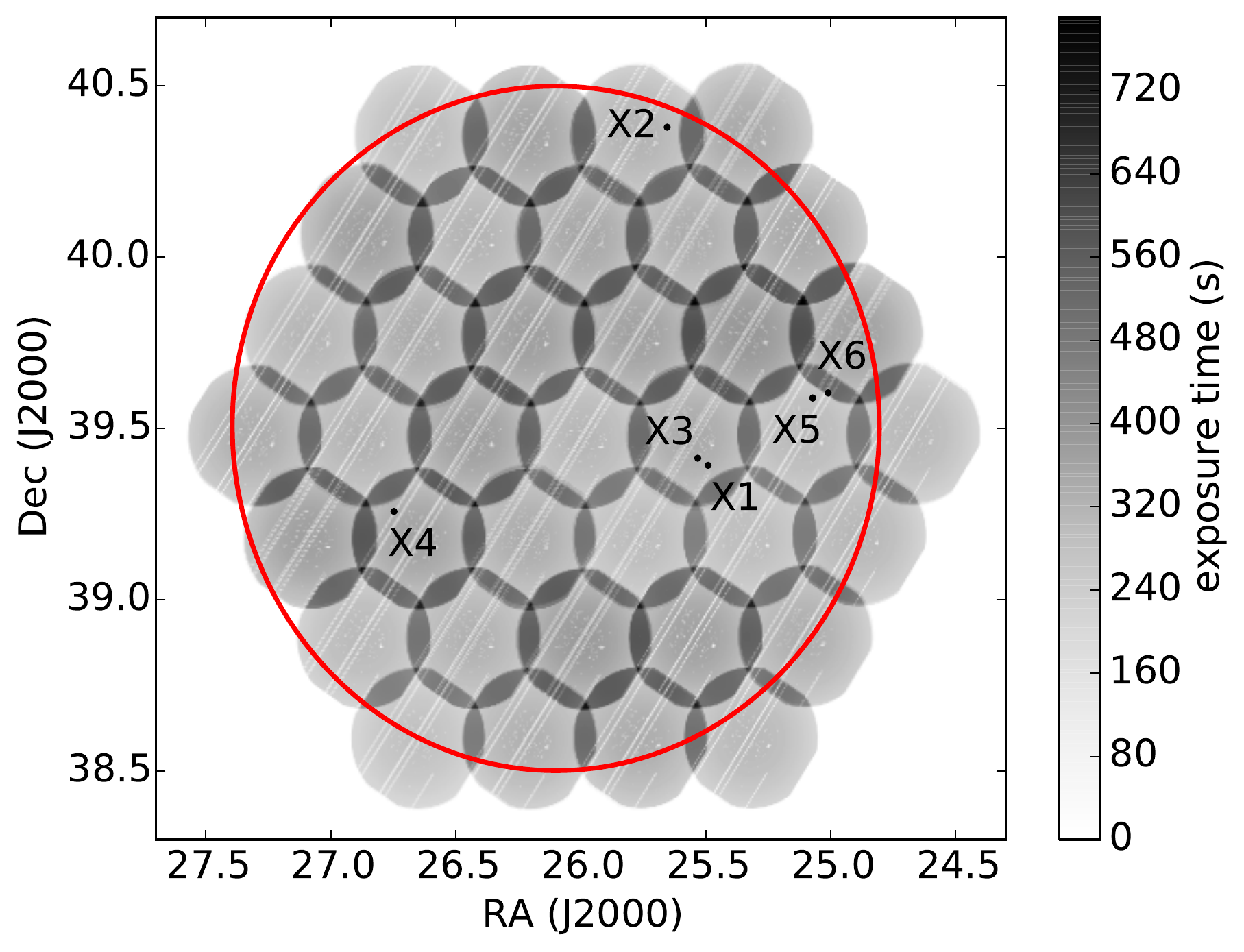}
\caption{Exposure map of the 37 \emph{Swift} XRT pointings averaging 320\,s per tiling. The red circle shows the 50\% confidence bound to the triplet position. XRT sources (compare Table~\ref{tab:sources}) are shown as black points.}
\label{fig:exp}
\end{figure}

The \emph{Swift} XRT is an X-ray imaging spectrometer sensitive to the energy range $0.3-10$\,keV. The telescope's FoV has a diameter of $0.4^\circ$.
To search for possible \xray\ counterparts to the neutrino triplet over the largest feasible region, we requested a 37-pointing mosaic of \emph{Swift} observations. These observations began at 2016-02-18 17:57:42 (22.6\,h after the neutrino detection; Target IDs 34342 to 34379), with the resulting exposure map shown in Fig.~\ref{fig:exp}. The achieved exposure per pointing is 0.3--0.4\,ks. Data were analyzed as described in~\citet{evans2015}, leading to a single unified \xray\ image, exposure map, and list of \xray\ sources. The \emph{Swift} XRT observations cover nearly the complete 50\% containment region.


Six \xray\ sources were identified (Table~\ref{tab:sources}) with the detection flag {\it good} which means that their probability of being spurious is $<0.3\%$ \citep{evans2015}. As revealed from searches of the NASA Extragalactic Database\footnote{NASA Extragalactic Database: \url{https://ned.ipac.caltech.edu/}} and examination of archival optical images, X1 is spacially coincident with a known Seyfert 1 galaxy; X2, X3, X4, and X5 correspond to cataloged stars and X6 remains unidentified. We note that X-rays associated with a bright star were detected when \emph{Swift} followed up a neutrino candidate detected by the ANTARES neutrino telescope \citep{antares2015, smartt2015}. The large number of stars detected in our observations shows that such chance coincidences are frequent. We do not consider the stars as potential sources of high-energy neutrinos.

The X-ray source X1 is classified as a Flat Spectrum Radio Quasar by \citet{healey2007} and is located at a redshift of $z=0.08$ \citep{wills1986}; it has been detected several times by ROSAT, XMM-\emph{Newton} and the \emph{Swift} XRT. Compared to the previous detections, X1 was not flaring during these XRT observations.

Among the identified sources, X6 is unique in not having an obvious counterpart within its 90\% error circle. To refine the localization and study the \xray\ variability, X6 was followed up with 1\,ks and 8.6\,ks \emph{Swift} observations on 2016-03-18 and 2016-07-23 (Target ID 34429). The source was re-detected in the deepest XRT observation; it faded by a factor of nine within five months. The XRT light curve, shown in Fig. \ref{fig:x6_xrt_lc}, is consistent with a $t^{-0.5}$ decay over five months which is too shallow for a GRB afterglow (see Sect. \ref{sec:grb}) or a typical tidal disruption event which fades with $t^{-5/3}$ in the X-ray regime \citep{komossa2015}. The latter detection rules out the possibility that X6 is a GRB.

In archival PTF images we find two bright stars, hereafter referred to as S1 and S2, located close to the 90\% error circle of X6. To look for fainter optical sources we obtained a Keck image in which a third object, O3, is detected (see Fig.~\ref{fig:x6_keck}). The properties of the three potential optical counterparts are specified in Table~\ref{tab:x6_counterpart}.

To search for short-lived optical emission, we analyze simultaneous UVOT observations. During the first XRT observation the UVOT observed in the U band (Target ID 34357). We use the \swpkg{heasoft} tool \swtool{uvotdetect} to measure the aperture flux within a circle with a radius of 3\arcsec centered around the best fit location of X6. This small radius was chosen to avoid contamination from the star S2. No source is detected and the $3\,\sigma$ limit is $17.39\,\text{mag}_\text{AB}$ which corresponds to a flux upper limit of $10^{-15}\,\text{erg\, s}^{-1}\text{cm}^{-2}\text{\AA}^{-1}$ at a wavelength of $3501\AA$.

Considering all available observations, we identify two possible scenarios: X6 could either be an extreme stellar flare or it could be an obscured and distant AGN. We discuss the nature of X6 in more detail in Appendix~\ref{sec:x6}, where we come to the conclusion that it is not likely associated with the neutrino triplet.


Except for X-ray source X6, the \emph{Swift} follow-up observations identified no unknown \xray\ sources within the 50\%-containment region of the neutrino triplet. Our upper limits on any source over this region are derived from the 0.3--1.0\,keV, 1--2\,keV, 2--10\,keV, and 0.3--10\,keV (full band) background maps. Background count rates for each bandpass were estimated from three regions, sampling the on-axis, off-axis, and field-overlap portions of the total exposure pattern; these provide a $3\,\sigma$ count-rate upper limit following the Bayesian method of \citet{Kraft91}. The upper limits were then multiplied by a factor of 1.08 to correct for the finite size of the aperture (a 20 pixel radius). The rate upper limits are converted to fluxes for each of two spectral models: a typical AGN spectrum in the X-ray band (a power law with photon index $\Gamma= -1.7$, $N_{\rm H} = 3\times 10^{20}$ cm$^{-2}$) and a GRB spectrum (a power law with $\Gamma=-2$, $N_{\rm H} = 3\times 10^{21}$ cm$^{-2}$). The range of 
resulting upper limits is listed in 
Table~\ref{tab:xrtUL}. In Sect. \ref{sec:grb} we compare the limits to detected GRB afterglows. 


\subsection{Gamma-ray observations}
\label{sec:gamma}

The position of the triplet was observed by the \emph{Fermi} LAT about 30\,min after the neutrino detection (see Sect. \ref{sec:lat}). Bad weather conditions in La Palma did not allow immediate observations with either MAGIC \citep{magic2016} or FACT \citep{fact2013} and the position is not observable for HESS. VERITAS observed the direction with a delay of one week (see Sect. \ref{sec:veritas}) and the position was within HAWC's FoV at the arrival time of the triplet (see Sect. \ref{sec:hawc}). 

\subsubsection{Fermi Large Area Telescope}
\label{sec:lat}

\begin{center}
\begin{figure*}[htb]
\subfloat[The \emph{Fermi} LAT likelihood ratio test statistic (TS) within the region of interest. The significance of fluctuations above the expected background scales roughly with $\sqrt{\text{TS}}$.\label{fig:fermi_ts}]{\includegraphics[width=88mm]{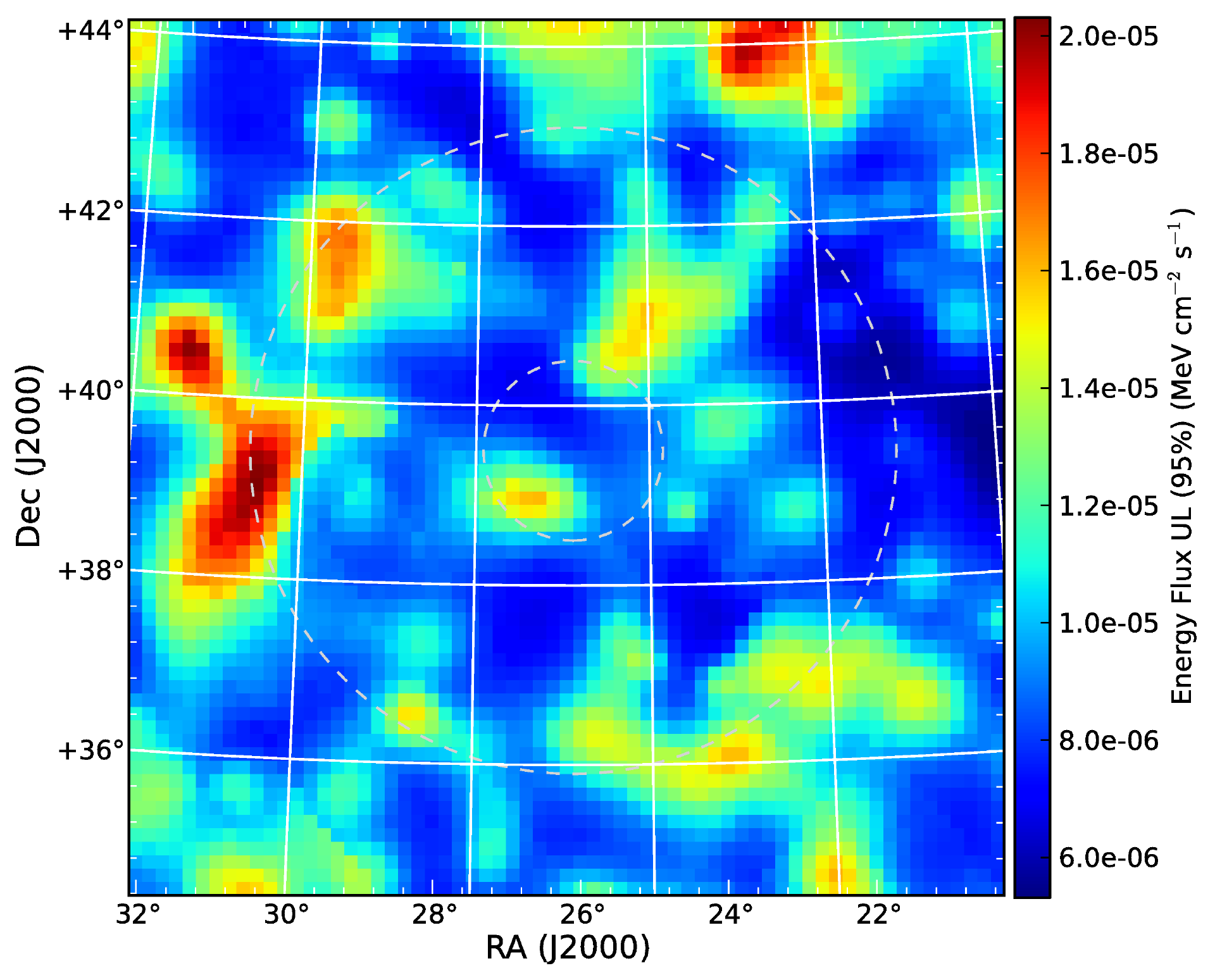}}
\hfill
 \subfloat[\emph{Fermi} LAT 95\% upper limits on the flux in the 100\,MeV to 100\,GeV energy range. Crosses indicate the locations of known \emph{Fermi} sources. \label{fig:fermi_ul}]{\includegraphics[width=88mm]{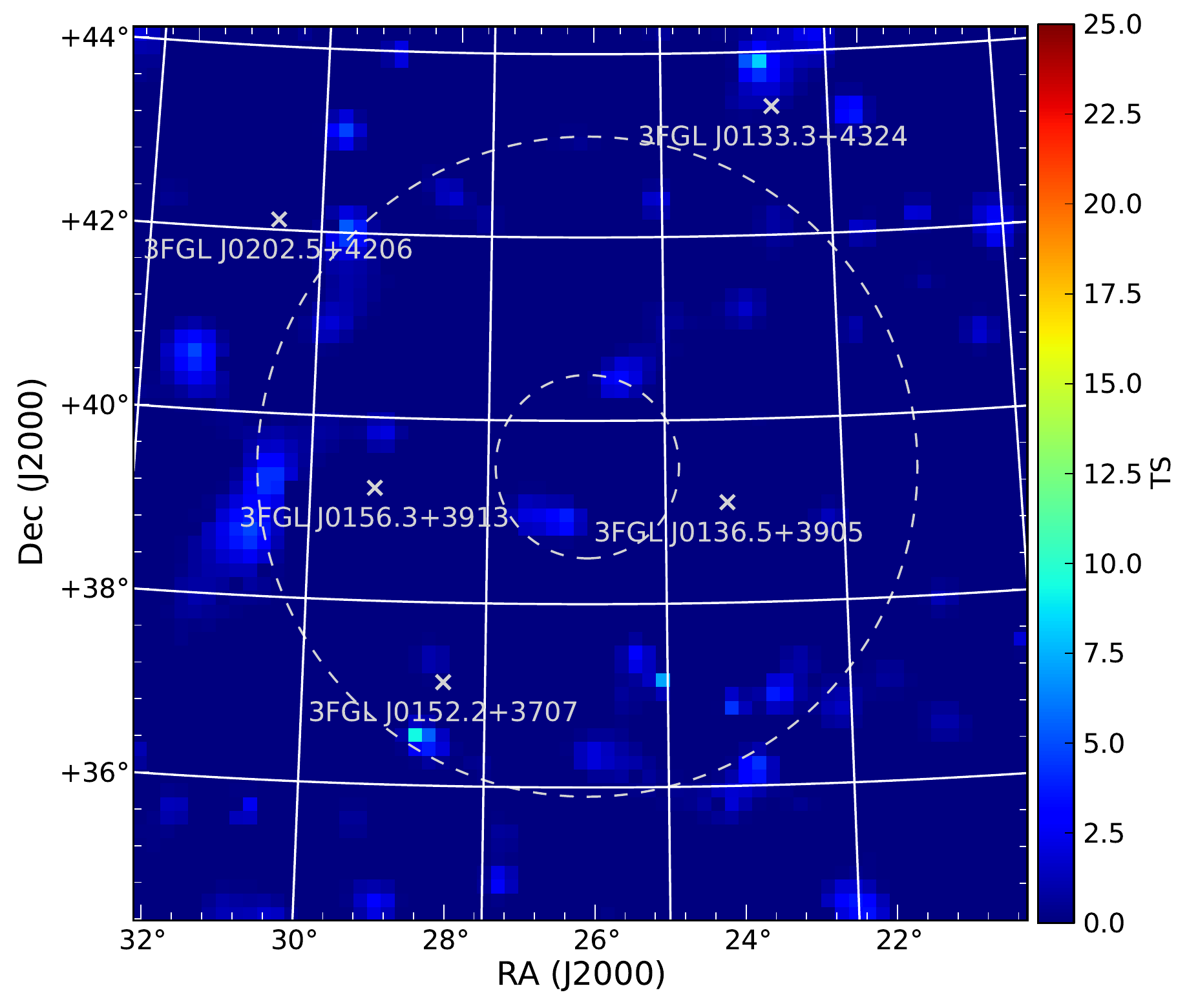}}
\caption{\label{fig:fermi}\emph{Fermi} LAT results from the unbinned likelihood analysis within the region of interest using all data within 14\,days of neutrino detection. The dashed circles show the 50\% and 90\% error circles of the neutrino triplet.}
\end{figure*}
\end{center}

\begin{table*}
{\small
\hfill{}
\caption{\emph{Fermi} LAT flux upper limits. \label{tab:widgets}}
\begin{center}
\begin{tabular}{llllll}
\hline\hline     
Interval & Duration & Start date & End date & Median U.L. (95) & Median U.L. (95) \\
 & & (UTC) & (UTC) & (ph cm$^{-2}$ s$^{-1}$) & (GeV cm$^{-2}$ s$^{-1}$) \\
\hline
$T_{~\rm FAVA1}$ & 24 h & 2016-02-17 19:21:32 & 2016-02-18 19:21:32 & -- & -- \\
$T_{~\rm FAVA2}$ & 24 h & 2016-02-16 19:21:32 & 2016-02-17 19:21:32 & -- & -- \\
$T_{~\rm FAVA3}$ & 24 h & 2016-02-17 07:21:32 & 2016-02-18 07:21:32 & -- & -- \\
$T_{~\rm FAVA4}$ & 7 days & 2016-02-15 15:43:35 & 2016-02-22 15:43:35 & -- & -- \\
\hline
$T_{~\rm Like1}$ & 6 h  & 2016-02-17 19:21:32 & 2016-02-18 01:21:32 & $3.32 \times 10^{-7}$ & $1.82 \times 10^{-7}$ \\
$T_{~\rm Like2}$ & 12 h & 2016-02-17 19:21:32 & 2016-02-18 07:21:32 & $1.86 \times 10^{-7}$ & $1.01 \times 10^{-7}$ \\
$T_{~\rm Like3}$ & 24 h & 2016-02-17 19:21:32 & 2016-02-18 19:21:32 & $1.27 \times 10^{-7}$ & $6.96 \times 10^{-8}$\\
$T_{~\rm Like4}$ & 24 h & 2016-02-16 19:21:32 & 2016-02-17 19:21:32 & $1.15 \times 10^{-7}$ & $6.30 \times 10^{-8}$ \\
$T_{~\rm Like5}$ & 24 h & 2016-02-17 07:21:32 & 2016-02-18 07:21:32 & $1.11 \times 10^{-7}$ & $6.08 \times 10^{-8}$ \\
$T_{~\rm Like6}$ & 14 days& 2016-02-17 19:21:32 & 2016-03-02 19:21:32 & $1.73 \times 10^{-8}$ & $9.48 \times 10^{-9}$\\
\hline                                  
\end{tabular}
\end{center}
}
\tablefoot{A summary of the FAVA and likelihood analysis timescales. FAVA does not provide flux upper limit estimates. The upper limit estimates quoted for the likelihood analysis are the median 95$\%$ C.L. considering all upper limits within the 90\% error circle. They have been obtained for the energy range from 100\,MeV to 100\,GeV and a spectral index of $\Gamma=-2.1$ has been assumed.}
\hfill{}
\end{table*}

The \emph{Fermi Gamma-ray Space Telescope} consists of two primary instruments, the Large Area Telescope (LAT) and the Gamma-Ray Burst monitor (GBM). The LAT is a pair-conversion telescope comprising a $4\times4$ array of silicon strip trackers and cesium iodide (CsI) calorimeters. The LAT covers the energy range from 20\,MeV to more than 300\,GeV with a FoV of $\sim\!2.4$\,steradian, observing the entire sky every two orbits ($\sim\!3$\,h) while in normal survey mode \citep{atwood2009}.
The GBM is comprised of 12 sodium iodide (NaI) and two bismuth germanate (BGO) scintillation detectors that have an instantaneous view of 70\% of the sky. The NaI and BGO detectors are sensitive to emission between 8\,keV and 1\,MeV, and 150\,keV and 40\,MeV, respectively \citep{meegan2009}.

The triplet location was occulted by the Earth at the detection time of the first neutrino event (T0). As a result, the GBM and LAT can place no constraints on the existence of a prompt gamma-ray transient coincident with the detection of the neutrino events. Within the period of 24\,h before and after T0, there were a total of four reported GBM detections\footnote{\url{http://gcn.gsfc.nasa.gov/fermi_grbs.html}}. They were all separated by more than $50^\circ$ from the triplet location and an association can be excluded.

The region of interest entered the LAT field-of-view after roughly 1600\,s and in the following we analyze the LAT data recorded within the days before and after the detection of the neutrino alert. We focussed on limiting the intermediate (hours to days) to long (weeks) timescale emission from a new transient source or flaring activity from a known gamma-ray emitter in the LAT energy range. We employed two different techniques to search for such emission in the LAT data; the \emph{Fermi} All-sky Variability analysis (FAVA; \citealt{fermi2013}) and a standard unbinned likelihood analysis. FAVA is an all-sky photometric analysis in which a region of the sky is searched for deviations from the expected flux based on the mission-averaged data. The unbinned likelihood analysis is the standard method of detecting and characterizing sources in the LAT data and is described in more detail in \citet{fermi2009}. We additionally employed a profile likelihood method described in \citet{fermi2012} to calculate upper limits in situations when no significant excess emission is detected.

The FAVA search was performed on 24\,h timescales bracketing T0, covering the periods of [T0$-$24\,h to T0], [T0$-$12\,h to T0$+$12\,h], and [T0 to T0$+$24\,h] (see Table~\ref{tab:widgets}). A single week-long timescale was also searched, covering the period of [T0$-$2.15\,days to T0$+$4.85\,days]. The FAVA analysis selects flares that have a significance of $6\,\sigma$ above the mission average emission at the location. Within the analyzed time windows no such flare was detected at the triplet location.

An examination of the second FAVA catalog (2FAV, paper in preparation), which lists all flaring sources detected in the LAT data on weekly timescales over the course of the entire mission, shows only one period of flaring activity within the 90\% error circle of the triplet location\footnote{\url{http://fermi.gsfc.nasa.gov/ssc/data/access/lat/FAVA/LightCurve.php?ra=26.1\&dec=39.5}}. This period of activity was between 2009-08-31 and 2009-09-07 and was associated with 3FGL\,J0156.3+3913 which is a blazar candidate of uncertain type \citep{acero2015}. No further activity from this source has been detected by FAVA.

The unbinned likelihood analysis was performed using the standard LAT analysis tools (ScienceTools version v10r01p0)\footnote{\url{http://fermi.gsfc.nasa.gov/ssc/}} by modeling all photons within a region of interest (ROI) with a radius of $12^\circ$, covering an energy range of 100\,MeV to 100\,GeV, and utilizing the \texttt{P8R2\_TRANSIENTR020\_V6} event class and the corresponding instrument response functions. For the purposes of this analysis, all modeled sources were fixed to their catalog values, while the normalization of the Galactic and diffuse isotropic components of the fit were allowed to vary. Because of the uncertainty in the triplet location, this analysis was repeated over a $10^\circ \times 10^\circ$ grid of coordinates with $0.15^\circ$ binning.

This search was performed over a variety of timescales, ranging from 6\,h to 14\,days (Table~\ref{tab:widgets}). The resulting significance maps show no emission in excess of the expected background on any of the timescales considered. For each bin in the coordinate grid, we calculated the 95$\%$ confidence levels (C.L.) upper limit on the photon flux of a candidate point source with a fixed spectral index of $\Gamma = -2.1$. This value is appropriate for both AGN (compare \citealt{fermi2015}) and GRBs \citep{ackermann2013, gruber2014} and is used as the standard value when searching for GRBs.

An example of the significance and energy upper limit maps for the T0$+$14\,day timescale is shown in Fig.~\ref{fig:fermi}. The median photon flux and energy flux upper limits calculated for each timescale are listed in Table \ref{tab:widgets}.

\subsubsection{Very Energetic Radiation Imaging Telescope Array System}
\label{sec:veritas}
The Very Energetic Radiation Imaging Telescope Array System (VERITAS) is a ground-based instrument for VHE gamma-ray astronomy with maximum sensitivity in the 80\,GeV to 30\,TeV range. 
It is located at the Fred Lawrence Whipple observatory in southern Arizona (31$^{\circ}$ 40\arcmin\, N, 110$^{\circ}$ 57\arcmin\, W) at an altitude of 1.3\,km above sea level. The array consists of four 12-m-diameter imaging air Cherenkov telescopes each equipped with a camera containing 499 photomultiplier tubes (PMTs) covering a 3.5$^\circ$ FoV. Full details of the VERITAS instrument performance and sensitivity are given in \cite{VERITAS}.

At the time the triplet detection was communicated to VERITAS, the Moon was approaching its full phase and the night sky was too bright to safely operate the sensitive PMT cameras. It is, however, not uncommon for some variable VHE sources such as AGN to exhibit extended periods of intense flaring activity that can be detected days after the source has reached its peak flux \citep{dermer2016}. Observations were started eight days after the detection of the neutrino events on 2016-02-25, when VERITAS observed the triplet location between 02:32 and 03:20 UTC. Additional observations were taken on 2016-02-26 between 02:36 and 03:43 UTC. The combined exposure time during these two nights was 62.8 min, after quality cuts were applied.
These observations were carried out in the normal ``wobble'' mode, where the pointing direction of the telescopes is offset from the source position to allow for simultaneous measurement of the background \citep{berge2007}. A wobble offset of 0.7$^\circ$ was selected to cover a larger region of sky given the uncertainty in the averaged triplet position.

\begin{figure}[tb]
        \centering
        \includegraphics[width=88mm]{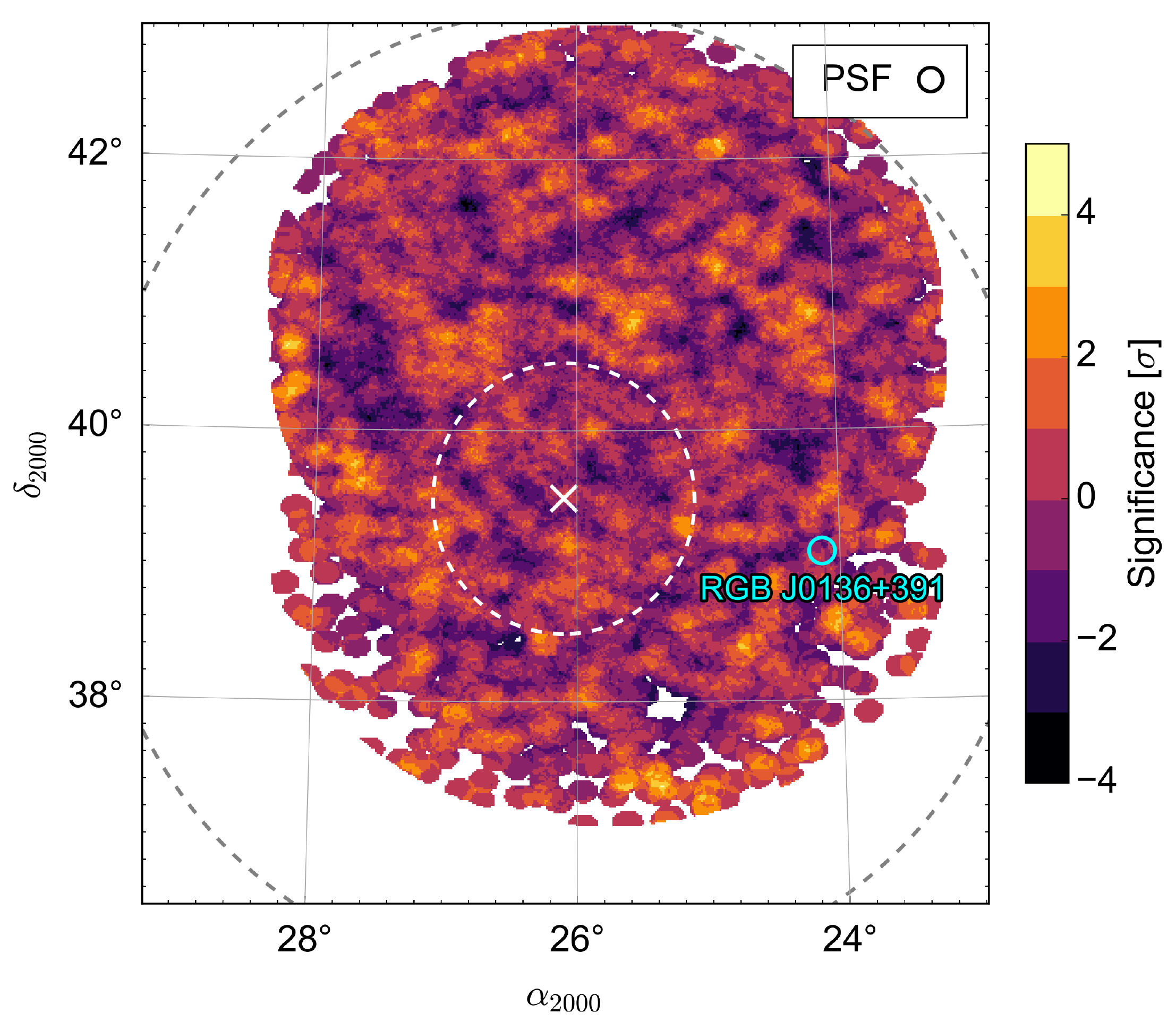}
        \caption{Significance sky map for the VERITAS observations of the neutrino triplet region. The dashed white (gray) line indicates the 50\% (90\%) error circle for the triplet. No gamma-ray excess was detected in the FoV. The known VHE source RGB\,J0136+391 (also known as 3FGL J0136.5+3905; compare Fig.~\ref{fig:fermi_ts}) is located approximately 1.6$^\circ$ away from the triplet central position. }
        \label{fig:veritas}
\end{figure}

An analysis of the VERITAS data showed no significant gamma-ray excess in the triplet region of interest (see Fig.~\ref{fig:veritas}). Consequently, differential flux upper limits were calculated at the 95\% confidence level in four energy bins for a gamma-ray point source located at the averaged triplet position and are given in Table~\ref{tab:veritas}. Furthermore, no new gamma-ray sources were detected anywhere within the triplet 50\% error region or within the VERITAS FoV. 

The only known VHE source in the vicinity of the triplet is the high-synchrotron-peaked blazar RGB J0136+391\footnote{\href{http://tevcat.uchicago.edu/?mode=1;id=244}{http://tevcat.uchicago.edu/?mode=1;id=244}} (also 3FGL\,J0136.5+3905; see Fig.~\ref{fig:fermi_ts}). It has an approximate angular distance of $1.6^\circ$ from the triplet central position and was not detected during the VERITAS observations (see Sect.~\ref{sec:agn} for further discussion of this source). Therefore, the data show no indication of a persistent VHE gamma-ray source, or a high state of RGB\,J0136+391, which could be associated with the neutrino events.

\subsubsection{The High Altitude Water Cherenkov observatory}
\label{sec:hawc}

The High Altitude Water Cherenkov (HAWC) observatory is an array of 
300 detectors, each filled with approximately 200\,000 liters of purified water and instrumented with four 
photo-multiplier tubes. Light-tight bladders provide optical isolation. The observatory is 
optimized to detect Cherenkov light 
from extensive air showers produced by gamma-ray primaries at energies between 
100\,GeV and 100\,TeV. HAWC is located in the state of Puebla, Mexico at an altitude of 4\,100\,m (97.3$^\circ$W, 19.0$^\circ$N). HAWC operates continuously and 
has an average down time due to maintenance of only $\sim\!5$\%. A 
wide FoV, approximately defined by a cone with an 
opening angle of 45$^\circ$ from zenith, spans the declination range of 
$-26^\circ$ to $+64^\circ$ and rotates with the Earth through the full range of
right ascension every day. For a detailed description of the 
array and analysis methods see~\citet{hawcCrab}. 

\begin{figure}[t]
\begin{center}
\includegraphics[width=88mm]{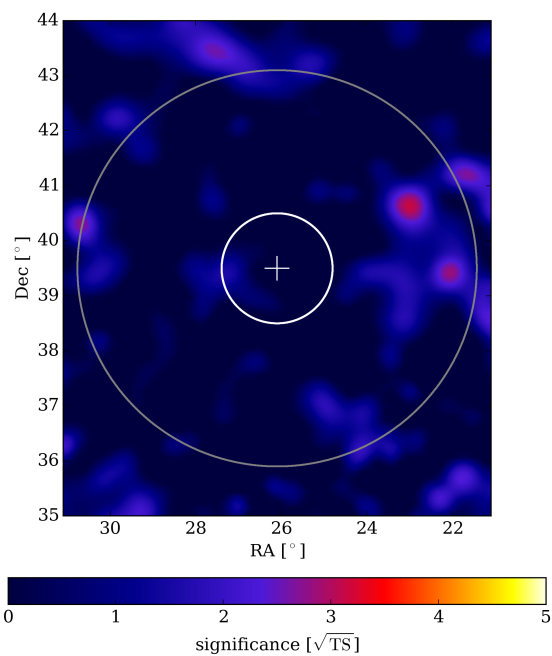}
\noindent
\caption{\small HAWC 500\,GeV to 160\,TeV significance sky map for data collected over one transit 
between 19:18 UTC on 2016-02-17 and 01:31 UTC on 2016-02-18, centered at 
RA~$= 26.1^\circ$, Dec~$= 39.5^\circ$. The IceCube 50\% (white) and 
90\% (gray) error circles are also shown.}
\label{fig:hawc-skymap}
\end{center}
\end{figure}

At the detection time of the neutrino triplet, its position 
had just entered HAWC's FoV. HAWC was operating normally and 
observed the full transit ($\sim\!6$\,h at zenith angles $<45^\circ$) of 
the triplet location between 19:15 UTC on 2016-02-17 and 01:30 UTC on 
2016-02-18. HAWC data are being continuously 
reconstructed on computers at the array site with an average time lag of approximately 4\,s and were 
immediately available for a follow-up analysis when the IceCube alert was 
received.

A scan of the region around 
the triplet coordinates was performed with the standard HAWC maximum-likelihood 
technique, using nine energy-proxy analysis bins that sort data 
according to the air shower size \citep{hawcGalacticPlane}. 
The analysis bins 
account for the varying angular resolution and background suppression efficiency.  
For each bin, the event count in each pixel of a 
HEALPix \citep{healpix} map is 
compared to a prediction composed of the average, smoothed background of cosmic 
rays measured from data and the simulated expectation of gamma-ray events from 
a point-like source. The signal expectation includes the modeling of the 
angular resolution, which improves with energy from $\sim\!1^\circ$ to 
$<0.2^\circ$ in the range from 1 to 100\,TeV. The differential flux in each analysis bin is described by a power law with a photon index of $\Gamma = -2.7$, which is the standard value used for HAWC point-source searches. This index also corresponds to the average of detected TeVCat sources \citep{hawcCat2017}.
Leaving only the normalization $N_0$ as a free parameter, a likelihood 
maximization over all bins and pixels was performed for all locations in a 
$9^\circ\times9^\circ$ area with a grid spacing of $0.06^\circ$. This scan 
revealed no significant excess with a pre-trial significance above
$5~\sigma$ and the results are fully compatible with a pure background
hypothesis. The resulting sky map is presented in 
Fig.~\ref{fig:hawc-skymap}, showing significance in standard 
deviations calculated as $\sqrt{\text{TS}}$, where TS is the standard test statistic from the likelihood 
ratio test.

Given the lack of a source candidate, we derived gamma-ray flux 
limits for the combined average neutrino direction, RA~$=26.1^\circ$, Dec~$=39.5^\circ$. 
The resulting limits are listed in Table \ref{tab:hawc} and shown in Sect.~\ref{sec:discussion}.
These upper limits were calculated separately for 
five intervals of width $0.5$ in $\log{(E/\mbox{TeV})}$ by modeling a 
flux that is non-zero only 
within each interval and using a scan of the likelihood space to determine the one-sided 95\% C.L. 
value. The limits correspond 
to the normalization $N_0$ of a power law with a photon index of 
$\Gamma=-2$. We checked that the normalization in the center of each 
interval did not change when varying $\Gamma$ between $0$ and $-3$ and 
conclude that the limits are independent of any spectral assumption. The energy 
range covered by these limits extends from 500\,GeV to 160\,TeV. A discussion of systematic uncertainties of HAWC flux measurements can
be found in \citet{hawcCrab}. These systematic
uncertainties are not incorporated into the limits.

For better comparison to other, non-coincident observations in this paper, we also analyzed
the 14 day period starting with the transit during the alert and ending on 2016-03-01, 00:30 UTC. 
Detector down time and quality cuts led to the exclusion of three transits (February 22, 25, and 26)
due to marginal coverage. No significant excess was found in the combined data for the eleven full
transits of the multiplet location and we also calculated limits for this period.

Since HAWC had been operating for more than a year before the alert and continues to provide daily 
monitoring, we also analyzed the integrated data from 508.2 transits of the triplet location 
between 2014-11-26 and 2016-06-02. No significant excess was found within the IceCube 
90\% error radius and we derived a quasi-differential limit for the average flux 
at the central location during this period, included in Table~\ref{tab:hawc}. 


\section{Discussion}
\label{sec:discussion}

\begin{center}
\begin{figure*}[tb]
\subfloat[Limits on short transients.]{\includegraphics[width=88mm]{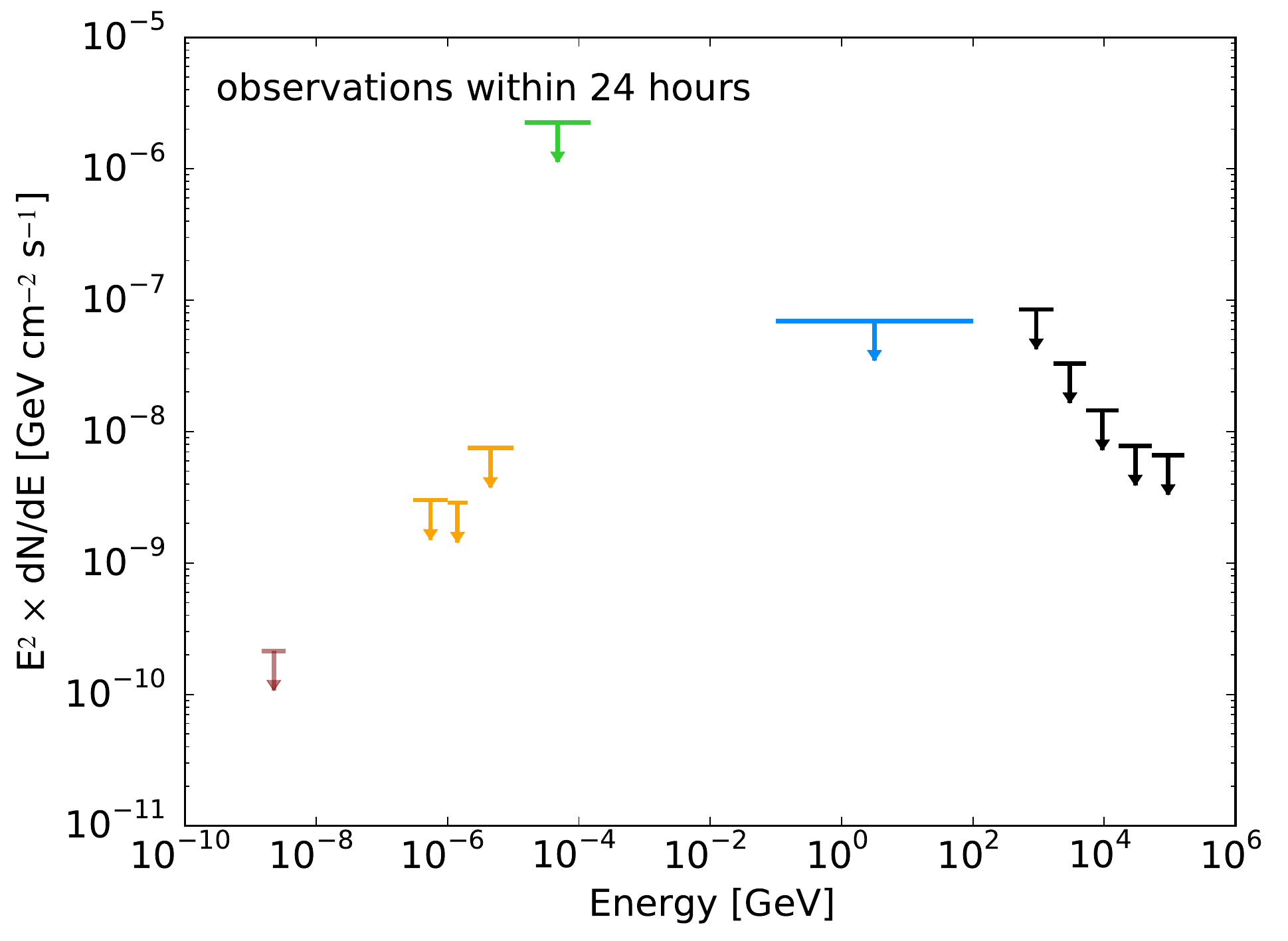}}
\hfill
 \subfloat[Limits on longer lasting transients.]{\includegraphics[width=88mm]{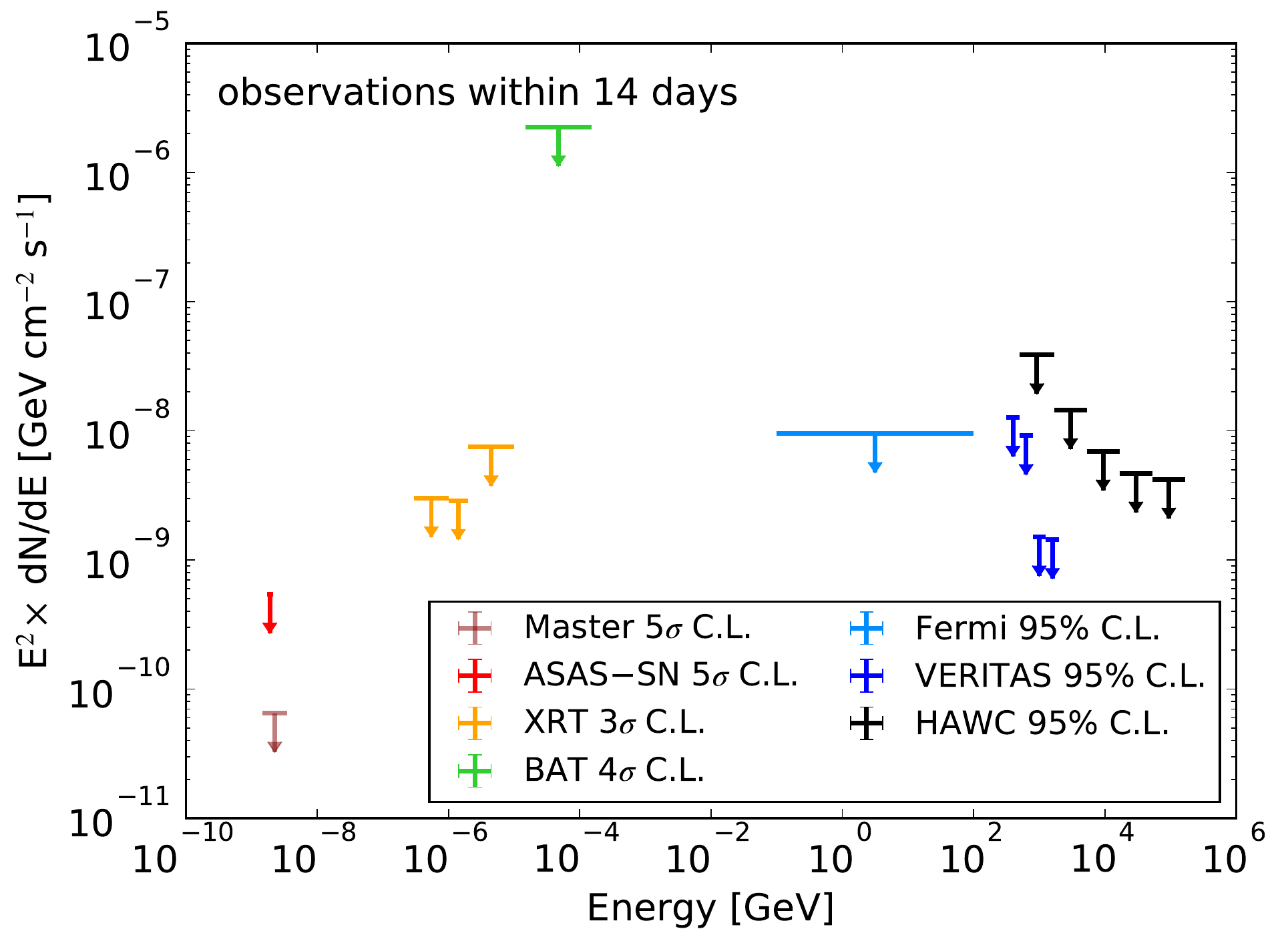}}
\caption{\label{fig:limits} Flux upper limits from the multiwavelength observations. The confidence level varies between the different observations as indicated in the legend and some limits depend on the assumed source spectrum (\emph{Swift} XRT and BAT $\Gamma=-2$ and \emph{Fermi} LAT $\Gamma=-2.1$; see Sect. \ref{sec:observations}). For the optical telescopes, the limit corresponding to the deepest observation is shown, while for the other instruments, all analyzed data were combined. The limit for the \emph{Swift} BAT is purely based on the observation taken 100\,s after the detection of the first neutrino (compare Sect. \ref{sec:bat}) and hence applies to prompt gamma-ray emission. Follow-up observations were triggered 22\,h after the detection of the neutrino triplet.}
\end{figure*}
\end{center}

We now draw conclusions from the non-detections during the follow-up observations and discuss the sensitivity of our program to a potential astrophysical multiplet source. An overview of the obtained limits is shown in Fig. \ref{fig:limits}.

As shown in Sect. \ref{sec:SignficanceCalc}, the detection of a neutrino triplet is expected once every $\sim\!13.7\,$yr from random coincidences of atmospheric background events and we cannot exclude such a chance alignment as the source of the triplet.
However, the neutrino multiplet could also stem from a transient neutrino source which emitted a $\sim\!100$\,s burst of TeV neutrinos. Since three neutrinos are detected, a potential source has to be either close-by or extremely energetic. Possible transient source classes include CCSN with an internal jet, GRBs, or AGN flares.

\subsection{Distance of an astrophysical neutrino source}
\label{sec:triplet_source}

We used a simulated population of transient neutrino sources to estimate their typical distances, which is important for the interpretation of the follow-up observations.
The astrophysical neutrino flux, detected at TeV/PeV energies, is best described by an $E^{-2.5}$ spectrum~\citep{icecube2015}\footnote{We note that a significantly shallower power law index of $E^{-2.13}$ was measured at energies above $\sim\!100\text{\,TeV}$ by \citet{icecube2016c}. The astrophysical neutrino spectra detected in both analyses are however consistent at those high energies. Like \citet{icecube2016c} we therefore interpret this apparent discrepancy as an indication of a break in the neutrino spectrum. The steep spectral index of $E^{-2.5}$ measured in \citep{icecube2015} is more relevant for this work because it extends to lower energies, down to $\sim\!10\text{\,TeV}$.}. We adopt this spectral shape as well as the measured normalization and consider simulated neutrino events which passed the event selection of the follow-up program. We expect the detection of 600 astrophysical muon neutrinos per year from the Northern sky. For this calculation, we extrapolated the measured neutrino spectrum down to 10\,GeV, below the IceCube sensitivity threshold. If we were only to consider events above $10\,\text{TeV}$ where the astrophysical flux has been measured~\citep{icecube2015}, we would expect the detection of 200 events per year. The large number of expected astrophysical neutrino events results from the broad, inclusive event selection of the follow-up program which aims to include all well-reconstructed track events.

\begin{figure}[tb]
\begin{center}
\includegraphics[width=88mm]{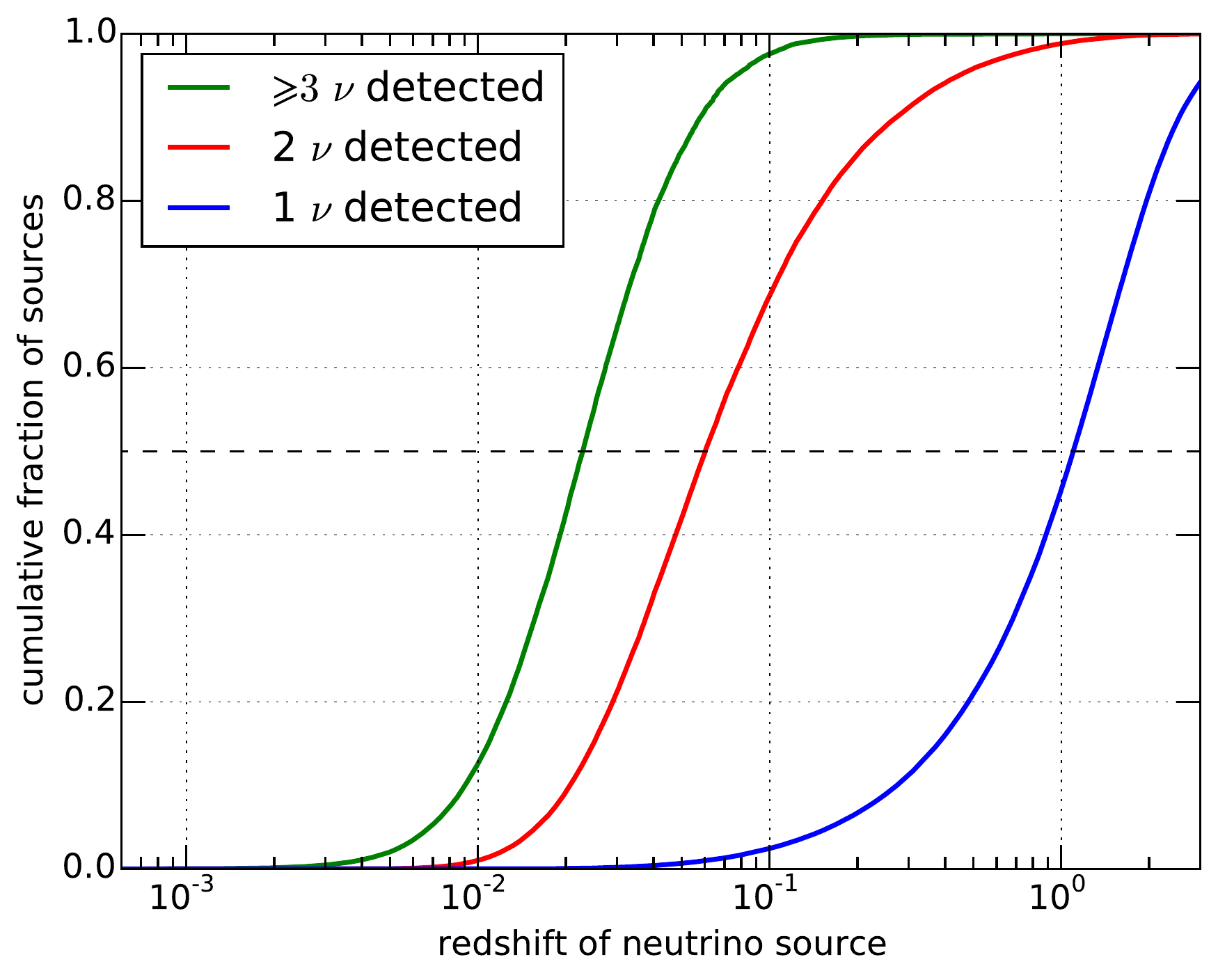}
\noindent
\caption{\small Probability of detecting a neutrino source within a certain redshift. The figure was generated by simulating a population of transient neutrino sources with a density of $10^{-6}\,\text{Mpc}^{-3}\,\text{yr}^{-1}$ distributed in redshift according to the star-formation rate and normalized to produce the detected astrophysical neutrino flux. Sources detected with only one single neutrino are on average far away (median redshift of $1.1$), while sources detected with three or more neutrinos must be located nearby.}
\label{fig:distances}
\end{center}
\end{figure}

We simulate a population of transient neutrino sources that accounts for the complete astrophysical neutrino flux. The cosmic star-formation rate approximately describes the redshift distributions of several potential neutrino sources, like CCSNe~\citep{cappellaro2015} and GRBs~\citep{wanderman2010, salvaterra2012, kruhler2015} which however tend to be located at slightly larger redshifts. We simulated a source population using the star-formation rate of~\citet{madau2014} and calculated for each source the probability of detecting it with a certain number of neutrinos after applying the event selection of the follow-up program. We find that a source detected with a single neutrino is located at a median redshift of $z = 1.1$, as shown in Fig.~\ref{fig:distances}.

To calculate the distance to a source detected with multiple neutrinos, we have to simulate how bright the individual sources are. We assume a population with a local source rate of $10^{-6}\,\text{Mpc}^{-3}\,\text{yr}^{-1}$, which corresponds to $\sim\!1\%$ of the CCSN rate (see e.g., \citealt{strolger2015}). If this population accounts for the astrophysical neutrino flux, we expect the detection of one neutrino triplet (or higher multiplet) per year. The rate of multiplet alerts, however, strongly depends on the spectral shape and considered energy range of the neutrino flux. We further assumed that the luminosity fluctuations between the neutrino sources follow a log-normal distribution with a width of one astronomical magnitude, which is comparable to the luminosity spread of CCSNe in optical light at optical wavelengths.

Figure \ref{fig:distances} shows that the source of a neutrino doublet has a median redshift of $z = 0.06$ and the median redshift of a triplet source is $z = 0.023$. We note that these results strongly depend on the spectral shape of the astrophysical neutrino flux. Considering only neutrino events with an energy above 10\,TeV, the source rate that yields one triplet per year is $3\times10^{-8}\,\text{Mpc}^{-3}\,\text{yr}^{-1}$ and the median redshift of a triplet source increases to $z=0.07$. If we would adopt the spectral index of $E^{-2.13}$ \citep{icecube2016c}, the source rate would be $2\times10^{-9}\,\text{Mpc}^{-3}\,\text{yr}^{-1}$ which would result into a median distance of $z=0.17$ for a triplet source.

Similar calculations apply to a population of GRBs, AGN, or blazars, which, however, have different source densities, redshift distributions, and luminosity functions. We also note that the duration of 100\,s to which our search is sensitive, does not enter these estimates and the distance calculation applies equally to steady sources.

In summary, we estimate that a CCSN detected with three neutrinos has a median redshift of $z = 0.023$ or less assuming that CCSNe account for the complete astrophysical neutrino flux. 
Typical CCSNe below this redshift are easily detected with optical telescopes if they are not unusually faint or strongly affected by absorption. Even without extrapolating the astrophysical neutrino spectrum to lower energies or when adopting the hard spectral shape measured at high energies the SN would likely still be detectable (compare Sect.~\ref{sec:sn}).

\subsection{Supernovae}
\label{sec:sn}

\begin{figure}[tb]
\begin{center}
\includegraphics[width=88mm]{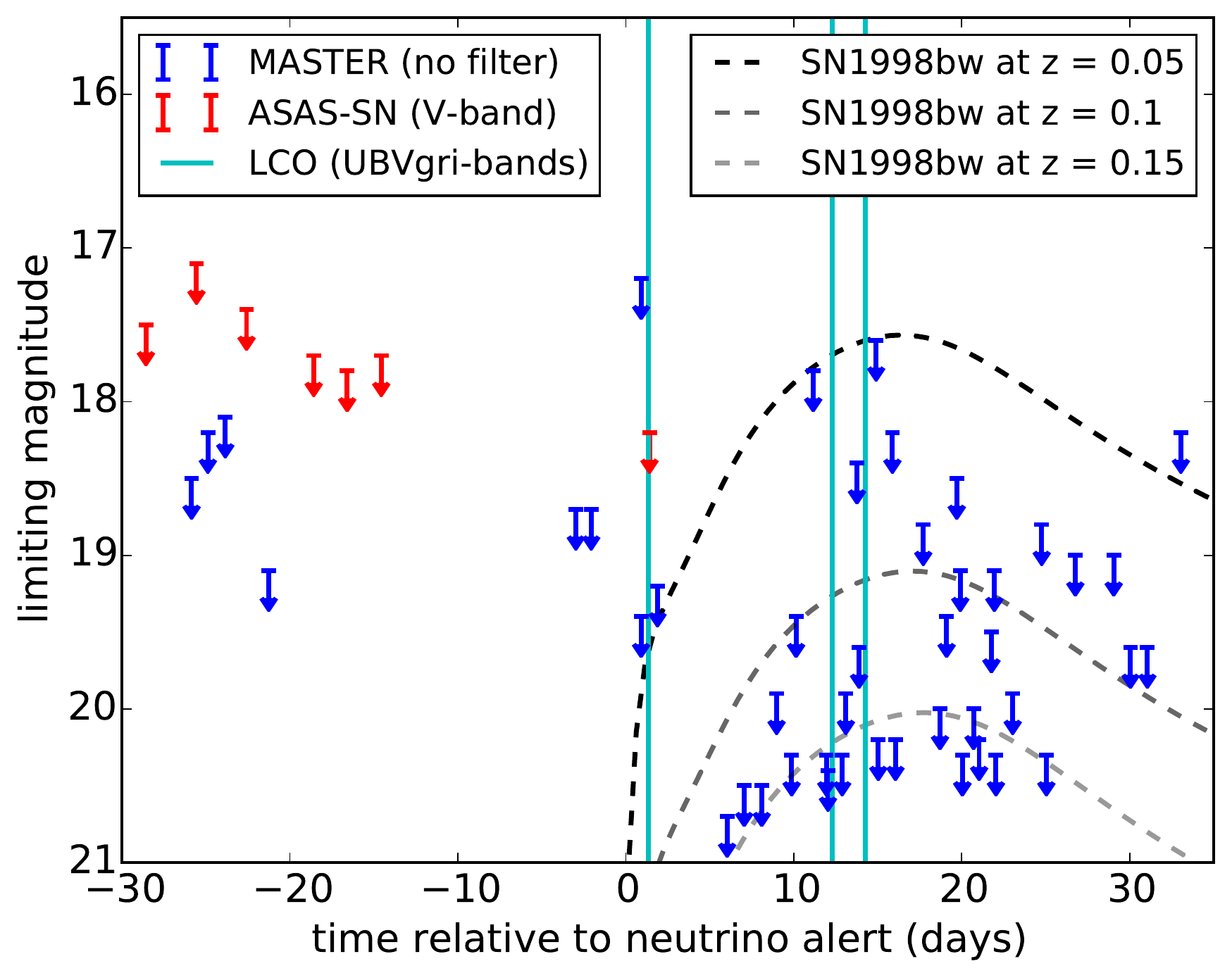}
\noindent
\caption{\small Optical $5\,\sigma$ limiting magnitudes from Table \ref{tab:optical_obs} and described in Sect. \ref{sec:optical}. LCO epochs (from Table \ref{tab:optical_obs_lcogt}) are shown as vertical lines. At these times, observations in the UBVgri bands were obtained, however no image subtraction was done. We overplot, as an example, the $V$-band light curve of SN\,1998bw, which was associated with GRB\,980425. The synthetic light curves of SN 1998bw have been created using the method presented in \citet{cano2014}.}
\label{fig:optical_ul}
\end{center}
\end{figure}

Figure \ref{fig:optical_ul} shows the constraints derived from the optical observations before and after the alert. As a comparison we plot the light curve of the bright Type Ic broadlined supernova SN\,1998bw which accompanied GRB\,980425 \citep{galama1998}. A similar supernova would be detectable out to a redshift of $\sim\!0.15$ which is much further than the expected redshift of a triplet source (compare Fig. \ref{fig:distances}). 

In follow-up observations of the most significant neutrino doublet detected so far, a fading Type IIn supernova was found \citep{icecube2015b}. A comparable event can be ruled out with the optical observations shown in Fig. \ref{fig:optical_ul}. We hence can exclude a nearby supernova unless it was unusually dim or heavily obscured.

\subsection{Gamma-ray bursts}
\label{sec:grb}

\begin{center}
\begin{figure*}[tb]
\subfloat[\label{fig:grb_limits}Gamma-ray and X-ray GRB light curves.]{\includegraphics[width=88mm]{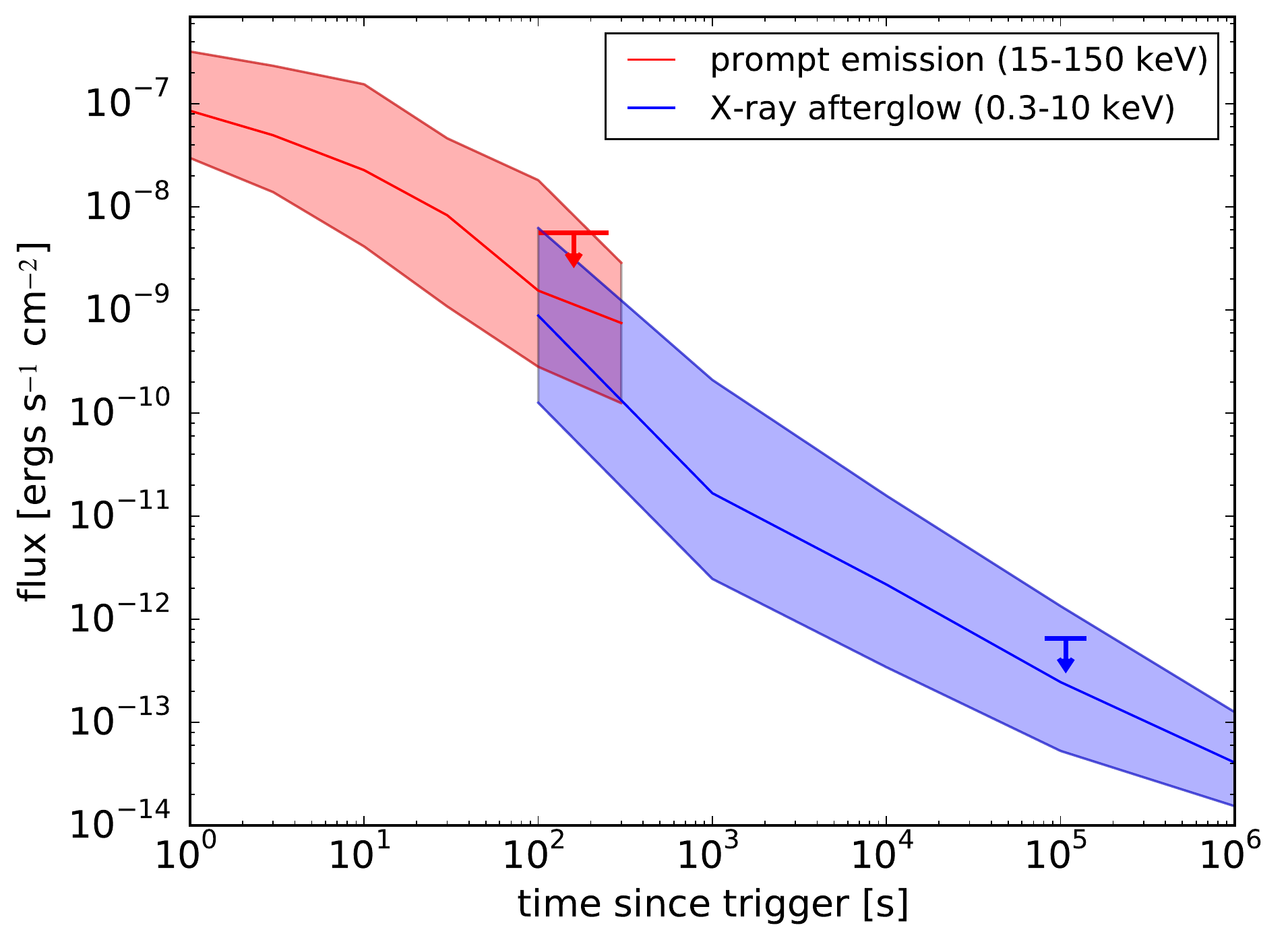}}
\hfill
 \subfloat[\label{fig:optical_afterglow}Optical GRB light curves.]{\includegraphics[width=88mm]{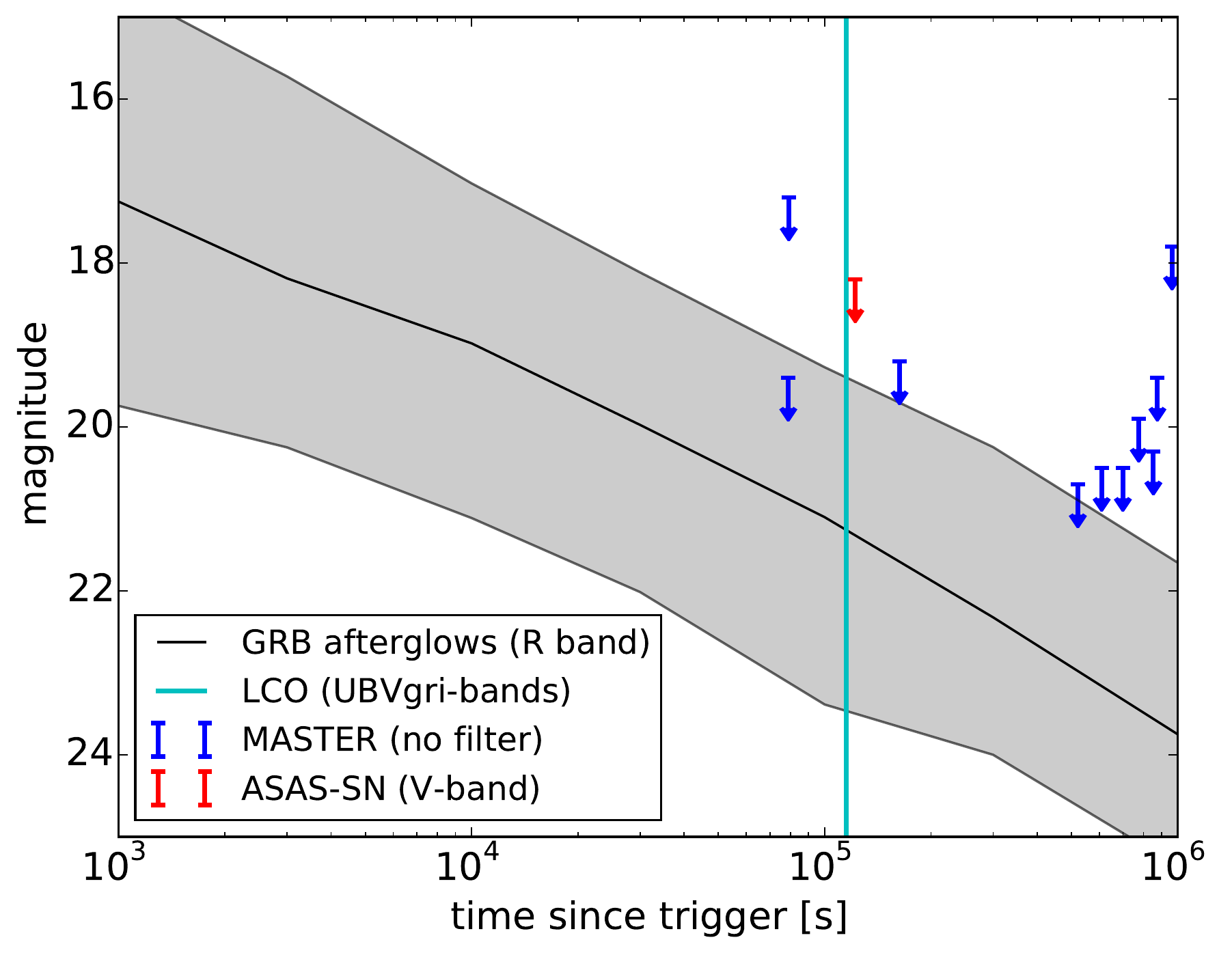}}
\caption{The shaded bands show the gamma-ray and X-ray light curves of detected GRBs (left) and optical afterglow light curves (right). The central line shows the median flux at the indicated time and the shaded bands include 80\% of all GRBs (i.e., the 10\% brightest and faintest afterglows are above or below the band, respectively). The arrows show the flux upper limits from the X-ray and optical follow-up observations (see Sects. \ref{sec:optical} and \ref{sec:xrays} for details).}
\end{figure*}
\end{center}

For CCSNe, we assumed that the source of a triplet must be close-by, following calculations in Sect. \ref{sec:triplet_source}. GRBs are much less frequent than CCSNe which means that they are on average located at larger distances. Another difference is that the luminosity differences between individual GRBs can be extreme in gamma-rays (see e.g., \citealt{wanderman2010}) which makes it likely that the neutrino luminosities also differ widely. Both effects boost the probability of finding a burst that is brighter (in neutrinos) than any burst that happened since the start of the follow-up program. We therefore do not restrict our search to very close-by GRBs.

To estimate whether or not a GRB would be detectable in the follow-up observations, we compare the upper limits to \emph{Swift} gamma-ray light curves and the X-ray afterglows in Fig. \ref{fig:grb_limits}. The light curves in the 15--50\,keV energy band were obtained from the UK \emph{Swift} Science Data Centre\footnote{\url{http://www.swift.ac.uk/burst_analyser/}} \citep{evans2010}. The median fluence deposited in this band is 41\% of the total fluence for GRBs in the Swift GRB catalog\footnote{\url{http://heasarc.gsfc.nasa.gov/W3Browse/swift/swiftgrb.html}}. We use this average factor to scale the fluxes to the full energy range of 15--150\,keV for which the BAT limit was calculated in Sect. \ref{sec:bat}. The central line corresponds to the median flux and the band contains 80\% of all GRB. The light curves are not corrected for the redshift and non-detections have been removed. The distribution is hence heavily biased and provides only a rough estimate for typical GRB light curves.

The limits from the \emph{Swift} BAT and XRT observations (see Sect. \ref{sec:xrays}) are comparable to the fluxes of bright GRBs. A brighter-than-average GRB would have been detected, but most GRBs are fainter than the limits. Neutrino multiplet alerts are usually sent to the XRT without delay and the XRT observations typically start within half an hour of the neutrino signal being detected \citep{evans2015} when GRBs are on average more than two orders of magnitude brighter.

We checked the archival data of the InterPlanetary Network (IPN; \citealt{hurley2010}) for a burst in temporal coincidence with the triplet. No confirmed\footnote{\url{http://heasarc.gsfc.nasa.gov/w3browse/all/ipngrb.html}} or unconfirmed\footnote{\url{http://www.ssl.berkeley.edu/ipn3/cosmic1.txt}} GRB was detected on the day of the triplet alert \citep{hurley2016}.

GRB afterglows are also detectable in optical observations. In Fig. \ref{fig:optical_afterglow} we compare our observations to a large sample of optical GRB afterglows \citep{kann2010,kann2011,kann2016}. As before, the shaded band includes 80\% of all GRBs in the sample. Only the brightest afterglows are detectable in the earliest optical observations. Nearby GRBs have been found to be accompanied by a Type Ic broadlined SN \citep{cano2016} and as shown in Sect. \ref{sec:sn} a nearby SN is disfavored. GRBs with a slightly misaligned jet might in addition produce orphan afterglows which could be detectable in optical (see e.g., \citealt{zou2007, ghirlanda2015, kathirgamaraju2016}) or in X-ray observations (see e.g., \citealt{evans2016, sun2017}).

Correlation analyses of detected GRBs with IceCube neutrino events show that gamma-ray bright GRBs are not the main sources of the astrophysical neutrino flux \citep{icecube2012b,icecube2015c, icecube2016}. These limits however only apply to gamma-ray bright sources which are routinely detected with current gamma-ray satellites. To gain sensitivity to low-luminosity GRBs, which might be missed in gamma rays, quick X-ray and optical observations are essential. 
In addition, early optical follow-up observations can be used to look for rapidly fading transients without associated gamma-ray emission (like the object found by \citealt{cenko2013}) or for GRBs that were missed by gamma-ray detectors \citep{cenko2015}.

In summary we conclude that a bright GRB likely would have been detected by both the BAT and the \emph{Swift} XRT while a typical GRB is too faint. Moreover, there is a class of low-luminosity GRBs \citep{liang2007} which could be below the detection threshold of existing instruments even when occurring at low redshifts. The accompanying SNe of such objects might however be detectable (compare Sect. \ref{sec:sn}).

\subsection{Active galactic nuclei}
\label{sec:agn}

The durations of typical AGN flares observed in gamma rays range from minutes to several weeks. The time scale of 100\,s is hence short and implies that the neutrinos have to be emitted from a very small region of the jet even when taking into account relativistic beaming.
The dedicated gamma-ray follow-up program of IceCube searches for neutrino emission on time scales of up to three weeks \citep{kintscher2016, icecube2016f}. Currently the gamma-ray follow-up program searches for emission from sources on a predefined source list and none of those sources is consistent with the triplet direction.

The \emph{Swift} XRT observations resulted in the detection of one known AGN (X1) and one AGN candidate (X6) within the 50\% error circle (see Sect. \ref{sec:xrt} and Appendix \ref{sec:x6}). X1 is a blazar but
does not exhibit flaring compared to X-ray observations taken in 2010 and 2011. X6 fades away following the neutrino alert, but is not very bright overall (see Appendix \ref{sec:x6}) and remains undetected in gamma rays.

No flares were detected in gamma rays by the \emph{Fermi} LAT, VERITAS, or HAWC. The three \emph{Fermi} LAT sources within the 90\% error circle of the event did not show a significant flux excess within the weeks before and after the alert. 3FGL\,J0156.3+3913 underwent flares in 2009, but was inactive at the time of the neutrino alert and 3FGL\,J0152.2+3707 has been classified as a blazar candidate of uncertain type \citep{acero2015}.

The third source, RGB\,J0136+391 (or 3FGL\,J0136.5+3905), is a high frequency peaked BL Lac object at a redshift of $>0.4$ (inferred from the non-detection of its host galaxy by \citealt{nilsson2012}). It was detected in VHE gamma rays by MAGIC in November 2009 with an observation time of 6.5\,h\footnote{\url{https://www.mpi-hd.mpg.de/hd2012/pages/presentations/Mazin.pdf}} (see also the non-detection by VERITAS at a similar time; \citealt{veritas2012}). During the VERITAS observation eight days after the neutrino alert the source was not detected with $\sim\!1$\,h of observation time (see Sect. \ref{sec:veritas}). The source hence did not undergo a very bright and long-lasting flare. A shorter or less luminous flare is not excluded, even though no variability was detected by the \emph{Fermi} LAT during this period (see Sect.~\ref{sec:lat}).

To estimate how likely it is to find an unrelated VHE source within the 90\% error circle of this neutrino alert we consider all AGN in the Northern sky that are detected in VHE gamma rays. The 60 sources in the TeVCat\footnote{status of 2017-01-04} yield a probability of $\sim\!6$\% of finding a source within $3.6^\circ$ of a random position. This rough estimate does not consider that neither the neutrino alerts nor the detected VHE sources are distributed randomly over the sky. It indicates, however, that the presence of RGB\,J0136+391 could be a coincidence.

We conclude that there is no evidence for AGN flares within the region of interest. We derived flux upper limits for two time ranges using observations taken within a period of 24\,h and 14\,days after the neutrino detection. The limits in the different wavelength regimes are shown in Fig. \ref{fig:limits}. It is unclear whether or not an AGN flare below the derived limits can yield a large neutrino flux.

\section{Summary}

For the first time, the IceCube follow-up program was triggered by three neutrinos within 100\,s and with reconstructed directions consistent with a point source origin. Such an alert is expected from the coincidence of background events once every $13.7\,$yr.
Considering that the program has been running since December 2008 in different configurations, the probability of detecting one or several triplets from atmospheric background is 32\%. When an alternative event reconstruction algorithm (Spline MPE) is applied, the event directions have larger angular separations and the multiplet would not have been considered interesting. This is an additional indication that the multiplet probably is not astrophysical.

Even so, the triplet is the most significant neutrino multiplet detected since the beginning of the follow-up program and follow-up observations were obtained in different wavelength regimes to search for a potential electromagnetic counterpart (see Sect. \ref{sec:observations}). No transient source was detected in the optical or gamma-rays regimes. The \emph{Swift} XRT detected one highly variable X-ray source whose nature remains unknown (see Appendix \ref{sec:x6} for a detailed discussion). As described in Sect. \ref{sec:xrt} this source is not consistent with a GRB. It could be a flaring AGN which however would not be very bright and is not detected in gamma rays. We therefore conclude that this X-ray source is most likely  not connected to the neutrinos.

Our optical observations are sufficient to rule out a nearby CCSN (see Sect. \ref{sec:sn}). A bright GRB would likely have been detected both in the \emph{Swift} XRT observations and by the \emph{Swift} BAT which serendipitously observed the location within minutes of the alert (see Sect.~\ref{sec:grb}). However, low-luminosity GRBs might be too dim to be detectable even if they are located at low redshifts. No flaring AGN were found in either X-rays, gamma rays, or very-high-energy gamma rays. We conclude that no likely counterpart was identified in follow-up observations. Since the neutrino alert is consistent with background (see Sect.~\ref{sec:SignficanceCalc}) we cannot place new constraints on astrophysical models for neutrino emission. 

This work demonstrates that the IceCube follow-up program is able to trigger observations in near real-time to search for transient neutrino sources. While this alert was not triggered automatically, causing a delay of 22\,h, the system typically issues alerts within $\sim\!1$\,min, such that even rapidly fading transients are observable. Using additional serendipitous observations we demonstrate in Sect.~\ref{sec:discussion} that the program is well suited to testing several suggested source classes.

We are planning to replace the fixed cuts used currently in the optical follow-up program (compare Sect.~\ref{sec:multi_filter}) with a likelihood search. This will increase the sensitivity and allow us to search for sources that last longer than 100\,s. A global network of optical telescopes, including ASAS-SN, LCO, MASTER, and the upcoming Zwicky Transient Facility \citep{bellm2014}, will moreover result into much better data coverage compared to previous years.

In the case of an astrophysical multiplet detection, the follow-up network employed here and in its future extension should enable the detection of its electromagnetic counterpart and hence identification of a neutrino source. Moreover, some of the methods presented here are readily generalizable to searches for counterparts of high-energy single neutrino events or for follow-up observations of gravitational wave events.

\begin{acknowledgements}

Neil Gehrels sadly died during the late stage of the production of this paper. As Swift PI he was an enthusiastic supporter of multi-messenger observations; he will be sorely missed.\\

The IceCube collaboration acknowledges the support from the following agencies: U.S. National Science Foundation-Office of Polar Programs, U.S. National Science Foundation-Physics Division, University of Wisconsin Alumni Research Foundation, the Grid Laboratory Of Wisconsin (GLOW) grid infrastructure at the University of Wisconsin - Madison, the Open Science Grid (OSG) grid infrastructure; U.S. Department of Energy, and National Energy Research Scientific Computing Center, the Louisiana Optical Network Initiative (LONI) grid computing resources; Natural Sciences and Engineering Research Council of Canada, WestGrid and Compute/Calcul Canada; Swedish Research Council, Swedish Polar Research Secretariat, Swedish National Infrastructure for Computing (SNIC), and Knut and Alice Wallenberg Foundation, Sweden; German Ministry for Education and Research (BMBF), Deutsche Forschungsgemeinschaft (DFG), Helmholtz Alliance for Astroparticle Physics (HAP), Research Department of Plasmas with Complex Interactions (Bochum), Germany; Fund for Scientific Research (FNRS-FWO), FWO Odysseus programme, Flanders Institute to encourage scientific and technological research in industry (IWT), Belgian Federal Science Policy Office (Belspo); University of Oxford, United Kingdom; Marsden Fund, New Zealand; Australian Research Council; Japan Society for Promotion of Science (JSPS); the Swiss National Science Foundation (SNSF), Switzerland; National Research Foundation of Korea (NRF); Villum Fonden, Danish National Research Foundation (DNRF), Denmark

This work made use of data supplied by the UK \emph{Swift} Science Data Centre at the University of Leicester. Funding for the Swift project in the UK is provided by the UK Space Agency. Part of this work was facilitated by the GROWTH project, a partnership in international research and education, NSF PIRE Grant No 1545949.

ASAS-SN is supported by NSF grant AST-1515927. Development of ASAS-SN
has been supported by NSF grant AST-0908816, the Center for Cosmology
and AstroParticle Physics at the Ohio State University, the Mt. Cuba
Astronomical Foundation, and by George Skestos.

The \Fermi\ LAT Collaboration acknowledges generous ongoing support
from a number of agencies and institutes that have supported both the
development and the operation of the LAT as well as scientific data analysis.
These include the National Aeronautics and Space Administration and the
Department of Energy in the United States, the Commissariat \`a l'Energie Atomique
and the Centre National de la Recherche Scientifique / Institut National de Physique
Nucl\'eaire et de Physique des Particules in France, the Agenzia Spaziale Italiana
and the Istituto Nazionale di Fisica Nucleare in Italy, the Ministry of Education,
Culture, Sports, Science and Technology (MEXT), high-energy Accelerator Research
Organization (KEK) and Japan Aerospace Exploration Agency (JAXA) in Japan, and
the K.~A.~Wallenberg Foundation, the Swedish Research Council and the
Swedish National Space Board in Sweden. Additional support for science analysis during the operations phase is gratefully
acknowledged from the Istituto Nazionale di Astrofisica in Italy and the Centre National d'\'Etudes Spatiales in France.

The HAWC Collaboration acknowledges the support from: the US National Science Foundation
(NSF); the US Department of Energy Office of High-Energy Physics; the
Laboratory Directed Research and Development (LDRD) program of Los
Alamos National Laboratory; Consejo Nacional de Ciencia y
Tecnolog\'{\i}a (CONACyT), M{\'e}xico (grants 271051, 232656, 260378,
179588, 239762, 254964, 271737, 258865, 243290, 132197), Laboratorio
Nacional HAWC de rayos gamma; L'OREAL Fellowship for Women in Science
2014; Red HAWC, M{\'e}xico; DGAPA-UNAM (grants IG100317, IN111315,
IN111716-3, IA102715, 109916, IA102917); VIEP-BUAP; PIFI 2012, 2013,
PROFOCIE 2014, 2015;the University of Wisconsin Alumni Research
Foundation; the Institute of Geophysics, Planetary Physics, and
Signatures at Los Alamos National Laboratory; Polish Science Centre
grant DEC-2014/13/B/ST9/945; Coordinaci{\'o}n de la Investigaci{\'o}n
Cient\'{\i}fica de la Universidad Michoacana. We thank Luciano D\'{\i}az
and Eduardo Murrieta for technical support of the HAWC detector.

Support for I.~Arcavi was provided by NASA through the Einstein Fellowship Program, grant PF6-170148. D.~A.~Howell, C.~McCully, and G.~Hosseinzadeh are supported by NSF-1313484. This work makes use of observations from the LCO network.

The MASTER collaboration was supported by the Russian Science Foundation 16-12-00085 (also RSF 1612-00085) and RFBR 15-02-07875.

VERITAS research is supported by grants from the U.S. Department of Energy Office of Science, the U.S. National Science Foundation and the Smithsonian Institution, and by NSERC in Canada. VERITAS acknowledges the excellent work of the technical support staff at the Fred Lawrence Whipple Observatory and at the collaborating institutions in the construction and operation of the instrument. 

E.~O.~Ofek and A.~Gal-Yam acknowledge a Minerva grant.

\end{acknowledgements}



\bibliographystyle{aa}
\bibliography{OFUTriplet}


\clearpage

\begin{appendix}
\section{The nature of the \xray\ source X6}
\label{sec:x6}

\begin{figure}[b]
\begin{center}
\includegraphics[width=88mm]{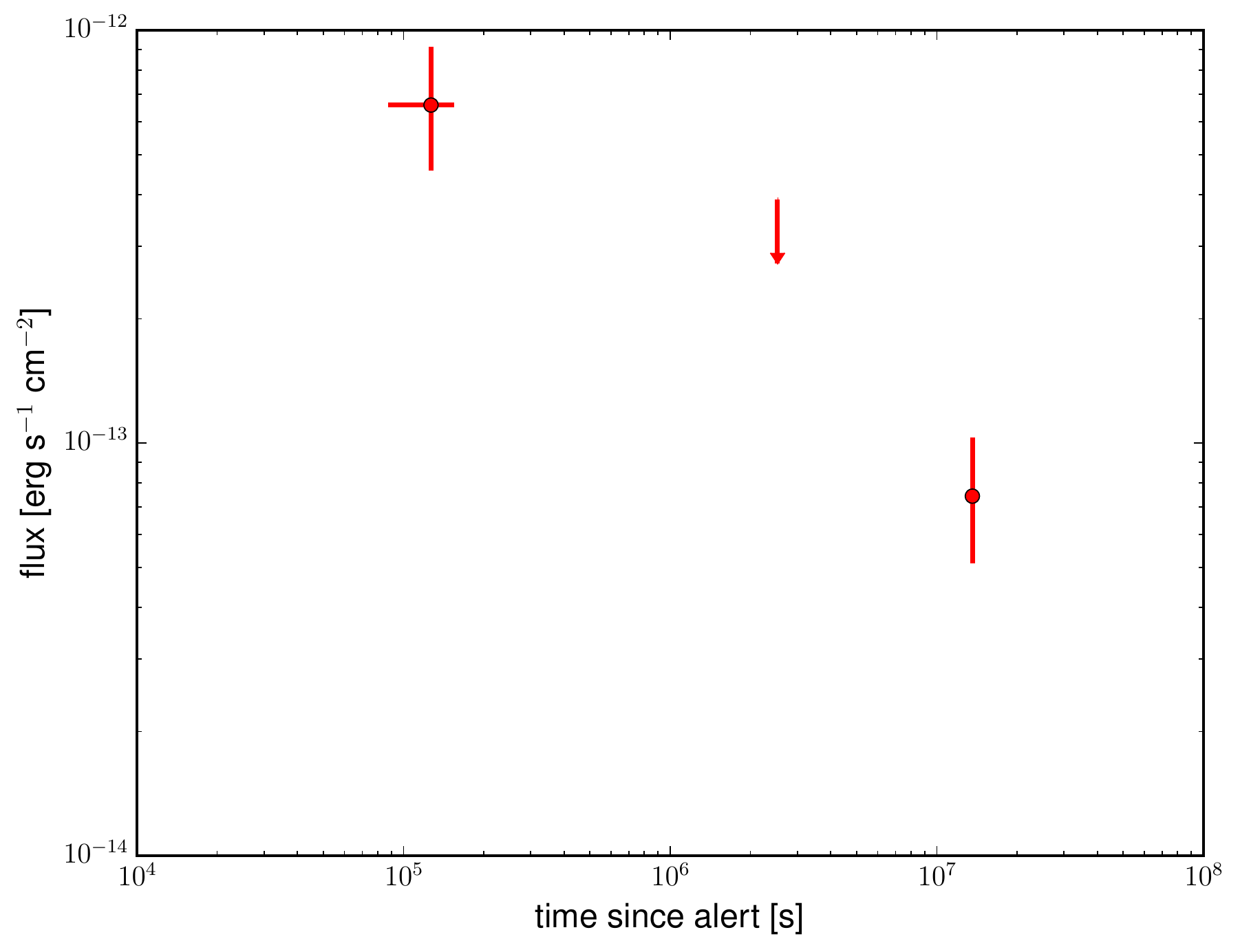}
\noindent
\caption{\small XRT light curve of X6 in the 0.3--10\,keV range. The error bars are at the $1\,\sigma$ level and the upper limit is at $3\,\sigma$ confidence.}
\label{fig:x6_xrt_lc}
\end{center}
\end{figure}

As described in Sect. \ref{sec:xrt}, we detected a highly variable \xray\ source (see Fig. \ref{fig:x6_xrt_lc}) without an obvious optical counterpart. The first \emph{Swift} detection of X6 has a probability of a spurious detection of $<0.3$\% \citep{evans2014}. Since it was later re-detected with high confidence, we consider X6 a genuine astrophysical source.

The X-ray light curve of X6 is consistent with a $t^{-0.5}$ decay over a period of 5 months. During this time, its X-ray flux in the 0.3 to 10.0\,keV energy range fades by a factor of nine.
As an aside, we note that the source appears to exhibit variability during the third XRT observation, where 9 out of 11 counts are detected during the first 45\% of the exposure time. However, closer investigation revealed that this was due to the source being placed near a bad column on the detector, which leads to lost counts during the second half of the exposure. The flux estimate for the source takes this loss of exposure into account.

The probability of detecting a serendipitous X-ray source at the flux level of X6 is 5\% when considering the covered area as well as the exposure time of the tiled XRT observation \citep{evans2015}. \citet{voges1999} systematically studied the variability of X-ray sources detected by ROSAT in the $0.07$-- $2.4$\,keV energy range. They find that 9\% of the sources are variable by a factor of more than three. Out of those sources 57\% are unidentified, 30\% are stars and the remaining 13\% are extragalactic sources; mostly AGN. Only 0.7\% of the sources in their sample are variable by a factor of 10 or more. The detection of X6 is hence unexpected.

\begin{figure}[tb]
\begin{center}
\includegraphics[width=88mm]{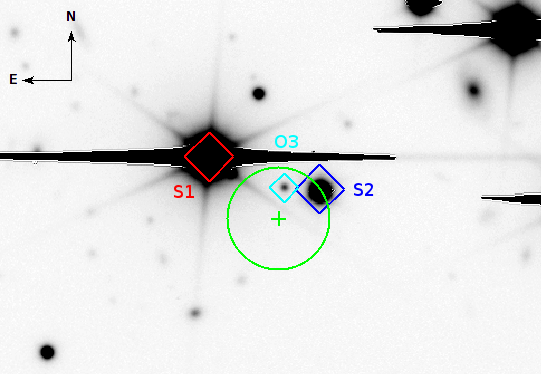}
\noindent
\caption{\small Keck/LRIS image. Shown in green are the position of X6 and the 90\% error circle which has a radius of $6.2\arcsec$. Three potential optical counterparts are marked with diamonds: S1 in red, S2 in blue and O3 in cyan (see Table \ref{tab:x6_counterpart} for details). While S1 and S2 are Sun-like stars (see Fig. \ref{fig:x6_spec}), the nature of O3 is unknown.}
\label{fig:x6_keck}
\end{center}
\end{figure}

We identified two possible scenarios that are consistent with all obtained observations. The X-rays could be emitted by a distant and obscured highly variable AGN. Alternatively they could be associated with one of two nearby stars, S1 or S2 (see Fig. \ref{fig:x6_keck}), or an \xray\ bright binary companion of one the stars. Neither scenario yields a detectable neutrino flux on Earth.

\begin{figure}[tb]
\begin{center}
\includegraphics[width=88mm]{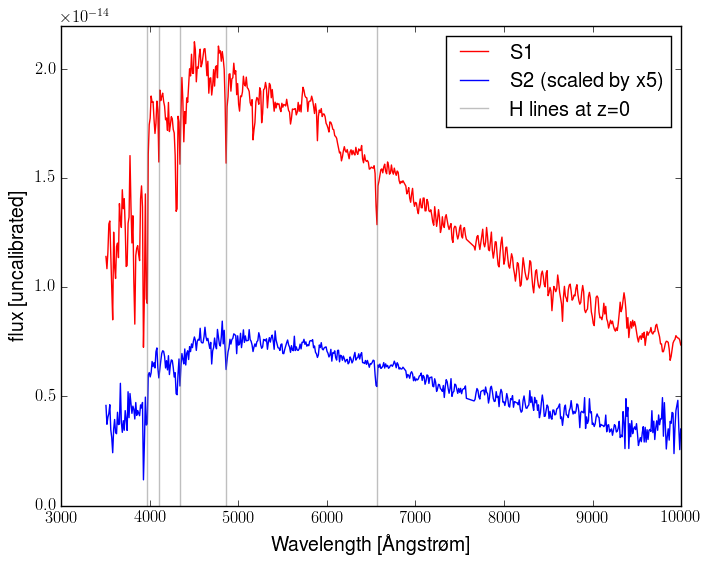}
\noindent
\caption{\small LCO spectra of S1 and S2 (compare Fig.~\ref{fig:x6_keck}). Hydrogen absorption lines show that they are F or G stars in our Galaxy. Telluric bands at 6870\,\AA\ and 7600\,\AA\ were removed from the spectra.}
\label{fig:x6_spec}
\end{center}
\end{figure}

\begin{table*}
{\small
\hfill{}
\caption{Possible optical counterparts of X6.}
\label{tab:x6_counterpart}     
\begin{center}
\begin{tabular}{lllllll}        
\hline\hline       
Name & Object type & RA & Dec & Ang. sep. from X6 & Distance & $R$ band magnitude \\
     &             & ($^\circ$) & ($^\circ$) & (\arcsec) & (pc) & (mag) \\
\hline
S1 & F or G star  & $25.01375$ & $+39.60553$ & 11.6 & $\sim\!510$    & 13.0 \\
S2 & G star       & $25.00892$ & $+39.60431$ & 6.2  & $\sim\!1500$   & 15.8 \\
O3 & unknown      & $25.01044$ & $+39.60440$ & 3.9  & unknown      & 20.7 \\
\hline                                  
\end{tabular}
\end{center}
}
\tablefoot{The locations of the three objects are shown in Fig. \ref{fig:x6_keck} and the spectra of S1 and S2 are presented in Fig \ref{fig:x6_spec}. The magnitudes of S1 and S2 have been measured from PTF images and the one of O3 is from the Keck/LRIS image. All magnitudes are approximate because the point spread functions of the three objects overlap.}
\hfill{}
\end{table*}

\subsection{A distant active galactic nucleus}

The faint object, O3, is the only detected source within the 90\% error circle of X6 (compare Fig. \ref{fig:x6_keck} and Table \ref{tab:x6_counterpart}). Since we do not have a spectrum or additional photometric points we do not know whether it is a star, a compact galaxy, or an AGN. An AGN could easily account for the detected \xray\ flux even if it is located at a high redshift ($z\gtrsim1$; see e.g., \citealt{aird2015}). It is also possible that O3 is an unrelated object and that an even fainter AGN is located within the error circle of X6.

An AGN can be faint in the optical if the accretion disk and jet, if present, are obscured by dust. If it is located at a high redshift its host galaxy may not be detectable either. The absence of a bright optical counterpart is therefore not unusual, but it does indicate that the AGN likely is not close-by.

AGN typically have variable \xray\ luminosities due to perturbations in their accretion disk. However, large amplitude variability, as observed for X6, is only detected for a few percent of all AGN \citep{strotjohann2016}. Such bright \xray\ flares can, for example, be caused by changing jet activity in blazars. No gamma-ray emission is detected by the \emph{Fermi} LAT, VERITAS, or HAWC (compare Sect. \ref{sec:gamma}) and no known radio source is consistent with the position of X6. So there is no further evidence for a flaring blazar and if a jet is present, it does not emit a strong flux of GeV or TeV photons.

Even though blazars are promising candidates for the emission of high-energy neutrinos (see e.g., \citealt{padovani2016} and references therein), it seems unlikely that a rather faint \xray\ source that is not detected at higher energies emits a strong neutrino flux. We therefore do not consider this AGN candidate a possible counterpart for the detected neutrinos.

\subsection{Stellar \xray\ flares}

In addition to O3, the stars S1 and S2, are located close to the 90\% error circle of X6 as shown in Fig.~\ref{fig:x6_keck} and Table~\ref{tab:x6_counterpart}. Especially S2 is just at the edge of the error circle and has a reasonable chance to be associated with X6. Optical spectra taken with LCO are shown in Fig. \ref{fig:x6_spec}. The hydrogen absorption lines at redshift zero show that both sources are stars. The temperature of S2 is very similar to the Sun (class G2) while S1 has a higher temperature. It could either be a hot G star or a low-temperature F star\footnote{Standard spectra for comparison can be found at \url{http://classic.sdss.org/dr5/algorithms/spectemplates/}.}.


A rough estimate can show whether S2 can account for the detected X-ray flux. Assuming that the star has solar luminosity we estimate its distance to be $\sim\!1500\,\text{pc}$. Based on this distance the X-ray luminosity is $\sim\!10^{32}\,\text{erg\,} \text{s}^{-1}$, which is a factor of $10^4$ brighter than the flaring Sun. Assuming again solar luminosity S1 is located at $\sim\!500$\,pc and would have to emit an \xray\ luminosity of $\sim\!10^{31}\,\text{erg\,} \text{s}^{-1}$ to account for the detected X-ray flux. In the samples presented by \citet{agueros2009} and \citet{wright2010} less than one percent of the stars detected in X-rays reach luminosities above $10^{31}\,\text{erg\,} \text{s}^{-1}$ and only $\sim\!10$ such stars have ever been detected. If S1 or S2 is the source of the X-rays, the star underwent an extreme flare.

Extreme stellar X-ray flares can be emitted by close or active binary systems (see e.g., \citealt{wright2010}). It is possible S1 or S2 has a binary partner that is too faint to be detectable in the optical spectra in Fig. \ref{fig:x6_spec}. The spectra do not show Balmer emission lines thus there is no evidence for an accretion disk. However, a close binary without mass transfer would be consistent with our observations. To search for evidence for a binary system we analyze the forced photometry light curve of S2 which consists of 185 $g$ band PTF images acquired over more than three years. While there is evidence for variability at a low level of $0.05\,\text{mag}_{g}$, no significant period was detected. The optical light curve hence does not provide evidence for a binary partner, but neither can we rule out its presence. We cannot repeat this analysis for S1 which is saturated in most PTF observations. O3 is not detected in individual or stacked PTF observations and we hence do not know whether or not it is variable in the optical.

Another possibility is that O3 is a nearby faint star (e.g., an M dwarf) that undergoes a strong X-ray flare or it could be an X-ray binary. Due to the lack of an optical spectrum we cannot verify this scenario.

\subsection{Conclusion}

We detected a highly variable but faint X-ray source which could be associated with several potential optical counterparts. Five months after the initial detection the source was re-detected in X-rays at a flux level nearly ten times lower. This latter detection rules out a GRB or a typical tidal disruption event.

We cannot make a definitive conclusion about the nature of this source. The X-rays could be associated with one of the stars S1 or S2. In this case we have found a very bright and rare stellar flare. Another possible scenario is that the X-rays are emitted by O3 (or a fainter object undetected in the optical). O3 could either be a distant flaring AGN or it could be a nearby faint star exhibiting a strong X-ray flare.

X6 is quite faint in X-rays and not detected in gamma rays. We therefore do not consider it a likely source of the detected neutrinos even if it is a flaring AGN.


\section{Observations}

The following tables list the observations and resulting limits by the different telescopes. Table~\ref{tab:optical_obs} shows the observations by ASAS-SN and MASTER and Table~\ref{tab:optical_obs_lcogt} shows those by LCO. Table~\ref{tab:xrtUL} lists the limits obtained from \emph{Swift} observations. The limits calculated by VERITAS are shown in Table~\ref{tab:veritas} and the ones by HAWC in Table~\ref{tab:hawc}. An overview plot including the limits at different wavelengths is shown in Fig.~\ref{fig:limits}.

\begin{table*}
{\small
\hfill{}
\caption{Optical observations from MASTER and ASAS-SN}
\label{tab:optical_obs}    
\begin{center}
\begin{tabular}{llllll}
\hline\hline
Telescope     & Time, UTC     & Time-$t_0$ (days)& Filter     & Number of exposures and exposure time     & $5\,\sigma$ limiting mag.\\ 
\hline
ASAS-SN Brutus          & 2016-01-20.24     & $-28.57$  & V     & 3 (90\,s)             & $17.5$\\
MASTER-IAC          & 2016-01-22 22:56:34   & $-25.85$  &       & 3 (60\,s)       & 18.5\\
ASAS-SN Brutus          & 2016-01-23.25     & $-25.56$  & V     & 3 (90\,s)             & $17.1$\\
MASTER-IAC          & 2016-01-23 22:14:49   & $-24.88$  &       & 3 (60\,s)       & 18.2\\
MASTER-IAC          & 2016-01-24 23:09:39   & $-23.84$  &       & 3 (60\,s)       & 18.1\\
ASAS-SN Brutus          & 2016-01-26.23     & $-22.58$  & V     & 3 (90\,s)             & $17.4$\\
MASTER-Tunka        & 2016-01-27 13:12:46   & $-21.25$  &       & 3 (60\,s)       & 19.1\\
ASAS-SN Brutus          & 2016-01-30.23     & $-18.58$  & V     & 3 (90\,s)             & $17.7$\\
ASAS-SN Brutus          & 2016-02-01.22     & $-16.58$  & V     & 3 (90\,s)             & $17.8$\\
ASAS-SN Brutus          & 2016-02-03.25     & $-14.56$  & V     & 3 (90\,s)             & $17.7$\\
MASTER-IAC          & 2016-02-14 20:03:58   & $-2.97$   &       & 3 (60\,s)       & 18.7\\
MASTER-Kislovodsk   & 2016-02-15 17:56:50   & $-2.06$   &       & 6 (60\,s)       & 18.7\\
MASTER-Kislovodsk   & 2016-02-18 17:15:58   & $0.91$      &     & 25$\times$2 (180\,s)   & 19.4 (18.6)\\ 
MASTER-Tunka        & 2016-02-18 17:20:21   & $0.92$      &     & 3 (60\,s)       & 17.2\\
ASAS-SN Brutus          & 2016-02-19.22     & $1.41$    & V     & 20 (90\,s)        & $18.2$\\
MASTER-Kislovodsk   & 2016-02-19 16:37:32   & $1.89$       &    & 18$\times$2(180\,s)    & 19.2 (18.5)\\
MASTER-IAC          & 2016-02-23 20:11:37   & $6.03$      &     & 20$\times$2 (180\,s)   & 20.7 (19.5)\\
MASTER-IAC          & 2016-02-24 20:32:18   & $7.05$      &     & 4$\times$2 (180\,s)    & 20.5 (19.8)\\
MASTER-IAC          & 2016-02-25 21:36:18   & $8.09$      &     & 4$\times$2 (180\,s)    & 20.5 (19.7)\\
MASTER-Kislovodsk   & 2016-02-26 18:49:01   & $8.98$      &     & 12$\times$2(180\,s)    & 19.9 (19.2)\\
MASTER-Kislovodsk   & 2016-02-27 16:21:47   & $9.87$      &     & 20$\times$2 (180\,s)   & 20.3 (19.9)\\
MASTER-IAC          & 2016-02-27 22:40:13   & $10.14$      &    & 3$\times$2 (180\,s)    & 19.4 (18.9)\\
MASTER-IAC          & 2016-02-27 22:59:51   & $10.15$      & B  & 2 (180\,s)      & 19.0 (18.7)\\
MASTER-IAC          & 2016-02-27 22:59:51   & $10.15$      & I  & 2 (180\,s)      & 17.0\\
MASTER-IAC          & 2016-02-28 23:08:13   & $11.16$      &    & 6$\times$2 (180\,s)    & 17.8\\
MASTER-Kislovodsk   & 2016-02-29 17:51:45   & $11.94$      &    & 18$\times$2 (180\,s)   & 20.3 (19.8)\\
MASTER-IAC          & 2016-02-29 20:17:28   & $12.04$      &    & 4$\times$2 (180\,s)    & 20.4 (19.9)\\
MASTER-IAC          & 2016-02-29 20:28:52   & 12.05     & B     & 2 (180\,s)      & 20.2\\
MASTER-IAC          & 2016-02-29 20:28:52   & 12.05     & I     & 2 (180\,s)      & 18.0\\
MASTER-Kislovodsk   & 2016-03-01 16:31:39   & $12.88$   &       & 32 (180\,s)     & 20.3 (19.9)\\
MASTER-IAC          & 2016-03-01 21:51:21   & $13.10$   &       & 4$\times$2 (180\,s)    & 19.9 (19.3)\\
MASTER-IAC          & 2016-03-01 22:14:23   & 13.12     & B     & 2 (180\,s)      & 18.8\\
MASTER-IAC          & 2016-03-01 22:14:23   & 13.12     & I     & 2 (180\,s)      & 17.2\\
MASTER-Tunka        & 2016-03-02 13:41:01   & $13.76$   &       & 12 (60\,s)      & 18.4\\
MASTER-Kislovodsk   & 2016-03-02 16:40:35   & $13.89$   &       & 10 (180\,s)     & 19.6 (19.0)\\
MASTER-Kislovodsk   & 2016-03-03 17:04:55   & $14.90$   &       & 6 (180\,s)      & 17.6 (17.2)\\
MASTER-IAC          & 2016-03-03 20:11:40   & $15.03$   &       & 3$\times$2 (180\,s)    & 20.2 (19.7)\\
MASTER-IAC          & 2016-03-03 20:20:15   & 15.04     & B     & 2 (180\,s)      & 19.4\\
MASTER-IAC          & 2016-03-03 20:20:15   & 15.04     & I     & 2 (180\,s)      & 17.8\\
MASTER-Kislovodsk   & 2016-03-04 16:20:27   & 15.87     &       & 6 (180\,s)      & 18.2\\
MASTER-IAC          & 2016-03-04 20:41:12   & 16.06     &       & 12$\times$2 (180\,s)   & 20.2 (19.3)\\
MASTER-Tunka        & 2016-03-06 12:24:08   & 17.71     &       & 8 (60\,s)       & 18.8\\
MASTER-Tunka        & 2016-03-07 12:18:37   & 18.71     &       & 12 (60-180s)  & 20.0 (19.3)\\
MASTER-IAC          & 2016-03-07 21:44:32   & 19.09     &       & 3$\times$2 (180\,s)    & 19.4 (18.7)\\
MASTER-Tunka        & 2016-03-08 12:17:08   & 19.71     &       & 6 (180\,s)      & 18.5\\
MASTER-Kislovodsk   & 2016-03-08 17:19:59   & 19.92     &       & 6 (60\,s)       & 19.1\\
MASTER-IAC          & 2016-03-08 20:15:08   & 20.04     &       & 3$\times$2 (180\,s)    & 20.3 (19.6)\\
MASTER-Tunka        & 2016-03-09 12:18:41   & 20.71     &       & 6 (180\,s)      & 20.0 (19.3)\\
MASTER-IAC          & 2016-03-09 20:13:47   & 21.04     &       & 3$\times$2 (180\,s)    & 20.2 (19.6)\\
MASTER-Tunka        & 2016-03-10 13:49:52   & 21.77     &       & 6 (180\,s)      & 19.5 (19.0)\\
MASTER-Kislovodsk   & 2016-03-10 17:57:18   & 21.94     &       & 10 (60\,s)      & 19.1\\
MASTER-IAC          & 2016-03-10 20:16:12   & 22.03     &       & 4$\times$2 (180\,s)    & 20.3 (19.6)\\
MASTER-IAC          & 2016-03-11 20:11:23   & 23.04     &       & 4$\times$2 (180\,s)    & 19.9 (19.2)\\
MASTER-Tunka        & 2016-03-13 13:39:33   & 24.76     &       & 3 (180\,s)      & 18.8\\
MASTER-IAC          & 2016-03-13 20:18:08   & 25.04     &       & 3$\times$2 (180\,s)    & 20.3 (19.5)\\
MASTER-Tunka        & 2016-03-15 13:41:19   & 26.76     &       & 6 (180\,s)      & 19.0 (18.5)\\
MASTER-IAC          & 2016-03-17 20:31:50   & 29.05     &       & 3$\times$2 (180\,s)    & 19.0 (18.6)\\
MASTER-IAC          & 2016-03-18 20:31:42   & 30.05     &       & 4$\times$2 (180\,s)    & 19.6 (19.0)\\
MASTER-IAC          & 2016-03-19 20:35:02   & 31.05     &       & 3$\times$2 (180\,s)    & 19.6 (18.7)\\
MASTER-IAC          & 2016-03-21 20:30:07   & 33.05     &       & 3$\times$2 (180\,s)    & 18.2\\
\hline                                  
\end{tabular}
\end{center}
}
\tablefoot{The columns list the telescope, the start time of the observation, the time relative to the neutrino alert, the band if a filter was used, the number of exposures, the time per exposure and a typical limiting magnitude. A white filter was used for most MASTER observations. The factor $\times$2 indicates that both tubes of the MASTER twin telescopes pointed at the same position. The limiting magnitudes are for co-added images and the limits for individual images are given in parentheses. All limits correspond to the $5\,\sigma$ level.} 
\hfill{}
\end{table*}

\begin{table*}
{\small
\hfill{}
\caption{Optical observations by LCO}
\label{tab:optical_obs_lcogt}     
\begin{center}
\begin{tabular}{llllllll}     
\hline\hline       
RA          & Dec       & Obs. date and UTC     & Time$-t_0$ & Filter     & Exposure     & Airmass     & $5\,\sigma$ limiting mag.\\ 
($^\circ$)  & ($^\circ$)   &                       & (days) &            & (s     )     &             & (mag)\\ 
\hline
26.46854        &       39.48407        &                       2016-02-19      01:53:36 AM      &       1.272           &       g       &       200     &       1.27973 &       21.11\\                         
26.46854        &       39.48411        &                       2016-02-19      01:57:54 AM      &       1.275           &       r       &       120     &       1.29248 &       20.58\\                         
25.58188        &       39.48409        &                       2016-02-19      02:03:01 AM      &       1.279           &       g       &       200     &       1.32705 &       21.05\\                         
25.58188        &       39.48408        &                       2016-02-19      02:07:25 AM      &       1.282           &       r       &       120     &       1.34209 &       20.64\\                         
25.58188        &       39.4841 &                       2016-02-19      02:10:40 AM      &       1.284           &       i       &       120     &       1.35507 &       20.31\\                         
26.02522        &       39.4841 &                       2016-02-19      02:14:10 AM      &       1.287           &       U       &       300     &       1.369                                   &       NULL\\
26.02521        &       39.48409        &                       2016-02-19      02:26:07 AM      &       1.295           &       B       &       200     &       1.41944 &       21.04\\                         
26.02521        &       39.48409        &                       2016-02-19      02:30:09 AM      &       1.298           &       B       &       200     &       1.43915 &       21.03\\                         
26.02521        &       39.48408        &                       2016-02-19      02:34:26 AM      &       1.301           &       V       &       120     &       1.45737 &       20.66\\                         
26.02522        &       39.48409        &                       2016-02-19      02:37:13 AM      &       1.303           &       V       &       120     &       1.4718  &       20.72\\                         
26.02521        &       39.48407        &                       2016-02-19      02:40:16 AM      &       1.305           &       g       &       200     &       1.49206 &       21.01\\                         
26.02521        &       39.48409        &                       2016-02-19      02:44:17 AM      &       1.307           &       g       &       200     &       1.51466 &       20.99\\                         
26.02522        &       39.48409        &                       2016-02-19      02:48:31 AM      &       1.310           &       r       &       120     &       1.53538 &       20.46\\                         
26.02521        &       39.48409        &                       2016-02-19      02:51:13 AM      &       1.312           &       r       &       120     &       1.55162 &       20.53\\                         
26.02521        &       39.48407        &                       2016-02-19      02:54:13 AM      &       1.314           &       i       &       120     &       1.57035 &       20.21\\                         
26.0252 &       39.48408        &                       2016-02-19      02:56:57 AM      &       1.316           &       i       &       120     &       1.58775 &       20.14\\                         
26.0252 &       39.04076        &                       2016-02-19      03:00:18 AM      &       1.319           &       g       &       200     &       1.61753 &       20.86\\                         
26.02521        &       39.04075        &                       2016-02-19      03:04:34 AM      &       1.322           &       r       &       120     &       1.64238 &       20.40\\                         
26.0252 &       39.04074        &                       2016-02-19      03:07:37 AM      &       1.324           &       i       &       120     &       1.66441 &       19.91\\                         
26.46856        &       39.4841 &                       2016-02-19      03:12:49 AM      &       1.327           &       g       &       200     &       1.69135 &       20.45\\                         
26.46855        &       39.48409        &                       2016-02-19      03:17:07 AM      &       1.330           &       r       &       120     &       1.71916 &       20.15\\                         
26.46853        &       39.4841 &                       2016-02-19      03:20:52 AM      &       1.333           &       i       &       120     &       1.74817 &       19.47\\                         
26.02521        &       39.48407        &                       2016-03-01      02:01:02 AM      &       12.277          &       B       &       200     &       1.51474 &       21.85\\                         
26.0252 &       39.48408        &                       2016-03-01      02:05:04 AM      &       12.280          &       B       &       200     &       1.53826 &       21.94\\                         
26.02522        &       39.48413        &                       2016-03-01      02:09:23 AM      &       12.283          &       V       &       120     &       1.56018 &       21.52\\                         
26.02521        &       39.48418        &                       2016-03-01      02:12:04 AM      &       12.285          &       V       &       120     &       1.57727 &       21.56\\                         
26.0252 &       39.48415        &                       2016-03-01      02:15:02 AM      &       12.287          &       g       &       200     &       1.60098 &       22.06\\                         
26.02521        &       39.48412        &                       2016-03-01      02:19:03 AM      &       12.290          &       g       &       200     &       1.62804 &       22.29\\                         
26.02522        &       39.48428        &                       2016-03-01      02:23:17 AM      &       12.293          &       r       &       120     &       1.65314 &       21.43\\                         
26.02521        &       39.48412        &                       2016-03-01      02:25:58 AM      &       12.295          &       r       &       120     &       1.67264 &       21.52\\                         
26.0252 &       39.92745        &                       2016-03-03      01:57:31 AM      &       14.275          &       B       &       200     &       1.53699 &       20.49\\                         
26.02522        &       39.92748        &                       2016-03-03      02:01:35 AM      &       14.278          &       B       &       200     &       1.56253 &       21.11\\                         
26.02521        &       39.92747        &                       2016-03-03      02:06:19 AM      &       14.281          &       V       &       120     &       1.58814 &       21.03\\                         
26.0252 &       39.92742        &                       2016-03-03      02:10:00 AM      &       14.284          &       V       &       120     &       1.6115  &       20.71\\                         
26.02522        &       39.92741        &                       2016-03-03      02:13:04 AM      &       14.286          &       g       &       200     &       1.63762 &       21.93\\                         
26.0252 &       39.9274 &                       2016-03-03      02:17:06 AM      &       14.289          &       g       &       200     &       1.66663 &       21.77\\                         
26.0252 &       39.92746        &                       2016-03-03      02:21:22 AM      &       14.292          &       r       &       120     &       1.69315 &       20.60\\                         
26.0252 &       39.9274 &                       2016-03-03      02:24:03 AM      &       14.293          &       r       &       120     &       1.71387 &       20.89\\                         
26.0252 &       39.92744        &                       2016-03-03      02:26:58 AM      &       14.295          &       i       &       120     &       1.73687 &       20.52\\                         
\hline                                  
\end{tabular}
\end{center}
}
\tablefoot{The limiting magnitudes correspond to the images without running a discovery pipeline and so apply to a source at a known location. The limiting magnitude could not be calculated for the U band because not enough stars are detected in the image to calibrate it.}
\hfill{}
\end{table*}

\begin{table}
{\small
\hfill{}
\caption{XRT upper limits.}
\label{tab:xrtUL}     
\begin{center}
\begin{tabular}{llll}        
\hline\hline       
E$_{\mathrm{min}}$ & E$_{\mathrm{max}}$ & Flux upper limit AGN & Flux upper limit GRB \\
(keV) & (keV) & (erg cm$^{-2}$ s$^{-1}$) & (erg cm$^{-2}$ s$^{-1}$)\\
\hline
0.3 & 1 & (2.67--4.83) $\times$10$^{-13}$ & (2.53--4.56) $\times$10$^{-13}$ \\
1   & 2 & (2.55--4.61) $\times$10$^{-13}$ & (2.58--4.65) $\times$10$^{-13}$ \\
2   & 10 & (1.00--1.80) $\times$10$^{-12}$ & (0.92--1.67) $\times$10$^{-12}$ \\
0.3 & 10 & (6.28--8.92) $\times$10$^{-13}$ & (6.56--9.32) $\times$10$^{-13}$ \\
\hline                                  
\end{tabular}
\end{center}
}
\tablefoot{All values are in erg cm$^{-2}$ s$^{-1}$ in the specified band. The upper limits are at $3\,\sigma$ confidence level.}
\hfill{}
\end{table}

\begin{table}
{\small
\hfill{}
\caption{VERITAS flux upper limits}
\label{tab:veritas}     
\begin{center}
\begin{tabular}{lll}        
\hline\hline       
E$_{\mathrm{min}}$ & E$_{\mathrm{max}}$ & Flux upper limit \\
(TeV) & (TeV) & (cm$^{-2}$ s$^{-1}$ TeV$^{-1}$) \\
\hline
0.316 & 0.501  & $8.0 \times 10^{-11}$ \\
0.501 & 0.794  & $2.3 \times 10^{-11}$ \\
0.794 & 1.259  & $1.5 \times 10^{-12}$ \\
1.259 & 1.995  & $5.7 \times 10^{-13}$ \\
\hline                                  
\end{tabular}
\end{center}
}
\tablefoot{Differential flux upper limits for a gamma-ray point-source located at the averaged triplet position. The limits are at 95\% confidence level and do not depend on the spectral shape.}
\hfill{}
\end{table}

\begin{table*}
{\small
\hfill{}
\caption{HAWC flux upper limits}
\label{tab:hawc}     
\begin{center}
\begin{tabular}{lllll}        
\hline\hline       
E$_{\mathrm{min}}$ & E$_{\mathrm{max}}$ & Upper limit 1 transit & Upper limit 11 transits & Upper limit 508 transits\\
(TeV) & (TeV) & (cm$^{-2}$ s$^{-1}$ TeV) & (cm$^{-2}$ s$^{-1}$ TeV) & (cm$^{-2}$ s$^{-1}$ TeV) \\
\hline
    0.5 &      1.7 &           $8.50 \times 10^{-11}$ & $3.86 \times 10^{-11}$ & $3.57 \times 10^{-12}$\\
    1.7 &      5.3 &           $3.31 \times 10^{-11}$ & $1.45 \times 10^{-11}$ & $1.03 \times 10^{-12}$\\
    5.3 &     16.7 &           $1.45 \times 10^{-11}$ & $6.93 \times 10^{-12}$ & $5.81 \times 10^{-13}$\\
   16.7 &     52.9 &           $7.82 \times 10^{-12}$ & $4.68 \times 10^{-12}$ & $2.16 \times 10^{-13}$\\
   52.9 &    167.2 &           $6.61 \times 10^{-12}$ & $4.20 \times 10^{-12}$ & $1.15 \times 10^{-13}$\\
\hline                                  
\end{tabular}
\end{center}
}
\tablefoot{Flux upper limits at the 95\% confidence level are calculated for the night of the transient during which the neutrino alert was detected (third column), using all data recorded within 14\,days after the alert (fourth column) as well as using all recorded data (last column).\\ \\ }
\hfill{}
\end{table*}

\end{appendix}

\end{document}